\documentclass{aa}
\usepackage{amsmath,amssymb}
\usepackage[varg]{txfonts}
\usepackage{graphicx}
\usepackage{natbib}
\bibpunct{(}{)}{;}{a}{}{,}
\usepackage{rotating}
\usepackage{xspace}
\usepackage[utf8]{inputenc} 
\usepackage{wasysym}
\usepackage{textcomp}
\usepackage{hyperref}

\graphicspath{{figures/}}

\newcommand{\XMMNewton}{\textsl{XMM-Newton}\xspace}

\newcommand{\Swift}{\textsl{Swift}\xspace}

\newcommand{\Fermi}{\textsl{Fermi}\xspace}

\newcommand{\Planck}{\textsl{Planck}\xspace}

\newcommand{\ISIS}{\textsc{ISIS}\xspace}

\usepackage{pdflscape}
\makeatletter
\ifcase \@ptsize \relax
  \newcommand{\miniscule}{\@setfontsize\miniscule{6}{7}}
\or
  \newcommand{\miniscule}{\@setfontsize\miniscule{5}{6}}
\or
  \newcommand{\miniscule}{\@setfontsize\miniscule{5}{6}}
\fi
\makeatother

\title{The TANAMI Multiwavelength Program: Dynamic SEDs of Southern Blazars}
\author{
  F. Krau\ss{}\inst{\ref{affil:remeis},\ref{affil:wuerzburg}}
  \and J.~Wilms\inst{\ref{affil:remeis}}
  \and M.~Kadler\inst{\ref{affil:wuerzburg}}
  \and R.~Ojha\inst{\ref{affil:nasa_gsfc}, \ref{affil:umd}, \ref{affil:cua}}
  \and R.~Schulz\inst{\ref{affil:remeis},\ref{affil:wuerzburg},\ref{affil:astron}}
  \and J.~Tr\"ustedt\inst{\ref{affil:remeis},\ref{affil:wuerzburg}}
  \and P.G.~Edwards\inst{\ref{affil:csiro}}
  \and J.~Stevens\inst{\ref{affil:csiro}}
  \and E.~Ros\inst{ \ref{affil:mpe},\ref{affil:val},\ref{affil:valobs}}
  \and W.~Baumgartner\inst{\ref{affil:nasa_gsfc}}
  \and T.~Beuchert\inst{\ref{affil:remeis},\ref{affil:wuerzburg}}
  \and J.~Blanchard\inst{\ref{affil:alma}}
  \and S.~Buson\inst{\ref{affil:nasa_gsfc}}
  \and B.~Carpenter\inst{\ref{affil:cua}}
  \and T.~Dauser\inst{\ref{affil:remeis}}
  \and S.~Falkner\inst{\ref{affil:remeis}}
  \and N.~Gehrels\inst{\ref{affil:nasa_gsfc}}
  \and C.~Gr\"afe\inst{\ref{affil:remeis},\ref{affil:wuerzburg}}
  \and S.~Gulyaev\inst{\ref{affil:nz}}
  \and H.~Hase\inst{\ref{affil:kart}}
  \and S.~Horiuchi\inst{\ref{affil:csiro2}}
  \and A.~Kreikenbohm\inst{\ref{affil:remeis},\ref{affil:wuerzburg}}
  \and I.~Kreykenbohm\inst{\ref{affil:remeis}}
  \and M.~Langejahn\inst{\ref{affil:remeis},\ref{affil:wuerzburg}}
  \and K.~Leiter\inst{\ref{affil:remeis},\ref{affil:wuerzburg}}
  \and J.E.J.~Lovell\inst{\ref{affil:tasman}}
  \and C.~M\"uller\inst{\ref{affil:nijmegen}}
  \and T.~Natusch\inst{\ref{affil:nz}}
  \and R.~Nesci\inst{\ref{affil:inaf}}
  \and T.~Pursimo\inst{\ref{affil:not}}
  \and C.~Phillips\inst{\ref{affil:csiro}}
  \and C.~Pl\"otz\inst{\ref{affil:kart}}
  \and J. Quick\inst{\ref{affil:hart}}
  \and A.K.~Tzioumis\inst{\ref{affil:csiro}}
  \and S.~Weston\inst{\ref{affil:nz}}
}

\institute{
  Dr.~Remeis Sternwarte \& ECAP, Universit\"at Erlangen-N\"urnberg,
  Sternwartstrasse 7, 96049 Bamberg, Germany\\
  \email{Felicia.Krauss@fau.de} 
  \label{affil:remeis} 
  \and
  Institut f\"ur Theoretische Physik und Astrophysik, Universit\"at
  W\"urzburg, Emil-Fischer-Str.\ 31, 97074 W\"urzburg, Germany
  \label{affil:wuerzburg}
  \and
  NASA, Goddard Space Flight Center, Greenbelt, MD 20771,
  USA  \label{affil:nasa_gsfc}
  \and
  University of Maryland, Baltimore County, Baltimore, MD 21250,
  USA 
  \label{affil:umd} 
  \and
  Catholic University of America, Washington, DC 20064,
  USA \label{affil:cua}
  \and
  ASTRON, the Netherlands Institute for Radio Astronomy, PO Box 2, 7990 AA
  Dwingeloo, Netherlands
  \label{affil:astron}
  \and
  CSIRO Astronomy and Space Science, ATNF, PO Box 76, Epping,
  NSW 1710, Australia
  \label{affil:csiro}
  \and
  Max-Planck-Institut f\"ur Radioastronomie, Auf dem Hügel 69, 53121
  Bonn, Germany 
  \label{affil:mpe}
  \and
  Departament d'Astronomia i Astrof\'isica, Universitat de Val\`encia,
  C/ Dr. Moliner 50,
  46100 Burjassot, Val\`encia, Spain
  \label{affil:val}
  \and
  Observatori Astron\`omic, Universitat de Val\`encia, C/ Catedr\'atico Jos\'e Beltr\'an no.~2,
  46980 Paterna, Val\`encia, Spain
  \label{affil:valobs}
  \and Departamento de Astronom\'ia, Universidad de Concepci\'on,
  Casilla 160, Chile
  \label{affil:alma}
   \and
   Institute for Radio Astronomy and Space Research, Auckland University
   of Technology, Auckland 1010, New Zealand
   \label{affil:nz} 
   \and
   Bundesamt f\"ur Kartographie und Geod\"asie, 93444 Bad K\"otzting,
   Germany
   \label{affil:kart}
   \and
   CSIRO Astronomy and Space Science, Canberra Deep Space
   Communications Complex, P.O.\ Box 1035, Tuggeranong, ACT
   2901, Australia
   \label{affil:csiro2}
  \and
   School of Mathematics \& Physics, University of Tasmania, Private
   Bag 37, Hobart, Tasmania 7001, Australia
   \label{affil:tasman}
   \and
   Department ofAstrophysics/IMAPP, Radboud University Nijmegen,
   Heyendaalseweg 135, 6525 AJ Nijmegen, Netherlands
   \label{affil:nijmegen}
   \and
   INAF–IAPS, via Fosso del Cavaliere 100, 00033 Rome, Italy
   \label{affil:inaf}
  \and
   Nordic Optical Telescope, Apartado 474, 38700, Santa Cruz de La Palma,
   Spain
   \label{affil:not}
   \and
   Hartebeesthoek Radio Astronomy Observatory, Krugersdorp, South
   Africa
   \label{affil:hart}
}

\authorrunning{F. Krau{\ss}~et al.}
\titlerunning{Dynamic broadband SEDs of southern blazars}
\date{Received $<$date$>$ / Accepted $<$date$>$}

\abstract
    {Simultaneous broadband spectral and temporal studies of blazars are an
      important tool for investigating active galactic nuclei (AGN)
      jet physics.}
    {We study the spectral evolution between quiescent and
      flaring periods of 22 radio-loud AGN through
      multi-epoch, quasi-simultaneous broadband spectra. For
      many of these sources these are the first broadband studies.} 
    {We use a Bayesian block analysis of \Fermi/LAT light curves in
      order to determine time ranges of constant flux for constructing
      quasi-simultaneous SEDs. The shapes of the resulting 81 SEDs are
      described by two logarithmic parabolas and a blackbody spectrum
      where needed.}
    { For low states the peak frequencies and
      luminosities agree well with the blazar sequence, higher
      luminosity implying lower peak frequencies. This is not
      true for sources in a high state. 
      The $\gamma$-ray photon index in
      \Fermi/LAT correlates with the synchrotron peak
      frequency in low and intermediate states. No correlation is
      present in high states. The black hole mass cannot
      be determined from the SEDs. 
      Surprisingly, the thermal
      excess often found in FSRQs at optical/UV wavelengths can be
      described by blackbody emission  and not an accretion disk spectrum.}
    {The
      ``harder-when-brighter'' trend, typically seen in X-ray spectra of
      flaring blazars, is visible in the blazar sequence. Our results for
      low and intermediate states, as well as the Compton dominance, are in
      agreement with previous results. Black hole
      mass estimates using the parameters from Bonchi (2013) are
      in agreement with some of the more direct measurements. For two
      sources, estimates disagree by more than four orders of
      magnitude, possibly due to boosting effects. The shapes of the
      thermal excess seen predominantly in flat spectrum radio quasars
      are inconsistent with a direct accretion disk origin. 
    } 


\begin{document}
\maketitle
\section{Introduction}\label{sec-intro}
Active galactic nuclei (AGN) are supermassive black holes at the
center of galaxies that are thought to be powered by accretion
\citep[e.g., ][]{Antonucci1993,Urry1995,Abdo2010}. Radio-loud AGN
typically exhibit relativistic outflows of matter, called jets. 
Blazars constitute an ideal target for multiwavelength studies in
order to understand their acceleration mechanisms and their role as
potential cosmic-ray emitters.
Blazars are a subclass of radio-loud AGN,
with their jet oriented at a small angle to the line of sight
\citep{Blandford1978}. They emit non-thermal radiation across the
whole electromagnetic spectrum \citep{Urry1995} and show strong
variability. 
Since the possible relationships between their variability in
different bands is unclear, quasi-simultaneous observations are
required for such studies. The radio to $\gamma$-ray
spectral energy distributions (SEDs) of these sources
generally show two peaks in a log $\nu$ -- log $\nu F_{\nu}$
representation. The lower energy peak is generally attributed to
synchrotron radiation from
relativistic electrons in the magnetic field of the jet
\citep[see][for a review]{Ghisellini2013}.
While both leptonic and hadronic processes likely contribute to the
high energy peak, their relative contributions remain a deeply
interesting open question
\citep{Abdo2011,Boettcher2013,Mannheim1992,Finke2008,Sikora2009,
  Balokovic2015,Weidinger2015}.
In the leptonic scenario the relativistic electrons that produce the
synchrotron emission are assumed to up-scatter the photons to high
energies. This process is called Synchrotron-Self Compton (SSC). Seed
photons from the ambient medium can also contribute by being
upscattered to $\gamma$-ray energies \citep{sikora:94a}, this consitutes
the External Compton (EC) 
contribution. In the hadronic scenario \citep[e.g.,][]{Mannheim1993},
protons and electrons are assumed to be accelerated in the jet.
Protons interacting with a UV seed photon field (e.g. thermal emission
from the accretion disk) produce pions. Neutral pions decay into
high-energy $\gamma$-rays, explaining the high-energy emission.

Based on their optical emission lines, blazars can be subdivided into
flat-spectrum radio quasars (FSRQs) and BL Lacertae (BL Lac) objects.
FSRQs show broad emission lines (rest-frame equivalent width $>5$\AA),
while BL Lacs typically show none. A few well known exceptions include
OJ\,287 \citep{Sitko1985} and BL Lac \citep{Vermeulen1995}. Blazars can also be
categorized by their synchrotron peak frequency into low, intermediate
and high synchrotron peaked blazars \citep[LSP, ISP, HSP;
][]{Padovani1995,Abdo2010}, with the ISP blazar peak located between
$10^{14}$\,Hz and $10^{15}$\,Hz. FSRQs often exhibit a thermal excess
in the optical-UV range with a temperature of $\sim$30000\,K
\citep{Sanders1989,Elvis1994}. This peak, called the ``big blue
bump'' (BBB), is described as a broad peak, as expected from an
accretion disk with a wide range of temperatures
\citep{Shields1978,Malkan1982}. Its origin is disputed
\citep{Antonucci2002}. Some authors argue for it to stem from the
accretion disk \citep{Shields1978,Malkan1982}, alternatively, free-free
emission has been proposed \citep{Barvainis1993}. The observed
temperature of the feature, however, is lower than what is expected
from an accretion disk \citep{Zheng1997,Telfer2002,Binette2005}.
The origin of the BBB could be reprocessed accretion disk emission
from clouds near the broad line region (BLR) \citep{Lawrence2012}.

While it is generally recognized that the best way to study blazars
is from (near-)simultaneous broadband data
\citep{Giommi1995,vonMontigny1995,Sambruna1996,Fossati1998,Giommi2002,
  Nieppola2006,Padovani2006,Giommi2012}, the lack of available
simultaneous data often forces the use of time-averaged data. In
non-simultaneous SEDs, physical models can only be poorly constrained.

In addition, elevated levels of flux in the optical/UV, \Fermi/LAT, or
Very High Energy (VHE) instruments -- called ``flares'' or
``high states'' -- often trigger follow-up multiwavelength
observations, which lead to the availability of large amounts of
quasi-simultaneous data, with a paucity of comparable data in a
quiescent state.
VHE instruments generally have trouble detecting fainter sources
(particularly FSRQs) in quiescent states. An exception is the large
campaign on the low state of 1ES\,2344+514 \citep{Aleksic2013}.
Other campaigns involving a large number of instruments are only
available for few bright sources, such as 3C\,454.3
\citep{Giommi2006,Abdo2009,Vercellone2009,Pacciani2010}, Mrk 421
\citep{Bla2005,Donnarumma2009,Abdo2011,Bartoli2015,Balokovic2015}, Mrk 501
\citep{Bartoli2012,Aleksic2015,Furniss2015}, 3C279
\citep{Grandi1996,Wehrle1998,Hayashida2015,Paliya2015}, BL Lac
\citep{Abdo2011bllac,Wehrle2016}, S5\,0716+714
\citep{Rani2013,Liao2014,Chandra2015}, and PKS\,2155$-$304
\citep{Aharonian2009,Abdo2010}.

In this study we used data from the TANAMI multiwavelength project
\citep{Kadler2015} to construct quasi-simultaneous broadband SEDs for
high energy (HE) $\gamma$-ray bright southern blazars. These SEDs
include several epochs at different flux levels and have good spectral
coverage. 
We selected the 22 TANAMI blazars that were brightest in the \Fermi/LAT
band and constructed a total of 81 SEDs with good coverage across the
entire spectrum.
For several sources we obtained SEDs in low, intermediate, and
high states. We used this large sample of SEDs to study the spectral
evolution over time, the blazar sequence, Compton dominance,
fundamental plane of black holes, and the Big Blue Bump.

The paper is structured as follows. In
Sect.~\ref{sec-broadband} we introduce the sample used and its
limitations and describe the multiwavelength data and their extraction
and analysis. We also include the method of constructing the broadband
SEDs, how systematic uncertainties are treated and caveats of our
method. In Sect.~\ref{sec-results} we present the results from the
broadband fits including results pertaining to the blazar sequence,
the Compton dominance, the thermal excess, and the fundamental plane of
black holes. We summarize and discuss the results in
Sect.~\ref{sec-summary}.

Throughout the paper we use the standard cosmological model with
$\Omega_m=0.3$, $\Lambda=0.7$, $H_0 = 70$\,km\,s$^{-1}$\,Mpc$^{-1}$
\citep{Beringer2012}. 

\section{Generation of contemporaneous broadband Spectral Energy
  Distributions}\label{sec-broadband}
\subsection{Sample selection}\label{sec-sample}
The Tracking Active Galactic Nuclei with Austral Milliarcsecond
Interferometry
(TANAMI)\footnote{\url{http://pulsar.sternwarte.uni-erlangen.de/tanami/}}
\citep{Ojha2010} sample includes $\sim$100 AGN in the southern sky, at
declinations below $-30^\circ$. It is a flux-limited sample,
covering southern flat spectrum sources with catalogued flux
densities above 1 Jy at 5\,GHz, as well as \Fermi detected
$\gamma$-ray loud blazars in the region of interest. These sources are
monitored by TANAMI with Very Long Baseline Interferometry (VLBI) at
8.4\,GHz and 22\,GHz (X-band and K-band, respectively). In addition
to the VLBI monitoring, single dish observations are performed at
several additional radio frequencies with the ATCA and Ceduna. These
radio observations
are complemented with multiwavelength observations, primarily with
\Swift and \XMMNewton in the X-rays, and the Rapid Eye Mount (REM)
telescope at La Silla in the optical. The TANAMI sample is regularly
extended by adding bright sources newly detected by \Fermi/LAT
\citep{Boeck2016}. 

Due to the good coverage in wavelength and time, the TANAMI sample is
ideal for a study of the behavior of blazar SEDs. Previous studies
include detailed studies of the blazars 2142$-$758 \citep{Dutka2013},
0208$-$512 \citep{blanchard_phd}, and PKS 2326-502 (Dutka, ApJ,
submitted).

In this paper we study the multiwavelength evolution of the 22
$\gamma$-ray brightest TANAMI sources according to the 3FGL catalog
\citep{fgl3}. Our results are therefore representative of a
$\gamma$-ray flux-limited sample. The 22 sources are listed in
Table~\ref{tab-srcs}. We include the IAU B1950 name, the 3FGL
association, the 3FGL catalog name, the source classification that we
used, the
redshift, right ascension and declination, the Galactic absorbing
column in the direction of the source, and finally the number of SEDs
that we were able to construct for each of the sources. Our sample
includes 9 BL Lac type objects, 11 FSRQs, and 2 blazars of unknown
type. The brightness of these sources enabled us to extract \Fermi/LAT
light curves with 14-day binning. For some of these sources, these
are their first broadband SEDs in the literature. While an optical
classification of most sources is relatively easy, some sources have
contradictory classifications in different AGN catalogs. These are
labeled as blazar candidates of
unknown type (BCU). One example is 0208$-$512. In the CGRaBS survey of
bright blazars \citep{Healey2008}, this source was listed as a BL Lac
type object, in
agreement with the optical classification from the 12th catalog of
quasars and active nuclei \citep{VeronCetty2006}.
It was classified as a FSRQ however, based on optical emisison lines
by \citet{Impey1988}.
The 5th Roma
BZCAT lists the source as a BZU (blazar of uncertain or transitional
type), and describes it as a transition object, but lists it as an
FSRQ \citep{Massaro2009}.
Note that
possible misclassifications did not change any of our results, as we
generally did not treat the two populations differently and find many
of the results are not dependent on the source classification.
For 0332$-$403 a redshift of 1.45 is often used
\citep{Hewitt1987}, but \cite{Shen1998} point out that the origin of
this value is unknown. 
It is further worth noting that 0521$-$365 is often not considered a
blazar, but a transitional object between a broad line radio galaxy
and a steep spectrum radio quasar with a VLBI morphology similar to a
misaligned blazar \citep{DAmmando2015}.

\begin{table*}
 \caption{Sources used in the SED catalog}\label{tab-srcs} 
\begin{tabular}{clllllllcl} 
No & Source & Catalog & 3FGL & Class. & $z$ & $\alpha$ & $\delta$ & $N_{\mathrm{H}}$& \# SEDs \\ 
 &  &  &  &  & & [J2000] & [J2000]  & [$10^{20}\,\mathrm{cm}^{-2}$] & \\\hline 
1 & 0208$-$512 & PKS\,0208$-$512 & J0210.7$-$5101 & BCU $^\mathrm{a}$  &
0.999$^\mathrm{b}$ & 32.6925$^\mathrm{c}$ &
$-$51.0172$^\mathrm{c}$ & 1.84 & 8\\
2 & 0244$-$470 & PKS\,0244$-$470 & J0245.9$-$4651 & FSRQ$^\mathrm{d}$ & 1.385$^\mathrm{e}$ & 41.5005$^\mathrm{d}$ & $-$46.8548$^\mathrm{d}$ & 1.89 &
2\\ 
3 & 0332$-$376 & PMN\,J0334$-$3725 & J0334.3$-$3726 & BL Lac$^\mathrm{d}$ & ? & 53.5642$^\mathrm{a}$ & $-$37.4287$^\mathrm{a}$ & 1.54 &
2\\ 
4 & 0332$-$403 & PKS 0332$-$403 & J0334.3$-$4008 & BL Lac$^\mathrm{d}$ & ? & 53.5569$^\mathrm{c}$ & $-$40.1404$^\mathrm{c}$ & 1.48
& 4\\ 
5 & 0402$-$362 & PKS\,0402$-$362 & J0403.9$-$3604 & FSRQ$^\mathrm{d}$ & 1.423$^\mathrm{f}$ & 60.9740$^\mathrm{c}$ & $-$36.0839$^\mathrm{c}$ & 0.60 &
2\\ 
6 & 0426$-$380 & PKS\,0426$-$380 & J0428.6$-$3756 & BL Lac$^\mathrm{d}$ & 1.111$^\mathrm{g}$ &
67.1684$^\mathrm{c}$ & $-$37.9388$^\mathrm{c}$ & 2.09 & 5\\ 
7 & 0447$-$439 &  PKS\,0447$-$439 & J0449.4-4350 & BL Lac$^\mathrm{h}$ & 0.107$^\mathrm{i}$ &
72.3529$^\mathrm{a}$ & $-$43.8358$^\mathrm{a}$ & 1.24 &
3\\ 
8 & 0506$-$612 & PKS\,0506$-$61 & J0507.1$-$6102 & FSRQ$^\mathrm{h}$ & 1.093$^\mathrm{j}$ &
76.6833$^\mathrm{c}$ & $-$61.1614$^\mathrm{c}$ & 1.95 & 4\\ 
9 & 0521$-$365 & PKS\,0521$-$36 & J0522.9$-$3628 & BCU & 0.055$^\mathrm{f}$
& 80.7416$^\mathrm{c}$ & $-$36.4586$^\mathrm{c}$ & 3.58 &
6\\ 
10 & 0537$-$441 & PKS\,0537$-$441 & J0538.8$-$4405 & BL Lac$^\mathrm{d}$  &
0.892$^\mathrm{k}$ & 84.7098$^\mathrm{l}$ & $-$44.0858$^\mathrm{l}$ & 3.14
& 6\\ 
11 & 0637$-$752 & PKS\,0637$-$75 & J0635.7$-$7517 & FSRQ$^\mathrm{d}$ & 0.651$^\mathrm{m}$ & 98.9438$^\mathrm{c}$ & $-$75.2713$^\mathrm{c}$ & 7.82 &
4\\ 
12 & 1057$-$797 & PKS\,1057$-$79 & J1058.5$-$8003 & BL Lac$^\mathrm{d}$ & 0.581$^\mathrm{n}$ & 164.6805$^\mathrm{c}$ &
$-$80.0650$^\mathrm{c}$ & 6.34 & 2\\ 
13 & 1424$-$418 & PKS\,B1424$-$418 & J1427.9$-$4206 & FSRQ$^\mathrm{d}$ & 1.522$^\mathrm{o}$ & 216.9846$^\mathrm{c}$ & $-$42.1054$^\mathrm{c}$ & 7.71
& 7\\ 
14 & 1440$-$389 & PKS 1440$-$389 & J1444.0$-$3907 & BL Lac$^\mathrm{d}$ &
0.065$^\mathrm{p}$ & 220.9883$^\mathrm{d}$ &
$-$39.1445$^\mathrm{d}$ & 7.83 & 3\\ 
15 & 1454$-$354 & PKS\,1454$-$354 & J1457.4$-$3539 & FSRQ$^\mathrm{d}$ & 1.424$^\mathrm{q}$ &
224.3613$^\mathrm{r}$ & $-$35.6528$^\mathrm{r}$ & 6.60 &
3\\
16 & 1610$-$771 & PKS\,1610$-$77 & J1617.7$-$7717 & FSRQ$^\mathrm{d}$ & 1.710$^\mathrm{s}$ & 244.4551$^\mathrm{c}$ & $-$77.2885$^\mathrm{c}$ & 6.76
& 2\\
17 & 1954$-$388 & PKS\,1954$-$388 & J1958.0$-$3847 &
FSRQ$^\mathrm{d}$ & 0.630$^\mathrm{t}$ & 299.4992$^\mathrm{u}$ & $-$38.7518$^\mathrm{u}$ & 6.43
& 2\\
18 & 2005$-$489 & PKS\,2005$-$489 & J2009.3$-$4849 & BL Lac$^\mathrm{d}$ &
0.071$^\mathrm{v}$ & 302.3558$^\mathrm{c}$ & $-$48.8316$^\mathrm{c}$ &
3.93 & 2\\ 
19 & 2052$-$474 & PKS\,2052$-$47 & J2056.2$-$4714 & FSRQ$^\mathrm{d}$ &
1.489$^\mathrm{w}$ & 314.0682$^\mathrm{c}$ & $-$47.2465$^\mathrm{c}$ & 2.89 &
2\\
20 & 2142$-$758 & PKS\,2142$-$75 & J2147.3$-$7536 &
FSRQ$^\mathrm{d}$ & 1.139$^\mathrm{w}$ & 326.8030$^\mathrm{c}$ &
$-$75.6037$^\mathrm{c}$ & 7.70 & 2\\ 
21 & 2149$-$306 & PKS\,2149$-$306 & J2151.8$-$3025 & FSRQ$^\mathrm{h}$ & 2.345$^\mathrm{j}$ &
327.9813$^\mathrm{c}$ & $-$30.4649$^\mathrm{c}$ & 1.63 & 4\\ 
22 & 2155$-$304 & PKS\,2155$-$304 & J2158.8$-$3013 & BL Lac$^\mathrm{d}$ & 0.116$^\mathrm{x}$ & 329.7169$^\mathrm{y}$ & $-$30.2256$^\mathrm{y}$ & 1.48 & 6\\ 
\hline
\end{tabular} 
\tablefoot{Columns:
  (1) source number (2) IAU B1950 name, (3) 3FGL association, (4) 3FGL
  catalog name
  \citep{fgl3}, (5) classification, (6) redshift, (7) right
  ascension, (8) declination, (9) absorbing column
  \citep{Kalberla2005,Bajaja2005}, (10) number of SEDs\\
  $^\mathrm{a}$\,\cite{Skrutskie2006},
  $^\mathrm{b}$\,\cite{Wisotzki2000},
  $^\mathrm{c}$\,\cite{Johnston1995},
  $^\mathrm{d}$\,\cite{Healey2007},
  $^\mathrm{e}$\,\cite{Shaw2012},
  $^\mathrm{f}$\,\cite{Jones2009},
  $^\mathrm{g}$\,\cite{Heidt2004},
  $^\mathrm{h}$\,\cite{VeronCetty2006},
  $^\mathrm{i}$\,\cite{Craig1997},
  $^\mathrm{j}$\,\cite{Hewitt1987},
  $^\mathrm{k}$\,\cite{Peterson1976},
  $^\mathrm{l}$\,\cite{MacMillan2002},
  $^\mathrm{m}$\,\cite{Hunstead1978},
  $^\mathrm{n}$\,\cite{Sbarufatti2009},
  $^\mathrm{o}$\,\cite{White1988},
  $^\mathrm{p}$\,\cite{Jones2004},
  $^\mathrm{q}$\,\cite{Jackson2002},
  $^\mathrm{r}$\,\cite{Fey2006},
  $^\mathrm{s}$\,\cite{Hunstead1980},
  $^\mathrm{t}$\,\cite{Browne1975},
  $^\mathrm{u}$\,\cite{Ma1998},
  $^\mathrm{v}$\,\cite{Falomo1987},
  $^\mathrm{w}$\,\cite{Jauncey1984},
  $^\mathrm{x}$\,\cite{Falomo1993}, 
  $^\mathrm{y}$\,\cite{Fey2004}
} 
\end{table*}

Having selected the sources, we generated contemporaneous broadband
spectral energy distributions for observational periods where our
sources were determined to be at a relatively constant level of
$\gamma$-ray activity.
These periods are
determined using a Bayesian blocks analysis of \Fermi/LAT light curves
(Sect.~\ref{sec-blocks}), for which we then searched for contemporaneous
observations in other energy bands (Sect.~\ref{sec-periods}).

\subsection{\Fermi/LAT light curve analysis}\label{sec-blocks}

The lack of simultaneous observation campaigns on most sources means
that we often have to rely on quasi-simultaneous data when assembling
the SED for an AGN. These SEDs will only be representative of the true
SED if the data included are from times where the source emission did
not change appreciably.
With the launch of \Fermi in 2008, we have access to continuous
$\gamma$-ray light curves of blazars, which are ideal for identifying
flux states and applying a criterion to separate the data into time
ranges of similar flux.

We calculated \Fermi/LAT \citep{LAT2009} light curves for the time
period August 4, 2008 through January 1, 2015 using the reprocessed
Pass~7 data (v9r32p5) and the P7REP\_SOURCE\_V15 instrumental response
functions \citep[IRF;][]{Pass7} and a region of interest (ROI) of
$10^\circ$. The data were separated into time bins of 14\,d, on which we
perform a likelihood analysis. The input model is based on point
sources from the 3FGL catalog \citep{fgl3}, and further includes
spatial and spectral templates for the Galactic (gll\_iem\_v05\_rev1)
and isotropic (iso\_source\_v05) diffuse emission.
The first step was to define a criterion for the time ranges. 
A wide variety of methods are used for defining quasi-simultaneity in
multiwavelength studies.
Some studies utilize a flux or count rate threshold \citep{Bla2005},
other methods include fixed time bins
\citep{Giommi2012,Carnerero2015,Tagliaferri2015}, double exponential
forms that are fit to the light curve 
\citep{Valtaoja1999,Abdo2010d,Hayashida2015}, and ``by eye''
definitions \citep{Tanaka2011,DAmmando2013}. These methods are either
model dependent or do not take the amount of variation into account.
A source might show strong, non-discrete variations during a flare,
which are not separated. They are also not useful for studying quiescent
SEDs.

We decided to choose time ranges based on a statistical tool, the
Bayesian blocks algorithm. The Bayesian Block method is
non-parametric, i.e., the data are not described by a model and
evaluated. Local (non-periodic) variability in the light curve is
found with a maximum likelihood approach by determining \textsl{change
  points} where the flux is inconsistent with being at a constant
level \citep{Scargle1998,Scargle2013}. Using an Interactive Spectral
Interpretation System \citep{ISIS2000}
adaption of the code of \citet[ISIS;][M.\,K\"uhnel, available
  online\footnote{\url{http://www.sternwarte.uni-erlangen.de/isis/}}]{Scargle2013}, we found the global optimum division of the light curve into segments
of constant flux. While this assumption of states of constant flux is
in reality not correct, as sources will rarely vary in a discontinuous
way, this approach is still very powerful in identifying time ranges
of source ``states'' where the flux is at least statistically
constant. Here we adapted a significance of the change points at the
95\% confidence level. Such a relatively low value was chosen as we
want to keep the number of false negatives (where real changes in flux
are missed) low. Introducing a low number of false positives, where
constant flux is seen as a change point, however, does not harm our
analysis. If a constant flux is interpreted as a change point, it
segments the data more than necessary. In the worst case this could
lead to two missed broadband spectra (if through the segmentation the
multiwavelength data in either time range is not sufficient for
constructing a broadband SED). Based on the 95\% confidence level, we
estimate that out of the 81 SEDs, only $\sim$4 are based on a
false-positive detection of a change in flux. The \Fermi/LAT light
curves are shown in Appendix~\ref{ap-lc}. The \Fermi/LAT data points
are shown in black, while the segmentation by the Bayesian blocks is
shown in dark gray. The average flux across the whole light curve is
shown in light gray. We additionally show available multiwavelength
data above the light curve at the corresponding times of the
observations. Blocks with a sufficient amount of multiwavelength data
are marked in color and are labeled with Greek letters.

We ensure that the flux at $\gamma$-ray energies is
statistically constant, but no such criterion can be applied to other
wavelengths due to a lack of good cadence observations. It is possible
that variability in the X-ray, optical, or radio band is missed in
\Fermi/LAT and averaged over or completely absent. This effect might
contribute to the problems of broadband fitting. Typically blazar
monitoring has shown that often the largest, and fastest relative
changes in flux occur at high, and very high energy $\gamma$-rays.
Variability in the radio occurs on much longer time scales, consistent
with the outward traveling of material from the base of the jet and
becoming optically thin at different locations.

\subsection{Quasi-simultaneous time periods}\label{sec-periods}
Due to the large uncertainty of individual flux measurements in
fainter AGN, the Bayesian blocks analysis can yield segments longer
than a year during which the $\gamma$-ray flux is found to be
statistically constant. This behavior can hide true variations in 
flux. 
We therefore subdivided Bayesian blocks if they are longer than 1 year
into a new size, depending on its \Fermi/LAT flux in the time range.
The new blocks have a size of at least (2, 5, 10, 25, or 42) $\times$ 14\,d
bins, if the \Fermi/LAT flux in the time range is greater than
$1\times 10^{-6}$, $0.5\times 10^{-6}$, $1\times 10^{-7}$, or $1\times
10^{-8}\,\mathrm{ph}\,\mathrm{s}^{-1}\,\mathrm{cm}^{-2}$,
respectively. This selection of fluxes and time bins accounts for
longer integration time needed for a source with low flux in order to
obtain a \Fermi/LAT spectrum of good quality, and is based on
experience. For a time bin of 370\,d duration with a flux of $2\times 
10^{-7}\,\,\mathrm{ph}\,\mathrm{s}^{-1}\,\mathrm{cm}^{-2}$, for
example, the new time range would be $10\cdot14\,\mathrm{d}=140$\,d.
For this new block size we obtained
$370\,\mathrm{d}/140\,\mathrm{d}=2.64$ new bins, which means that we
subdivided the original interval into $\lfloor 2.64\rfloor=2$ bins with
a length of 185 days each.

Time periods that include $\gamma$-ray, X-ray, optical, and VLBI
observations are then used for quasi-simultaneous SEDs. Earlier works
have shown that the radio flux varies on longer time scales than the
$\gamma$-rays \citep{Soldi2008}. We therefore also included time
periods that have $\gamma$-ray, X-ray, and optical data in the same
block, as well as VLBI observations inside the block, or close to the
block start or end. 
Close to the block is defined as within a time range $
t^\ast_\mathrm{start} = t_\mathrm{block\_start}-c$ and
$t^\ast_\mathrm{stop} = t_\mathrm{block\_stop}+c$ where
$c=\mathrm{max}\{0.6\Delta t,50\,\mathrm{d}\}$ and where $\Delta t$ is
the length of the block. The smaller value, 50\,d, was chosen, as the
radio emission varies on much longer time scales, so even for a very
short block of e.g., 14 days, using radio data 50 days prior to the
start of the block is acceptable. For longer time periods of
quiescence it is acceptable to use VLBI data that is offset from the
start or stop of the block by 60\% of the block length. 60\% is an
arbitrary value, based on the variability time scales of the VLBI
flux.
In the case of the previous example, $\Delta
t=185$\,d and therefore $c=111$\,d, such that radio data from an
interval of
$111\,\mathrm{d}+185\,\mathrm{d}+111\,\mathrm{d}=407\,\mathrm{d}$
length would be considered. It is only a small number of sources where
the considered time range was this large. 
In sources with large error
bars considerable time averaging had to be performed in
order to obtain a good quality \Fermi/LAT spectrum. This is why the
original sample was limited to ensure that time-averaging is only
necessary in a few cases. Thus the time interval exceeds 365 days in
24 of the 81 SEDs.

Blocks can be divided according to their average flux ranges into
three categories: high, intermediate, and low flux states. We compared
the flux in a block with the average flux across the whole light curve
in order to determine its ``state''. Blocks with a flux between 0.8
and 1.5 of their average flux were labeled as
intermediate states. The number of SEDs with the source in the
low state is relatively small. 
As expected, sources were found to be close to their average flux most
of the time. 
In the high
state, the large number of triggers on such flaring blazars and the
higher overall source flux allow for better statistics.

\subsection{Fitting strategy}\label{sec-fitting}

Having selected the time intervals with sufficient data, we extracted
broadband spectra for each interval. Broadband fitting is 
generally performed on energy flux spectra in the $\nu
F_\nu$-representation. 
This approach is very problematic however, especially in the X-ray and
gamma-ray regime, as the low spectral resolution of the instruments
used in these bands makes it mathematically impossible to recover the
source spectral shape and flux in an unambiguous way by ``unfolding''
\citep[e.g.,][]{lampton:76a,Broos2010,Getman2010}. These ``unfolded''
flux densities are in general biased by the shape of the spectral
model that was used in obtaining them \citep{nowak:05a}. For very
broad energy bands and strongly energy-dependent spectra, which are
present in blazar spectra, the
unfolded flux densities can be in error by a factor of a few. 
To avoid these problems, we used the Interactive Spectral
Interpretation System \citep[\ISIS;][]{ISIS2000} and treat all data
sets in detector space. \ISIS allows us to use data with an assigned
response function (e.g. \Swift/XRT and \Swift/UVOT data) in
combination with data that are only available as flux or flux density
such as the radio data, some of the optical data sets, or \Fermi/LAT
data. A diagonal matrix was assigned to these
latter data sets. All data modeling was performed in detector space, we 
use unfolded data only for display purposes. For the unfolding we use
the model-independent approach discussed by \citet{nowak:05a}. As this
approach is still biased by assuming a constant flux over each
spectral bin, the residuals shown in our figures, which were
calculated in detector space, can disagree with the photon data
converted to flux values.

We further caution that the methods used to obtain the fluxes in the
different energy bands are not identical. The \Fermi/LAT fluxes and
most of the optical data points are model dependent, while the X-ray,
\Swift/UVOT, and \XMMNewton/OM fluxes are model independent. These
uncertainties should be covered by the added systematic uncertainties,
which are described in the following.

The data reduction approach performed for the instruments entering our
analysis is as follows:
\begin{description}
\item[\Fermi/LAT:] We calculate \Fermi/LAT spectra for
  the individual time periods as determined from the \Fermi/LAT light
  curve.
  The adopted systematic uncertainty of the flux is 5\%, due to
  approximations in the instrumental response function (IRFs) and
  uncertainties in the PSF shape and the effective
  area\footnote{\url{http://fermi.gsfc.nasa.gov/ssc/data/analysis/LAT_caveats_p7rep.html}}.
   In addition, in order to show the average $\gamma$-ray flux, our
   SED figures also show unfolded spectra from the 3FGL, which cover
   the time period 2008 August to 2012 July.

 \item[\Swift/XRT:] \Swift \citep{Swift2004} data are from a TANAMI
   fill-in program and were supplemented with archival data. The data
   were reduced with the most recent software package (\textsc{HEASOFT
     6.17})\footnote{\url{http://heasarc.nasa.gov/lheasoft/}} and
   calibration database. For the windowed timing/photon counting mode
   a systematic uncertainty of 5\%/10\% has been adopted following
   \cite{Romano2005}. 
 \item[\XMMNewton/pn and MOS:] Data from the three CCDs on \XMMNewton
   \citep{turner:01a,XMM2001} were reduced using the \textsc{SAS
     14.0.0}\footnote{\url{http://xmm.esac.esa.int/sas/}}. According
   to the official calibration
   documentation\footnote{\url{http://xmm2.esac.esa.int/docs/documents/CAL-SRN-0321-1-2.pdf}},
   uncertainties in the absolute flux calibration are up to 5\%, which
   we used as the systematic uncertainty for the pn and the MOS
   cameras. 
  \item[\Swift/UVOT:] The UVOT data are from the same observations as
    the \Swift/XRT data. They were reduced with the most recent
    version of \textsc{HEASOFT} using standard methods. The systematic
    uncertainty for the \Swift/UVOT detector is 2\%. Contributions to
    the uncertainty include the change in filter sensitivity, i.e.,
    the effective area. The uncertainty due to coincidence loss is
    less than 0.01\,mag \citep[less than
      1\%;][]{Breeveld2005,Poole2008,Breeveld2010,Breeveld2011}.
  \item[\XMMNewton/OM:] The systematic uncertainty of the
    \XMMNewton/OM has been determined to be $\sim$0.1\,mag. This value
    does not include the uncertainty in the zero points. We therefore
    used 3\% as an estimate of the combined
    systematics\footnote{\url{http://xmm2.esac.esa.int/external/xmm_sw_cal/calib/rel_notes/index.shtml}}.
  \item[SMARTS:] SMARTS is an optical/IR blazar monitoring program using
    the SMARTS 1.3\,m telescope, and ANDICAM at CTIO
    \citep{Bonning2012}. They monitor bright southern
    \Fermi/LAT blazars on a monthly basis. The photometric systematic
    uncertainty for the SMARTS program is $\sim$0.05\,mag, with
    deviations up to 0.1\,mag. We therefore used 0.07\,mag for the
    systematic uncertainty, but it does not include the uncertainty in
    the zero points.
    \citet{Bonning2012} uses the zero points given by
    \cite{Persson1998} and \cite{Bessell1998} for the J filter, which
    gives a value of 1589\,mJy. \citet{Buxton2012} use the value from
    \citet{Frogel1978} and \citet{Elias1982}, which is given as
    1670\,mJy. We used the latter.
  \item[REM:] Based on photometry, the systematic uncertainty is
    0.05\,mag (R. Nesci, priv.\ comm.). This value does not include the
    uncertainty of the zero points. 

  \item[VLBI:] TANAMI VLBI observations were performed with the
    Australian Long Baseline Array (LBA) in combination with
    telescopes in South Africa, Chile, Antarctica, and New Zealand at
    8.4\,GHz. Details of the correlation of the data, the subsequent
    calibration, imaging, and image analysis can be found in
    \citet{Ojha2010}. We used the TANAMI core fluxes in our
    multiwavelength analysis, which excludes flux contributions from
    the extended jet in the case of non-compact sources. Contributions
    to the SED at X-ray and $\gamma$-ray energies is expected to
    originate from the inner regions, close to the base of the jet.
    Core radio fluxes are therefore expected to be representative of
    the same region as the high-energy data.
    The statistical errors of VLBI flux measurements are currently not
    well determined. We added a conservative 20\% flux uncertainty that
    covers statistical as well as systematic errors.

    TANAMI VLBI observations are supported by flux-density
    measurements with the Australia Telescope Compact Array
    \citep[ATCA;][]{atca} and the Ceduna 30\,m telescope
    \citep{CedunaCul}.

\end{description}
For optical instruments with no estimate of the zero point
uncertainty, we added an additional 5\% uncertainty. 

Data from the following instruments are shown in the SED figures in
the Appendix to better illustrate the average spectral shape of the
sources. They were not included in the spectral fits since no time
selection was possible on these data sets.
\begin{description}
  \item[\textsl{INTEGRAL:}] Spectra for two of the 22 sources were
    included from the HEAVENS online tool
    \citep{INTEGRAL2003,2010heavens}. The data are dominated by
    Poisson statistics, no systematic errors had to be added.
\item[\Swift/BAT:] BAT data are based on updated 104-month BAT survey
  maps \citep[see][for a description of the BAT
    survey]{Baumgartner2013}. No calibration uncertainty for
  the flux values are given for the \Swift/BAT instrument. We added an
  uncertainty of 0.75\% to the \Swift/BAT data, following the
  uncertainty quoted by \cite{Baumgartner2013} for broadband BAT
  light curves.
    
  \item[\textsl{Wide-Field Infrared Survey Explorer (WISE)}:] Data
    from the ALLWISE catalog \citep{WISE2010} are in the
    infrared waveband. Contributions to the photometric uncertainty 
    of \textsl{WISE} data include source confusion (negligible outside
    the Galactic plane), uncertainty in zero points and in background
    estimation, and the uncertainty of the photometric calibration
    \citep[$\sim
      7\%$\footnote{\url{http://wise2.ipac.caltech.edu/docs/release/allsky/expsup/sec6_3b.html}};][]{WISE2010}.
    The uncertainty of the zero points depends strongly on the filter.
    It lies between 4 and 20\% (for 
    W4)\footnote{\url{http://wise2.ipac.caltech.edu/docs/release/allsky/expsup/sec4_4h.html.}}.
    We used an average uncertainty of 14.5\% and apply the correction
    factor appropriate for a $F_\nu \propto \nu^{-1}$ spectrum in the
    conversion of magnitudes to fluxes \citep{WISE2010}.
    
  \item[2MASS:] The 2MASS point source catalog
    \citep[PSC;][]{Skrutskie2006} photometric 
    uncertainty is hard to determine, as the data were taken over many
    months, with varying weather, seeing, atmospheric transparency,
    background, and moonlight contamination. The average uncertainty
    is quoted as 0.02\,mag for bright sources above a Galactic
    latitude of
    $75^\circ$\footnote{\url{http://www.ipac.caltech.edu/2mass/releases/allsky/doc/sec2_2.html}}.
    To account for other latitudes we used a systematic uncertainty of
    0.05\,mag. This value does not include systematic uncertainties of
    the zero points.

  \item[\Planck:] We included the aperture photometry values from the
    Planck Catalog of Compact Sources \citep{Planck2014} for
    information purposes only. Above 100\,GHz, sources outside the
    Galactic plane have a contamination from CO of up to 6\%. The
    photometric calibration uncertainty is less than $1$\% below
    $217$\,GHz and less than $10$\% at frequencies between 217\,GHz
    and 900\,GHz \citep{PCCS}. We added an uncertainty of 10\% to the
    Planck data to account for the CO contamination and the
    photometric calibration uncertainty.
\end{description}

\subsection{Fitting the broadband spectrum}

The aim of this paper is to obtain an overall understanding of the
spectral behavior of our source sample and how it depends on primary
source parameters. Physical models often have the problem of a
large number of unknown parameters such as the black hole mass, jet
properties, etc. \citep[e.g.,][]{Boettcher2013}, which lead to
significant correlations between individual parameters. We describe
the data with the empirical \texttt{logpar} model \citep{Massaro2004},
a parabola in $\log F_\nu$-$\log \nu$-space. The \texttt{logpar} model 
\begin{equation}
S(E) = K \left(\dfrac{E}{E_1}\right)^{-a+b\,\log_{10}(E/E_1)}\,\mathrm{ph/cm^2/s}
\end{equation}
is parametrized by its normalization $K$, the photon index $a$ at the
energy $E_1$, and the curvature of the parabola $b$ at energy $E_1$. 

Two parabolas were necessary to describe the low and high-energy hump.
This continuum is modified by absorption and extinction (\texttt{tbnew} and
\texttt{redden}, respectively), and by a blackbody component where necessary.
The final model in \ISIS-syntax is 
\begin{multline}\label{eq-model}
N_\mathrm{ph}(E)=  \left( \mbox{\tt logpar(1)}+\mbox{\tt logpar(2)}+
\mbox{\tt blackbody(1)}\right) \\
  \cdot \mbox{\tt tbnew(1)} \cdot \mbox{\tt redden(1)}
\end{multline}
where $N_\mathrm{ph}(E)$ is the photon flux. 
The curvature and slope of the logarithmic parabolas are strongly
correlated. When deriving the peak frequency and peak flux/luminosity
and their respective errors, error propagation overestimates the
resulting uncertainty of these parameters. The resulting errors are
often larger than the values, thus conveying no useful information, as
it is very unlikely that the peak error is larger than more than two
orders of magnitude. The error bars have therefore been omitted from
the plots in the results section where they are not useful.
We have estimated the true uncertainty by shifting the peak position
and comparing the $\chi^2$ values. For sources with good to average
coverage, the total uncertainty is small ($\sim$ half an order of
magnitude). For sources with missing coverage close to the
peak the total uncertainty is one order of magnitude. This is shown in
the lower left corner of the corresponding figures.
For the Compton dominance it is harder to determine the uncertainty
and we conservatively estimate an order of magnitude.

\begin{figure}
\centering
\resizebox{\hsize}{!}{\includegraphics[width=0.46\textwidth]{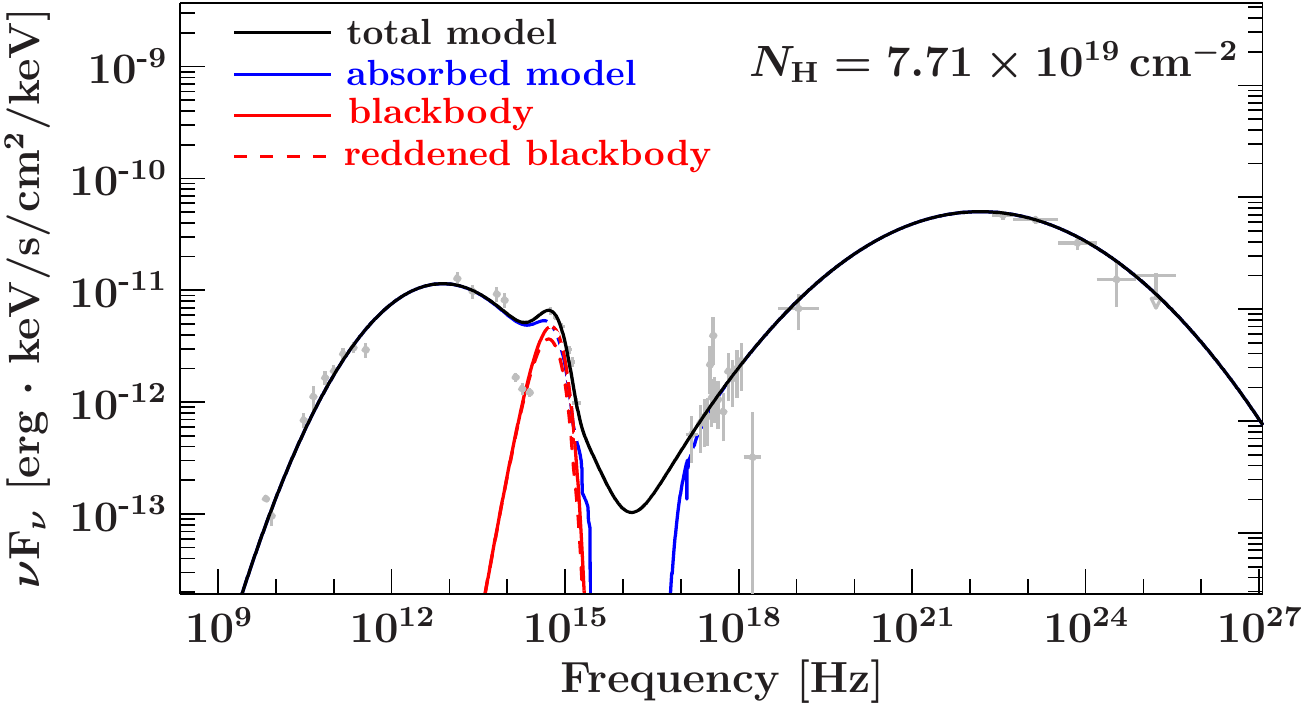}}
\caption{Broadband spectral model of 0402$-$362, with two
  logarithmic parabolas including reddening and absorption (blue), a
  dereddened and reddened blackbody (red, red dashed), and the total
  unabsorbed model
  (black).}\label{fig-sedmodel}
\end{figure}

The \verb|blackbody(1)|
component in Eq.~\eqref{eq-model} describes the ``big blue bump''
(BBB), an excess at optical to ultraviolet wavelengths that was first
seen in 3C 273 \citep{Shields1978}. 
In some sources (e.g., Seyfert galaxies, and some BL Lac objects) with
weak continuum emission, the emission of the host galaxy is not
outshone by the non-thermal continuum emission \citep[e.g., NGC
  4051,][]{Maitra2011}. This feature is very similar in shape to the
BBB, but located at lower energies, corresponding to lower
temperatures of $\sim 6000$\,K.
The origin of the BBB at higher temperatures of $\sim 30000$\,K is 
still debated. In many studies of blazar SEDs it is treated as
background to the non-thermal emission and is often assumed to be the
accretion disk. Typically this feature is visible in FSRQs \citep[and
  references therein]{Jolley2009}. In this work, we modeled the BBB
emission with a single-temperature blackbody. In a few cases,
a multi-temperature blackbody \texttt{diskbb} model, i.e., emission
from an accretion disk with $T(r)\propto r^{-3/4}$
\citep{mitsuda:84a,makishima:86a} is required to describe the BBB
shape.
Figure~\ref{fig-sedmodel} shows an example of the complete model.

Because of the very distinct features imposed by interstellar
absorption in the X-ray band, in our spectral fits we first determined the
hydrogen equivalent column, $N_\mathrm{H}$, from a power law fit to
the X-ray data only. Such a simple
absorbed power law fit worked well in almost all cases and no source
with a large excess above the Galactic $N_\mathrm{H}$ was found.
In the final broadband fits we fixed the absorbing column
$N_\mathrm{H}$ to the value determined by the best fit to the X-ray
data or to the Galactic 21\,cm value. The latter was used if the best
fit $\chi^2$ was high, or the best-fit value consistent with the
Galactic 21\,cm value. As the extinction at infrared, optical, and UV
wavelengths is 
due to the same material that absorbs X-rays we modeled the optical
extinction based on the $N_\mathrm{H}$ value that is used for the X-ray data,
converting it to $A_\mathrm{V}$ from X-ray dust scattering halo
measurements of \citet{Predehl1995} as modified by \citet{Nowak2012}
for the revised abundance of the interstellar medium. This approach
worked very well, contrary to many works which require an optical
extinction correction that is separate from the X-ray modeling. We
speculate that this is due to these papers using the original
\citet{Predehl1995} formulae and therefore obsolete abundances.

We caution that a possible uncertainty exists in the fit to the X-ray
data, which often have large errors ($\sim 50\%$) due to short
exposure observations by \Swift/XRT. The Galactic value from the LAB
survey has an uncertainty of $\sim$30\%, due to stray radiation,
unresolved structures, and the assumption of optical transparency
\citep{Kalberla2005}. An additional problem is that the X-ray modeling
assumes a fixed (Galactic) abundance, which might be the wrong
assumption for the absorption in the host galaxy.

Occasionally, the lack of data necessitated that some parameters are
fixed at a typical value in order to find a good fit, especially for
high-peaked BL Lac sources, where the peak of the high-energy hump
lies above \Fermi energies and is not covered by our data. It was not
possible in these cases to constrain the curvature of the parabola
well from the data. 
Further, due to scarce data around the peak frequencies (typically in
the sub-mm and the MeV range) the exact spectral shape of the two
bumps is unclear. Although the two log-parabolas work remarkably well
here and an averaged spectrum of 3C\,273 \citep{Tuerler1999} has shown
a shape that is remarkably parabola like, some physical models predict
steep bends or additional components
\citep{Mannheim1993,Boettcher2009}. 
In addition, in a few cases such as 1424$-$418 the parabola shape did
not describe the data -- especially the high-energy hump -- well, as
the X-ray spectrum is harder than what is expected from the parabola
fit.  

We use a $\chi^2$ approach to determine the goodness-of-fit. This
method is not statistically sound, as the errors on the VLBI data are only
estimated and likely too large. In relative terms the (reduced)
$\chi^2$ values still give a good estimate of the goodness-of-fit, but
are not indicative of an absolute goodness-of-fit, i.e., in the
probability of the model.

\section{Results \& Discussion}\label{sec-results}

Based on the methods outlined above, we fitted all 81 spectra with the
spectral model of Eq.~\eqref{eq-model}. 
Of the 22 sources 12 have more than two quasi-simultaneous SEDs, and
10 have only two quasi-simultaneous SEDs.
The fit results are listed in
Table~\ref{tab-fit}. The table shows that even though the logarithmic
parabolas are not a physical model, they can describe the broadband
behavior very well, reaching low $\chi^2$ values.
While this does not indicate the probability of the model, the
relative reduced $\chi^2$ values give an estimate of the
goodness-of-fit. It is surprising that they reach low values as many
instruments are not flux cross-calibrated.
We note that FSRQs tend to have an index which is too soft to describe
the \Swift/XRT and the \Fermi/LAT spectrum perfectly. In some sources
the LAT spectrum constrains the curvature of the parabola well, for
which the X-ray spectral indices are too soft (see e.g., 1424$-$418).
The reason for this behavior might be due to a spectral break in the
MeV energy range. Other possibilities include an accretion disk
component in the soft X-rays, or a pion decay signature at MeV
energies.

We find one source, 2005$-$489, with a peculiar excess in the hard
X-rays above 5\,keV, which can be described with a thermal blackbody,
but likely only due to a lack of data above 10\,keV. It might be
possible to explain this with a hadronic proton-synchrotron signature,
but the origin is as yet unclear (see Sect.~\ref{sec-2005-489}).

In the following sections we describe the behavior of individual
parameters in greater detail.
For some sources the redshift is unknown. While all broadband SEDs are
modeled without k-corrections, the analysis, e.g., of source fluxes or
peak positions often requires knowledge of the redshift. Sources
without redshifts are therefore not included in the results, unless
noted otherwise.

\subsection{The peak positions}
\subsubsection{Blazar sequence}\label{sec-blazseq}

The blazar sequence posits that more luminous blazars have lower peak
frequencies \citep{Fossati1998,Ghisellini1998}. While it is heavily
debated \citep[e.g.,][]{Giommiblazseq,Giommi2012}, it is generally 
observed for most sources with known redshift, although sources with low
luminosities at low peak frequencies have been found
\citep{Nieppola2006,Meyer2011,Giommiblazseq}. Sources at high
luminosities and high peak frequencies are still missing, however,
possibly due to the lack of redshift information.
\citet{Meyer2011} propose a modified blazar sequence, where more
luminous blazars are more efficient at accretion. Sources with lower
peak luminosities and higher peak frequencies than expected are
interpreted as being misaligned, leading to a shift in the peak.

\begin{figure}
\centering
\resizebox{\hsize}{!}{\includegraphics[width=0.46\textwidth]{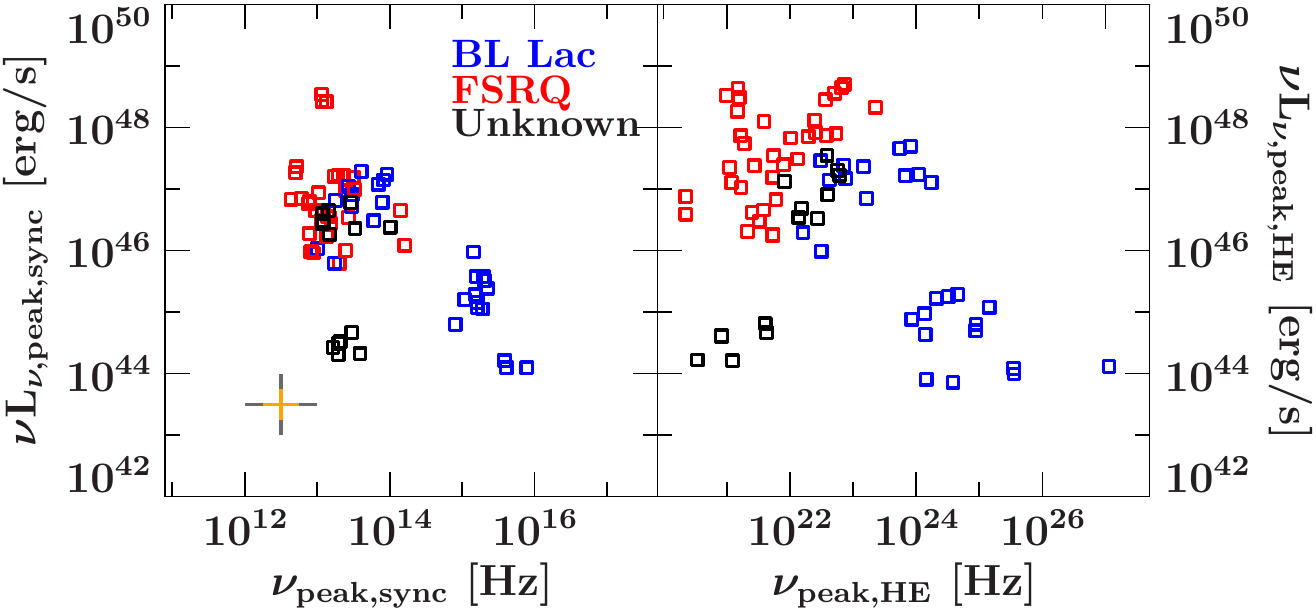}}
\caption{Peak frequency (k-corrected) and peak luminosity for the
  synchrotron peak (left) and the high energy peak (right). The
  estimated uncertainty is given in the lower left corner for sources
  with average coverage (orange) and for SEDs with a lack of data near
the peak position (gray).}\label{fig-allblazseq}
\end{figure}
\begin{figure}
\centering
\resizebox{\hsize}{!}{\includegraphics[width=0.46\textwidth]{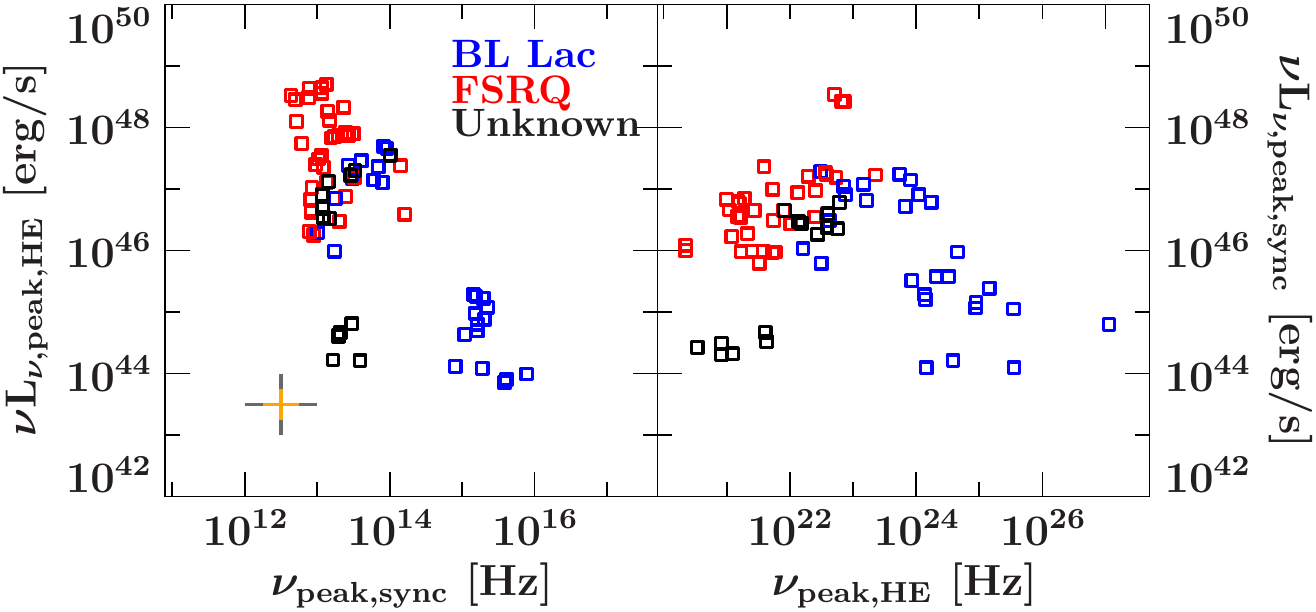}}
\caption{``Inverted blazar sequence'': HE peak luminosity (left)
  vs.\ peak synchrotron frequency (k-corrected) and vice versa
  (right). The estimated uncertainty is given in the lower left corner
  for sources with average coverage (orange) and for SEDs with a lack
  of data near the peak position (gray).
}
\label{fig-invallblazseq}
\end{figure}

Figure~\ref{fig-allblazseq} shows the k-corrected peak frequencies and
peak luminosities for all 21 sources in our sample for which a
redshift measurement is available. The synchrotron peak results are
consistent with the blazar sequence, with a gap between $10^{14}$ and
$10^{15}$\,Hz. This gap has also been seen in the 3LAC \citep{lac3}
and has been named the \Fermi blazar's divide \citep{Ghisellini2009}.
See Sect.~\ref{sec:divide} for a further discussion of this feature.

We also find one source, 0521$-$365 with a lower peak frequency
and peak luminosity than expected from the blazar sequence. It is
interesting to note, but likely a coincidence, that the peak of this
source is perpendicular to the blazar sequence at the location of the
gap. 

While the positions of the high energy peak seem to generally follow
the blazar sequence, the spread is much wider, consistent with
expectations from a SSC model. We note that when ``inverting'' the
blazar sequence, by looking at the synchrotron peak frequency versus
the HE peak luminosity, it still follows the blazar sequence. The
opposite is not true (Fig~\ref{fig-invallblazseq}). The HE peak
frequency vs. the synchrotron luminosity shows a rising and a falling
slope (or a \textsf{V}-shape flipped on the horizontal axis), which,
when going back to the regular blazar sequence, might also be visible
there. 

\begin{figure}
\centering
\resizebox{\hsize}{!}{\includegraphics{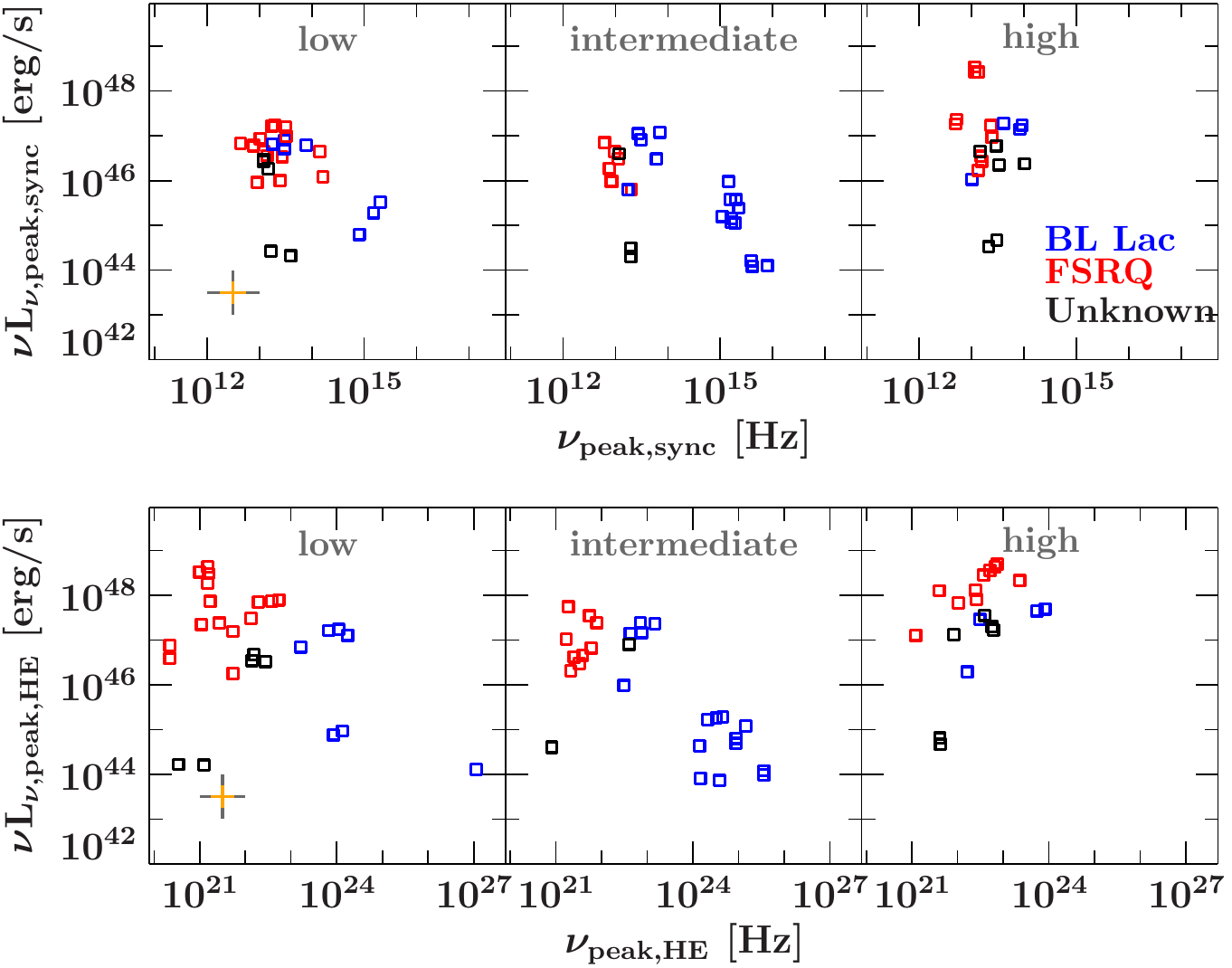}}
\caption{Peak frequencies and peak luminosities, separated into low,
  intermediate, and high states for the synchrotron peak (top row) and
  high energy peak (bottom row). While the low and intermediate states
  follow the blazar sequence for both peaks, the high energy peak in
  both states show a peculiar, almost inverted behavior, although the number
  of sources (especially BL Lacertae objects) is too low for any
  conclusive evidence. The
  estimated uncertainty is given in the lower left corner of the left
  panels for sources with average coverage (orange) and for SEDs with
  a lack of data near the peak position (gray).}
\label{fig-stateblazseq}
\end{figure}

Figure~\ref{fig-stateblazseq} shows the blazar sequence separated by
the activity of the source at the given time. The upper panel shows
the location of the synchrotron peak, the lower panel shows the
position of the high energy peak. Both panels are separated into low,
intermediate, and high states.
We find that in the intermediate state (and possibly in the low
state), the sources follow the blazar sequence
(Fig.~\ref{fig-allblazseq}). In the high state the synchrotron
 peak results are inconclusive, and seem to scatter. We find what has
 been seen previously, high-peaked BL Lac objects show a much lower
 occurrence of large outbursts in HE $\gamma$-rays, and our sample
 includes no high-peaked SED (above 10$^{14.5}$\,Hz) in a high state.
 Even when we take this lack of data into account, the blazar
 sequence slope of the high-energy peak in the high state is
 drastically different from the intermediate state, possibly showing
 an increase in peak frequency with peak luminosity.

To see whether this behavior is statistically significant, we have also
looked at the individual behavior. We find that in the intermediate
state the high energy peak tends to move towards lower frequencies,
while it moves towards higher frequencies in high states. This behavior
is discernible for the sources 0521$-$365, 0537$-$441, and
1454$-$354. For 0208$-$512, 0332$-$376, 0426$-$380, and 0402$-$362
only one of the effects is
visible, likely due to a lack of data (see Fig.~\ref{ap-sed}). 
For the other sources no disagreeing trends have been found, but some
SEDs lack information from all states, e.g., 0402$-$362 only has two
high state SEDs, so no 
information about the peak shift is available. While this behavior has 
not been documented for a large sample, a ``harder-when-brighter''
trend is often seen in the X-ray spectra of flaring blazars and other
AGN consistent with a peak shift to higher frequencies
\citep{Zamorani1981,Avni1982,Pian1998,Vignali2003,Emmanoulopoulos2012}.
For a number of flaring \Fermi/LAT sources a hardening of the spectral
index has also been observed \citep{Abdo2010a,Abdo2010c}, which might
be useful in the future for discriminating between intermediate and
flaring states, though no physical explanation is readily available.

\subsubsection{Spectral index and peak position}

\begin{figure}
\resizebox{\hsize}{!}{\includegraphics{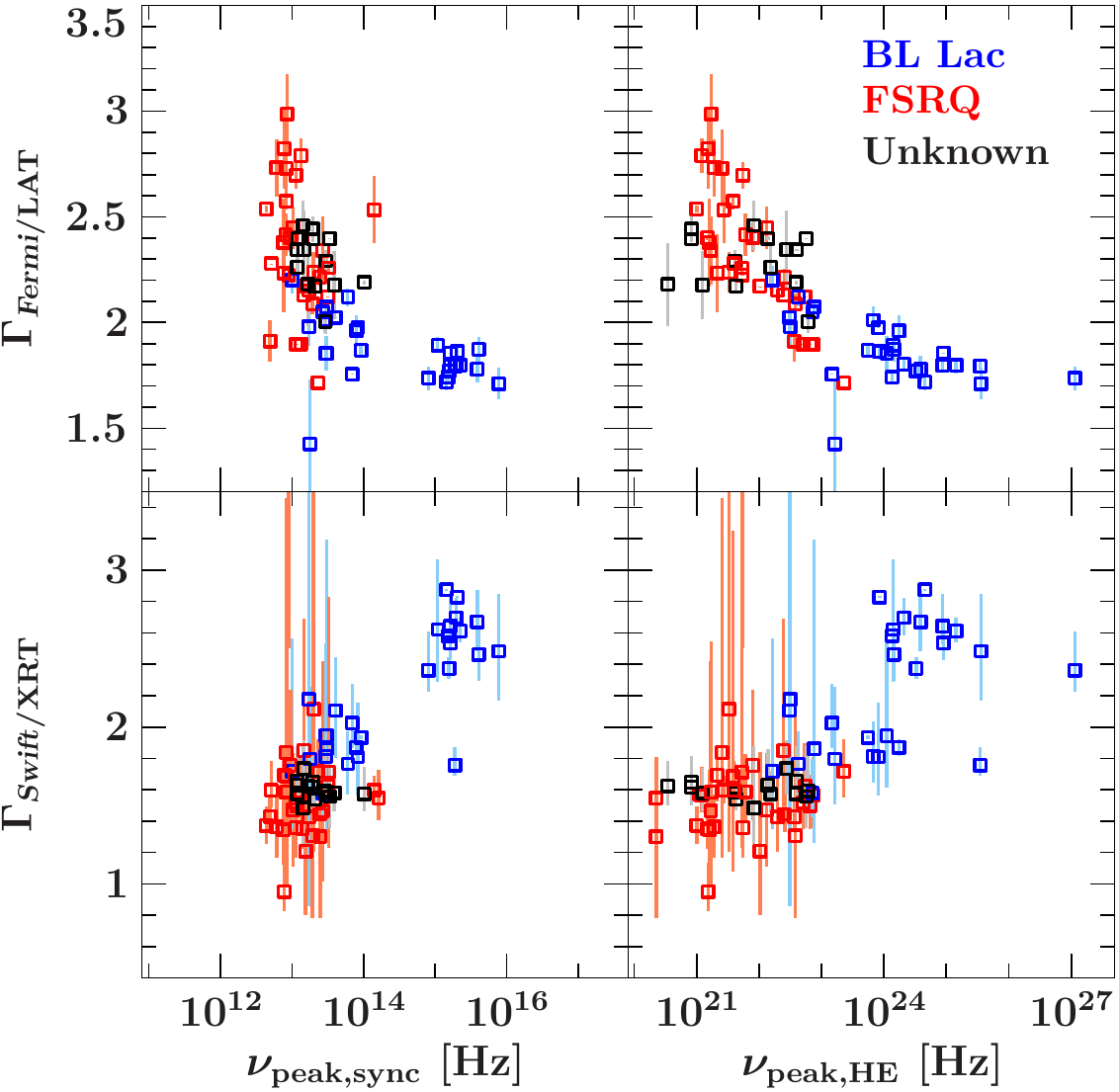}}
\caption{Behavior of the synchrotron (left) and HE (right) peak
  frequency as a function of the photon index seen in the \Fermi/LAT
  (top) and the \Swift/XRT bands (bottom).}\label{fig-index}
\end{figure}

\begin{figure}
\resizebox{\hsize}{!}{\includegraphics{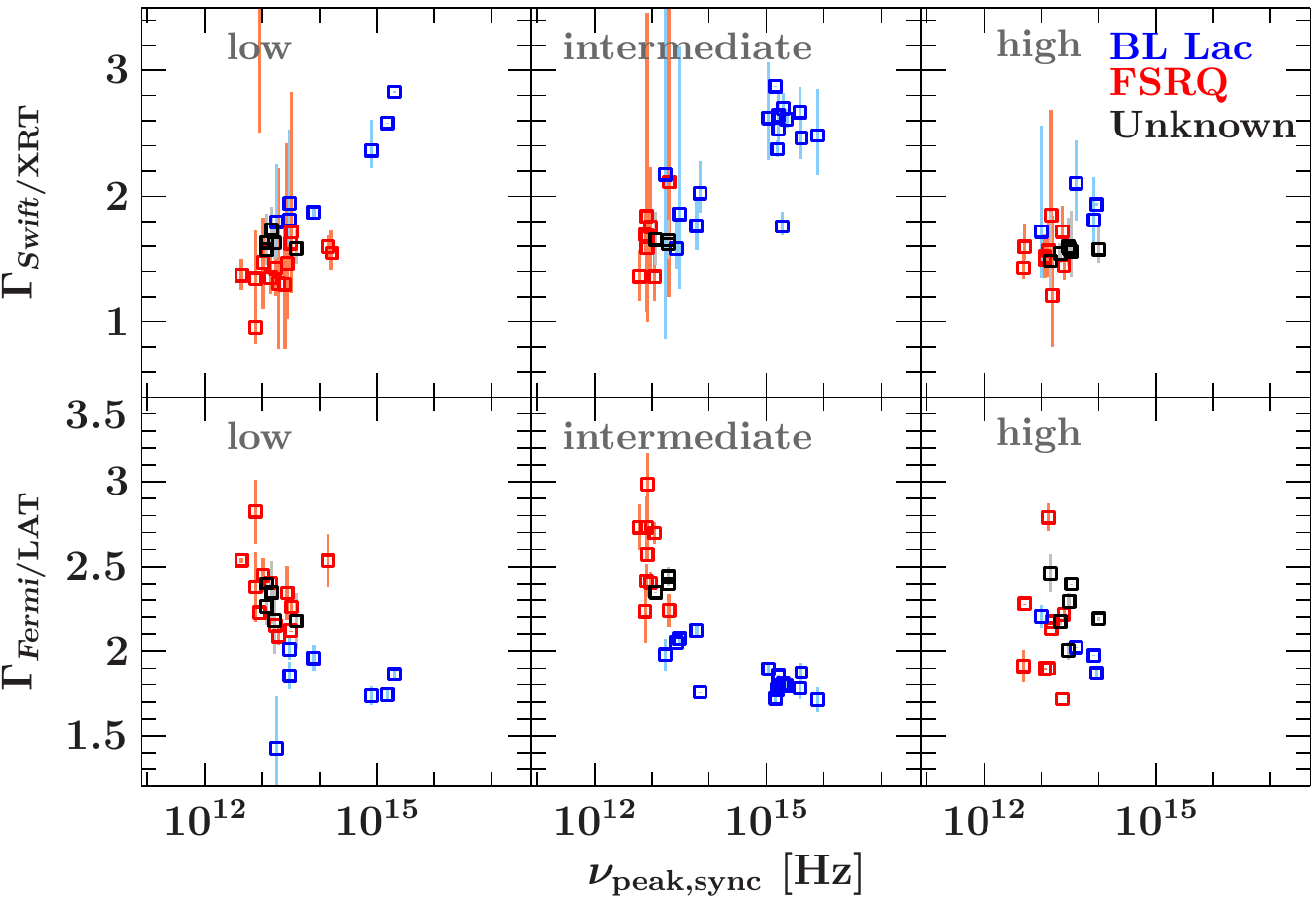}}
\resizebox{\hsize}{!}{\includegraphics{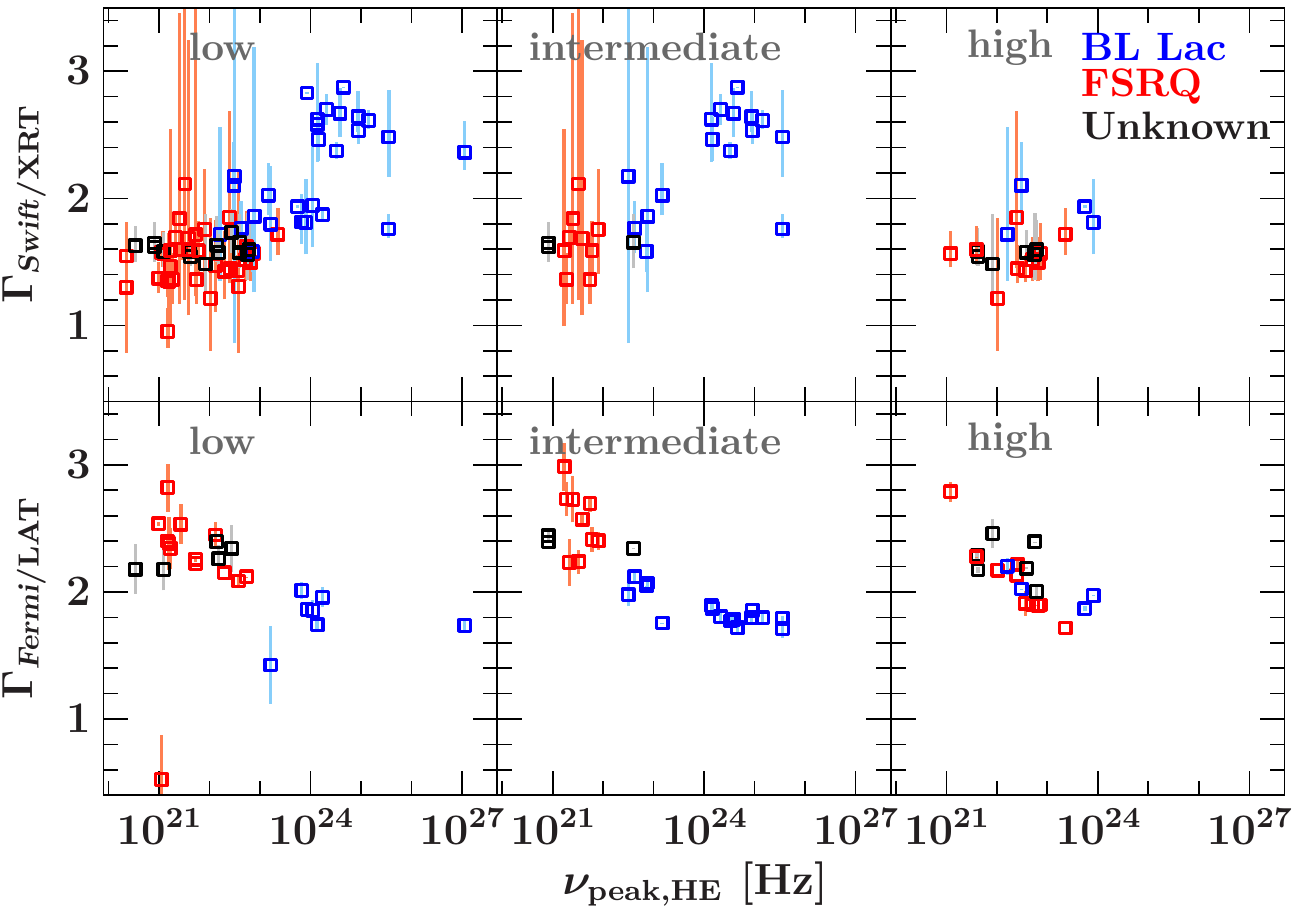}}
\caption{Synchrotron (top) and HE (bottom) peak frequency versus the
  photon index seen by \Swift/XRT (top) and \Fermi/LAT (bottom)
  separated by low, intermediate, and high state.}\label{fig-indexstate}
\end{figure}

The correlations between the spectral indices seen in \Fermi/LAT and
\Swift/XRT and the synchrotron peak frequency are well documented in
the 3FGL catalog \citep{fgl3}. Correlations with the high-energy peak
are less studied. All spectral indices are shown in
Fig.~\ref{fig-index}, and in Fig.~\ref{fig-indexstate} they are
separated into the low, intermediate, and high state. The top panel of
the figure shows the synchrotron peak
frequency versus the XRT and LAT indices, while the lower panel shows
the high-energy peak frequency versus the XRT and LAT indices. It is
interesting to note that the LAT index shows varying behavior in the
bottom panel of the top plot (synchrotron peak frequency) in
Fig.~\ref{fig-indexstate} depending on source state, but not in the
bottom panel of the bottom plot (HE peak frequency). This change in
the high state is consistent with a difference in synchrotron and 
high-energy peak behavior of the sources. In the low
and intermediate state the LAT index shows a correlation with the
synchrotron peak frequency, indicative of correlated processes. 
The data seems more scattered for SEDs in the high state. This change
is indicative of a change in the jet properties during a high state,
such as an acceleration of the jet flow \citep{Marscher2010}.

\subsubsection{Compton dominance and the blazar sequence}

\citet{Giommi2012} suggest that the blazar sequence is due to the
selection bias of the observed samples. The sources missing in the
blazar sequence are expected to peak in the optical/UV. These sources
should be the brightest among the optical-selected blazars.
\citet{Giommi2012} argue that these sources are dominated by jet
emission in the optical, making it nearly impossible to determine
their redshift spectroscopically. The argument is therefore that these
sources exist, and are known, but no luminosities are available.
Therefore, \citet{Finke2013} uses the Compton dominance,
$F_{\mathrm{peak,HE}}/F_{\mathrm{peak,sync}}$, a redshift independent 
quantity to verify the existence of the blazar sequence, and also
finds a lack of sources at high peak frequencies and luminosities. 

While we might miss low luminosity sources in the TANAMI sample, we
would expect to have found sources with high luminosities at high peak
frequencies if they exist. These are expected to be bright and have
hard spectral indices in \Fermi/LAT. As our sample is representative
of a $\gamma$-ray flux-limited sample it is possible that we miss
bright sources peaking in the optical if their Compton dominance is
low, i.e., if their high-energy peak is faint, possibly even fainter
than the synchrotron peak.

\begin{figure}
\resizebox{\hsize}{!}{\includegraphics{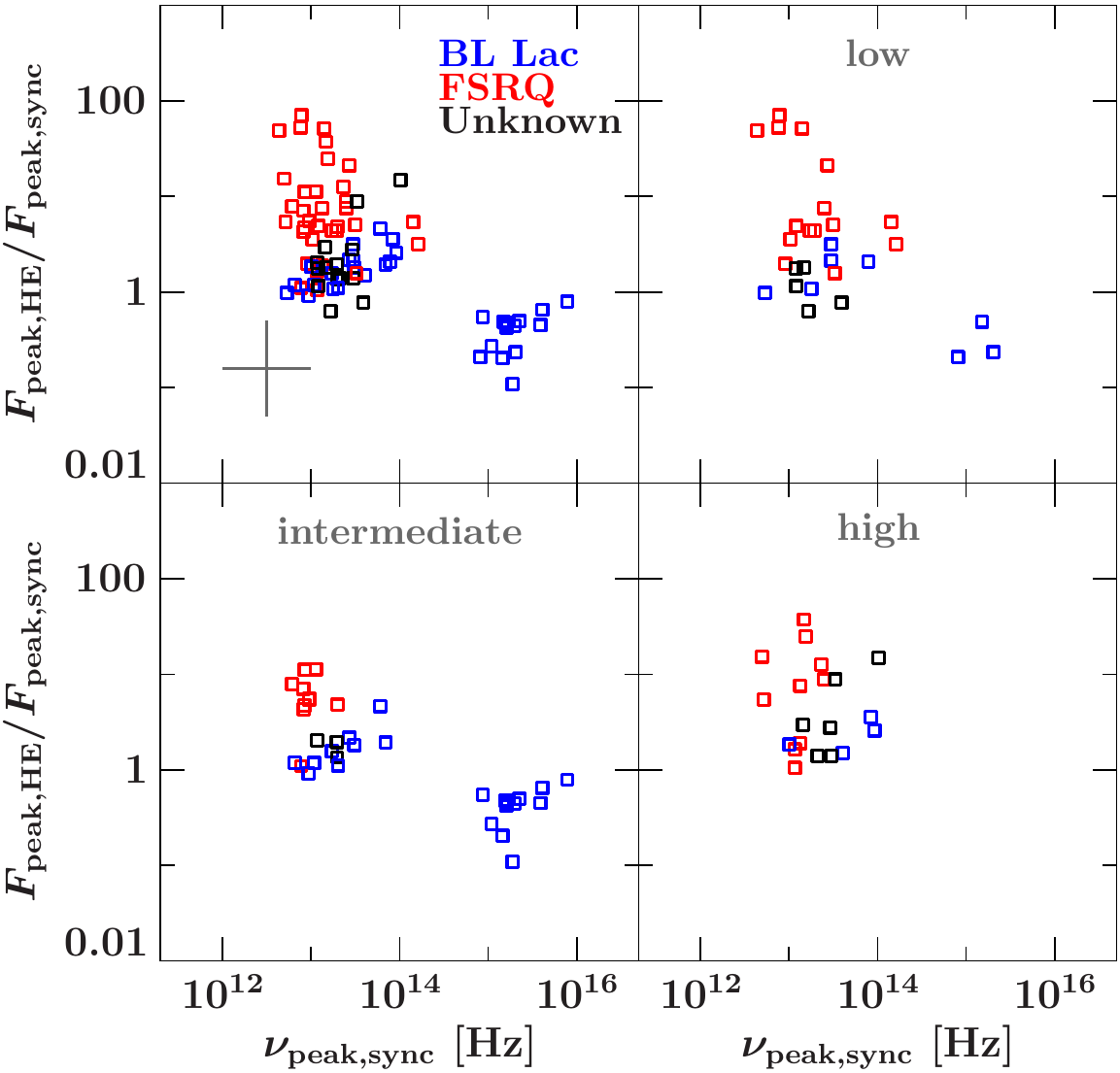}}
\caption{Top left: Compton dominance for all SEDs for all sources (no
  k-correction). It is interesting to note that the blazar
  divide is particularly strong; only few sources are found between
  $10^{14}$ and $10^{15}$\,Hz. Top right -- bottom: Same as above, but
  SEDs are separated into low, intermediate, and high states. The
  estimated uncertainty is given in the lower left corner of the top
  right panel in gray.} 
\label{fig-cd}
\end{figure}

Consistent with earlier findings \citep{Giommi2012,Finke2013},
Fig.~\ref{fig-cd} shows that there is a redshift-independent
correlation between the ratio of the peak fluxes and the peak
frequency. The sequence can be explained physically by increasing
power leading to larger external radiation fields and a larger Compton
dominance. Higher Compton scattering leads to faster cooling and a
lower cut-off of high-energy photons, possibly explaining the observed
blazar sequence.

Looking at the state separated behavior (Fig.~\ref{fig-cd}, bottom and
top right), while the number of SEDs in the high state is low, the
behavior during high states is different from the low, and
intermediate states. As for the blazar sequence, the low and
intermediate states are consistent with expectations from the blazar
sequence and FSRQs at higher Compton dominances. In the high state,
the Compton dominance shows a large scatter.
We further generate the Compton dominance for the bolometric fluxes,
instead of the peak flux.
\begin{figure}
\resizebox{\hsize}{!}{\includegraphics{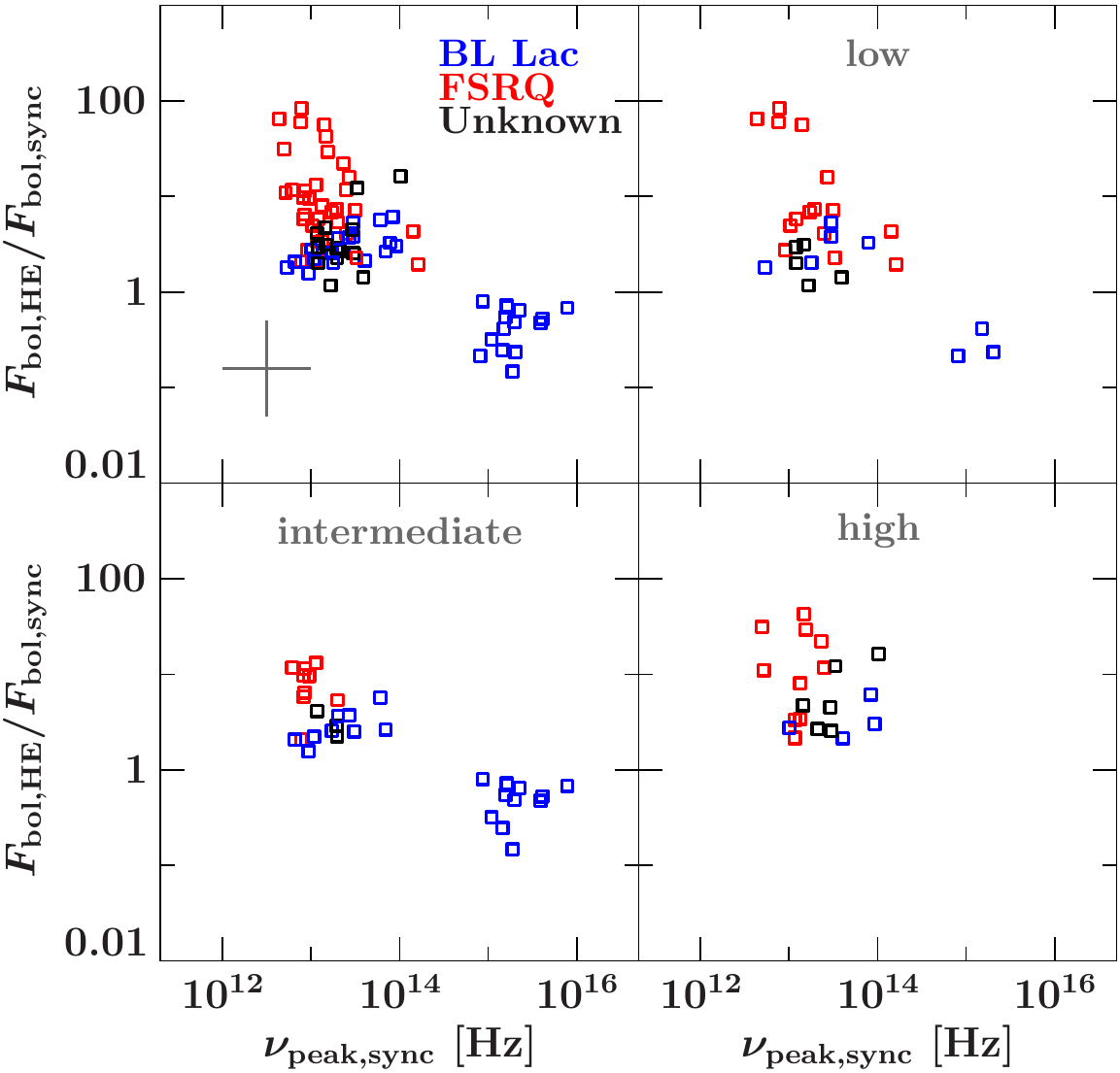}}
\caption{Top left: Bolometric Compton dominance for all SEDs for all sources (no
  k-correction). It is interesting to note that the blazar
  divide is particularly strong. Top right, bottom: Same as above, but SEDs are
  separated into low, intermediate, and high states. The
  estimated uncertainty is given in the lower left corner of the top
  right panel in gray.} 
\label{fig-cdb}
\end{figure}
The bolometric fluxes are calculated by integrating over each of the
two best-fit parabola functions separately (Fig.~\ref{fig-cdb}). 
The patterns in Fig.~\ref{fig-cd} and Fig.~\ref{fig-cdb} are very
similar. While the scatter is lower when using bolometric fluxes, it
shows that the peak position is a reliable tracer of the bolometric
flux.

\subsection{The \Fermi blazar's divide}\label{sec:divide}

In the blazar sequence and Compton dominance a large gap is visible,
which seems to separate FSRQs and BL Lac objects between $10^{14}$ and
$10^{15}$\,Hz. This gap has also been seen in the 3LAC \citep{lac3},
and is now named the \Fermi blazar's divide as first discussed by
\citet{Ghisellini2009}. These authors propose a physical difference in
these objects with a separation of objects into low and high
efficiency accretion flows. It is interesting to note, however, that
in our $\gamma$-ray flux limited sample this separation is much
stronger than in the 3LAC, suggesting a 
contribution of selection effects. These selection effects can
contribute in the same way as to the blazar sequence, i.e., we would
expect a lack of redshifts in objects peaking in the optical range
($10^{14}$--$10^{15}$\,Hz), which would show a featureless spectrum
due to a dominant jet component.
Further, the extinction in the UV and far-UV, as well as the
photoelectric absorption of soft X-rays in our Galaxy, hamper the
detection of blazars peaking in this energy range, exactly those peak
frequencies missing in the blazar's divide.
We expect that this can fully explain the \Fermi
blazar's divide and is also consistent with observations of black hole
binaries, which do not show a gap between accretion states.

Selection effects are able to explain the blazar's divide, while the
argument is less clear for the blazar sequence, which is found even in
the Compton dominance, which is redshift-independent. 
While selection effects can explain many of the observed features, it
is peculiar that no source has been found at high peak frequencies and
luminosities so far.

\subsection{The big blue bump}

It is generally believed that the thermal excess seen in many FSRQ
objects and in a small number of BL Lac objects is the thermal
emission from the accretion disk \citep{Shields1978,Malkan1982}. An
alternative model explains the BBB with free-free emission from the
hot corona of the supermassive black hole \citep{Barvainis1993}. While
no conclusive evidence for either theory has been presented, several
problems with the accretion disk scenario have been noted, namely the
temperature problem, the ionization problem, the time-scale problem
and the co-ordination problem (see \citet{Lawrence2012} for a summary).
The temperature problem states that the observed temperatures at
$\sim$30000\,K are too low for what would be expected
($\sim$76000\,K). \citet{Lawrence2012} proposes a reprocessing of the
accretion disk emission by clouds in the BLR and is able to explain
all four problems. 

\begin{figure}
  \resizebox{\hsize}{!}{\includegraphics{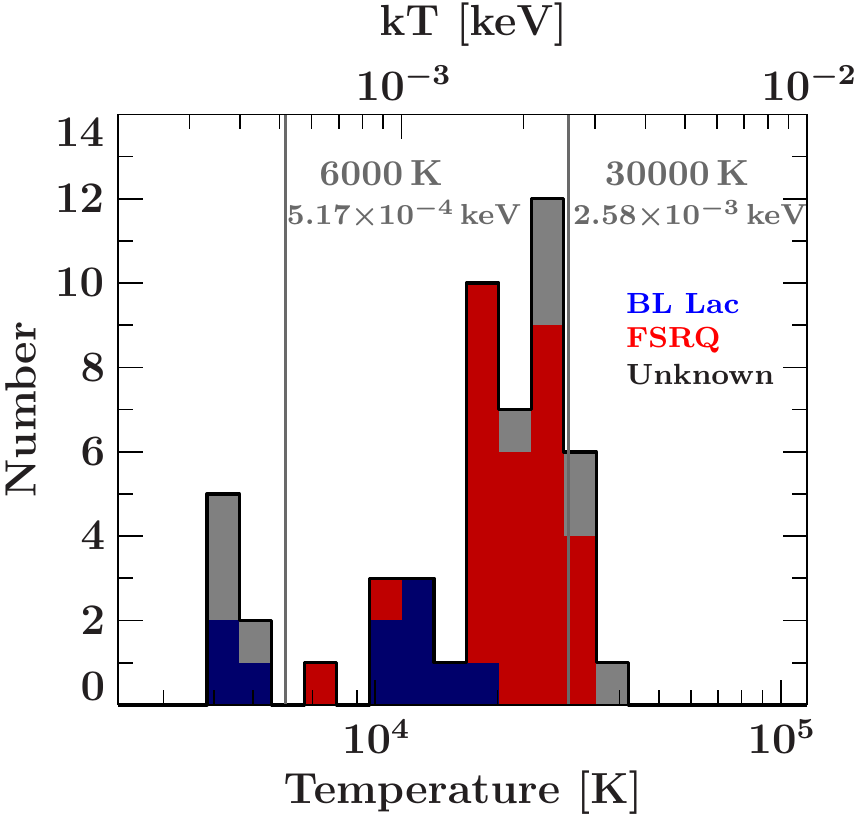}}
  \caption{Histogram of observed blackbody temperatures for all SEDs.
    Blackbody temperatures at $\sim$6000\,K (marked in gray with a
    vertical line) very likely represent a detection of the host
    galaxy. A temperature of 30000\,K is marked with another vertical
    gray line.}
  \label{fig-bbtemp}
\end{figure}

Concerning the temperature, our results are consistent with what has
been found previously \citep{Zheng1997,Telfer2002,Scott2004,Binette2005,Shang2005}.
For all sources the temperature remains below $\sim$32000\,K. Some BL
Lac objects exhibit temperatures of $\sim$6000\,K (see
Fig.~\ref{fig-bbtemp}, marked by a gray vertical line). Such cold
black bodies are very likely emission
from the host galaxy, which would support the theory of a weak disk
and inefficient accretion in BL Lac objects\footnote{Note that
  gravitational redshifting decreases the observed temperature, but
  even taking this effect into account would only slightly increase
  the temperatures by $\sim$3000\,K, still nowhere near the expected
  temperature for an accretion disk.}.

\begin{figure}
  \resizebox{\hsize}{!}{\includegraphics{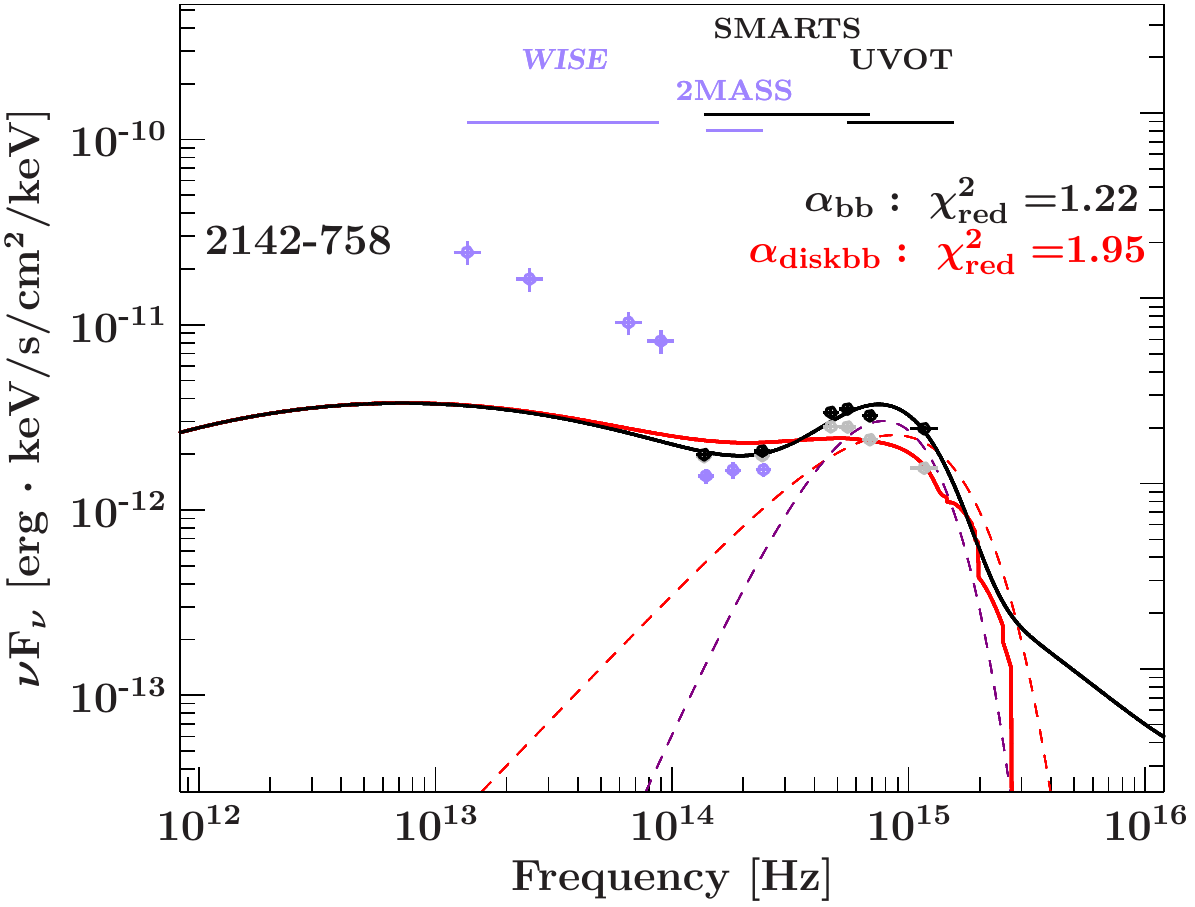}}
  \caption{SED of the $\alpha$ state of 2142$-$758, with the best fit
    single temperature blackbody (purple) and best fit multi temperature
    accretion disk spectrum (red).}
  \label{fig-bbsed}
\end{figure}
In general, the spectral shape of the thermal excess is also
inconsistent with an accretion disk origin. 
For all SEDs the thermal
excess can be well described by a single temperature blackbody. For
an accretion disk extending from a few to several hundreds or thousands
of gravitational radii a large range in temperature would be expected
due to the $r^{-3/4}$ temperature profile of accretion disks, with
further slight stretching of the spectrum by gravitational
redshifting. Figure~\ref{fig-bbsed} shows that the shape is reasonably
constrained by \Swift/UVOT. The red curve shows the spectrum expected
from a simple multi-temperature accretion disk. This \texttt{diskbb}
model is not able to describe the narrow shape as well as a single
temperature blackbody 
(purple line in Fig.~\ref{fig-bbsed}). While this evidence is not
conclusive due to the low spectral resolution of the UVOT, it
nevertheless is indicative of a more complex disk structure, which
might be puffed up and warped or truncated, leading to changes in the
thermal emission. Further theoretical and observational studies are
necessary to determine the origin and shape of the big blue bump.

\subsection{The black hole mass, $M_\mathrm{BH}$}

We study how the properties of the SED depend on the black hole mass.
The fundamental plane of black holes \citep[][and references
therein]{Merloni2003,Gallo2003,Falcke2004,Koerding2006,Gultekin2009,Plotkin2012,Bonchi2013,gultekin2014,Saikia2015,Nisbet2016}
relates the radio and X-ray luminosity to the black hole mass,
\begin{equation}
  \log\left(\dfrac{M_\text{BH}}{M_{\astrosun}}\right) =
d\,\log\left( \dfrac{L_\text{radio}}{\text{erg}\,\text{s}^{-1}}\right) -
e\log\left( \dfrac{L_\text{X-ray}}{\text{erg}\,\text{s}^{-1}} \right)- f
\end{equation}

The parameters $d$, $e$, and $f$ depend on the source populations.
Table~\ref{tab-fppar} lists typical recent values for AGN.
Here $L_\text{radio}$ is the radio flux density measured at the
frequency $\nu_\text{radio}$ listed in Table~\ref{tab-fppar}, while
$L_\text{X-ray}$ is the X-ray flux in the 2--10\,keV band.
We caution that the radio luminosities listed are not ``real''
luminosities, as the differential flux at the given radio frequency is
simply multiplied by $4\Pi d_L^2$, instead of using an integrated flux
in a waveband.

\begin{table*}

\caption{Parameters of the fundamental plane of black hole}\label{tab-fppar}
\begin{tabular}{llllll}
Reference & $d$ & $e$ & $f$ & $\nu_{\text{radio}}$ &
source population\\
& & & &[GHz] & \\
\hline
\citet{Merloni2003} & $1.28\pm0.15$ & $0.77\pm0.17$ & 9.40 & 5.0 & Quasars, LINERs,
Seyferts\\
\citet{Koerding2006} & $1.28\pm0.30$ & $0.73\pm0.20$ & 10.49 & 5.0 &
Quasars, LINERs, Seyferts\\
\citet{Gultekin2009} & $0.48\pm0.17$ & $0.24\pm0.16$ & 0.83 & 5.0 & Seyferts, Transition
Objects, Unclassified Objects\\
\citet{Bonchi2013} & 1.47 & $0.57\pm0.07$ & 24.43 & 1.4 & Type 1 \& Type 2\\
\citet{Nisbet2016} & $1.45\pm0.22$ & $0.94\pm0.18$ & 8.01 & 1.4 & LINERs \\
\hline
\end{tabular}
\end{table*}

Black hole mass measurements, based on measurements of the BBB (for
FSRQs), and variability arguments (BL Lacs), only exist for 8 of the
20 sources in our sample and are taken from \citet{Ghisellini2010}. We
note that for some sources different black hole mass measurement
exist \citep[e.g., 0208$-$512;][]{Stacy2003}, which vary by an order
of magnitude. We therefore use the
fundamental plane to estimate the black hole mass and compare the
estimates with measurements, where available. We use the distance
corrected radio flux density from the best-fit parabola model at the
same frequency as used in each of the studies. The X-ray 2--10\,keV
luminosity is taken from the separate fit to the X-ray data. We use
all SEDs from this work and the corresponding X-ray and radio
luminosities (where a redshift measurement is available, see
Tab.\ref{tab-srcs}) and calculated estimated black hole masses
following \citet{Merloni2003},
\citet{Koerding2006}, and \citet{Nisbet2016}. Our results are
presented in Table~\ref{tab-mbh}. For the results from
\citet{Gultekin2009} we use Eq.~4 in their paper, with the parameters
listed in
Eq.~6, where a linear regression was performed in order to find an
equation for an estimate of the black hole mass. For sources with more
than one SED, the black hole mass estimates are averaged. The masses
before averaging scatter depending on source state with a maximum
factor of 5 between the lowest and the highest estimate.

\begin{table*}
\caption{Black hole masses as measured and as estimated from the fundamental plane of black holes. All values are given as $\log_{10}(M)$.}
\label{tab-mbh}
\begin{tabular}{llllllll}
Source & M$_{\mathrm{BH}}$ & M$_{\mathrm{BH,Merloni}}$ &   M$_{\mathrm{BH,Koerding}}$ & M$_{\text{BH,G{\"u}ltekin}}$ &M$_{\text{BH,Bonchi}}$ &M$_{\mathrm{BH,Nisbet}}$&$L_{\mathrm{edd}}$\\
 & [\,M$_{\astrosun}$] & [M$_{\astrosun}$] & [M$_{\astrosun}$] & [M$_{\astrosun}$] & [M$_{\astrosun}$] & [M$_{\astrosun}$]& 10$^{46}$ [erg\,s$^{-1}$]\\\hline
0208-512 & $8.8$ &$5.7\pm0.9$ &$6.7\pm1.0$ &$7.4\pm0.4$ &$8.8\pm1.3$ &$4.06\pm0.04$ &9.1 \\
0244-470 &  &$5.1\pm2.6$ &$6.2\pm2.7$ &$7.2\pm1.2$ &$8\pm4$ &$3.08\pm0.24$ & \\
0402-362 &  &$5.2\pm2.7$ &$6.3\pm2.7$ &$7.3\pm1.3$ &$8\pm4$ &$2.9\pm0.4$ & \\
0426-380 & $8.6$ &$6.0\pm1.3$ &$7.0\pm1.5$ &$7.6\pm0.6$ &$9.1\pm1.9$ &$4.49\pm0.04$ &5.2 \\
0447-439 & $8.8$ &$2.5\pm2.0$ &$3.4\pm1.5$ &$6.2\pm1.0$ &$4.3\pm2.5$ &$0.3\pm0.5$ &7.8 \\
0506-612 &  &$5.7\pm1.5$ &$6.7\pm1.6$ &$7.4\pm0.7$ &$8.8\pm2.1$ &$4.14\pm0.05$ & \\
0521-365 &  &$3.6\pm1.2$ &$4.6\pm1.1$ &$6.6\pm0.6$ &$5.6\pm1.6$ &$1.67\pm0.17$ & \\
0537-441 & $9.3$ &$6.4\pm1.1$ &$7.4\pm1.4$ &$7.7\pm0.5$ &$9.8\pm1.7$ &$5.05\pm0.04$ &26.0 \\
0637-752 &  &$5.7\pm1.5$ &$6.7\pm1.6$ &$7.5\pm0.7$ &$9.2\pm2.1$ &$4.409\pm0.027$ & \\
1057-797 & $8.8$ &$5.8\pm2.5$ &$6.8\pm2.8$ &$7.5\pm1.1$ &$9\pm4$ &$4.27\pm0.04$ &7.8 \\
1424-418 &  &$6.0\pm1.1$ &$7.1\pm1.2$ &$7.6\pm0.5$ &$9.0\pm1.5$ &$3.98\pm0.06$ & \\
1440-389 &  &$1.9\pm2.0$ &$2.8\pm1.4$ &$5.9\pm1.1$ &$3.7\pm2.5$ &$-0.1\pm0.5$ & \\
1454-354 & $9.3$ &$6.0\pm1.8$ &$7.0\pm2.0$ &$7.6\pm0.8$ &$9.2\pm2.6$ &$4.567\pm0.009$ &26.0 \\
1610-771 &  &$5.9\pm2.6$ &$6.9\pm2.8$ &$7.5\pm1.1$ &$9\pm4$ &$4.35\pm0.08$ & \\
1954-388 &  &$5.7\pm2.5$ &$6.7\pm2.8$ &$7.4\pm1.1$ &$9\pm4$ &$4.408\pm0.017$ & \\
2005-489 & $8.7$ &$3.1\pm2.7$ &$4.0\pm2.2$ &$6.4\pm1.3$ &$5\pm4$ &$1.1\pm0.5$ &6.5 \\
2052-474 &  &$6.1\pm2.6$ &$7.1\pm2.9$ &$7.6\pm1.1$ &$9\pm4$ &$4.39\pm0.06$ & \\
2142-758 &  &$5.5\pm2.6$ &$6.5\pm2.7$ &$7.4\pm1.2$ &$9\pm4$ &$3.85\pm0.13$ & \\
2149-306 &  &$5.3\pm1.6$ &$6.3\pm1.6$ &$7.4\pm0.8$ &$8.9\pm2.2$ &$3.58\pm0.13$ & \\
\end{tabular}
\tablefoot{Columns: (1) IAU B1950 name, (2) $M_{\mathrm{BH}}$ from \cite{Ghisellini2009}, $M_{\mathrm{BH}}$ estimated after (3) \citet{Merloni2003}, (4) \citet{Koerding2006}, (5) \citet{Gultekin2009}, (6) \citet{Bonchi2013}, (7) \citet{Nisbet2016}, (8) Eddington luminosity for the measure black hole mass, assuming isotropic emission\\Note. The black hole mass estimates include the uncertainties from the parameters, not the uncertainties in luminosities, as these are much smaller.}
\end{table*}

All estimates except those using the parameters from
\citet{Bonchi2013} are lower than the measured values, with the
largest offset being that from the \citet{Merloni2003} parameters.
Applying the relation by \citet{Bonchi2013} gives a very good
agreement (less than a factor of 3) with the measurement values for
several sources such as 0208$-$512, 0537$-$441, 1057$-$797, and
1454$-$354. The largest difference is seen between the measurement and
the estimate for 0447$-$439, with four orders of magnitude between the
estimate using the \citet{Bonchi2013} parameters, and 6 orders of
magnitude using the \citet{Merloni2003} parameters.

While a large scatter is observed for the fundamental plane
\citep{Nisbet2016}, it probably does not explain a difference of four
or six orders of magnitude. A possibility is that the relativistic
boosting affects the observed masses in supermassive black holes, but
not the Galactic black holes. However, this would imply that the
intrinsic black hole masses in some of the AGN are much lower than
previously believed. We note that the uncertainties on the parameters
of the fundamental plane are large, which is represented in the large
uncertainties of the black hole mass estimates.

\subsection{The strange SED of 2005$-$489}
\label{sec-2005-489}

In general, all SEDs are well described by two log parabolas and a
blackbody to describe the excess. 2005$-$489, a well known VHE
emitter \citep{Aharonian2005}, is the only source with a strong
deviation from this model. VLBI data of the source has been
presented by \cite{Piner2014}.
The multiwavelength SED has been studied several times
\citep{Kaufmann2009,HESS2010}, with the latter arguing about a hard,
separate spectral component emerging in the X-ray observations in
September 2005. This is in agreement with our results of the source
during a high state.
While over most of the energy range it shows a non-thermal parabolic
behavior, its X-ray behavior in the high state ($\alpha$) seems to be
inconsistent with a leptonic model. In the low state ($\beta$), the
photon index $\Gamma = 2.28\pm0.12$ perfectly fits the parabolic
shape, we note that the 104-month averaged BAT data point seems to
indicate a small excess above a pure power law. While this is not
conclusive, the photon index $\Gamma= 1.70\pm0.04$ in the high state
is inconsistent with the parabolic model (and a synchrotron model as
well). The excess is reminiscent of a hadronic proton-synchrotron
signature in the spectrum, while the LAT data might also show a dip in
the spectrum, possibly due to a hadronic pion decay signature. While
this evidence is not conclusive it is the first source to show a clear
deviation, with a large difference in the photon index within a time
span of less than 2 years. A caveat of this SED is the long time range
over which the data were averaged in LAT, but it does not explain the
change in index and the inconsistency between the \Swift/UVOT and
\Swift/XRT data.

\section{Summary \& Conclusions}\label{sec-summary}

We have studied a mainly $\gamma$-ray selected sample of southern
blazars in the framework of the TANAMI project. We chose the 22
\Fermi/LAT brightest sources from the TANAMI sample. This approach
allowed us to use the LAT light curves with a Bayesian blocks
algorithm in order to determine states of statistically constant flux.
For time ranges with quasi-simultaneous data in the X-ray and radio
band we have constructed broadband SEDs. We show that a
``harder-when-brighter'' trend is observed in the high state of the
high energy peak, shifting it to higher frequencies. The Compton
dominance that we find is in agreement with previous results from the
literature. When separated by source state, the Compton dominance in
the high state shows a larger scatter and no discernible trend.
We further study the bolometric Compton dominance by using the
integrated fluxes of peaks fit with parabolas. The scatter in this bolometric
Compton dominance is lower, but it shows that the peak flux is a
reliable tracer of the bolometric flux.

We study the temperature range and shape of the BBB. We find that the
temperatures are consistent with previous results, showing
temperatures that are too low for the expected accretion disk
emission. It can possibly be explained by reprocessing the accretion
disk emission by BLR clouds, which is also able to solve other
problems. We also find 
that unexpectedly a single temperature model can best explain all
BBBs, which is inconsistent with an accretion disk origin. It is
unclear, whether this is true for all blazars. No detailed model
exists for a more realistic accretion disk which might be thick
and/or warped. It is unclear how this would change the expected
thermal emission.

We further study the fundamental plane of black holes as a tool for
estimating black hole masses. We find that the parameters by
\citet{Bonchi2013} for many sources are in good agreement with the
black hole masses from \citet{Ghisellini2010}, while this is not the
case for other parameter estimates \citep[those
  from][]{Merloni2003,Koerding2006,Gultekin2009,Nisbet2016}, however
the uncertainties are dominated by systematic effects and are very
large. It shows that choosing the source population introduces
selection effects. For 
a few sources such as 0447$-$439, the measured mass is not in agreement
with any of the parameters, with a very large offset of greater than
four orders of magnitude. We suggest that this might be due to
boosting effects. This result would imply, however, that some AGN
black hole masses are much lower than previously suspected.


\begin{acknowledgements}
  We thank the referee for the helpful comments.
  We thank S. Cutini for her useful comments. We thank S. Markoff for
  helpful discussions. 
  We thank J. Perkins, L. Baldini, and S. Digel for carefully reading
  the manuscript. We thank M. Buxton for
  her help with the SMARTS data. We acknowledge support and partial
  funding by the Deutsche
  Forschungsgemeinschaft grant WI 1860-10/1 (TANAMI) and GRK 1147,
  Deutsches Zentrum f\"ur Luft- und Raumfahrt grants 50\,OR\,1311 and
  50\,OR\,1103, and the Helmholtz Alliance for Astroparticle Physics
  (HAP). 
  This research was funded in part by NASA through Fermi Guest
  Investigator grants NNH09ZDA001N, NH10ZDA001N, NNH12ZDA001N, and
  NNH13ZDA001N-FERMI.
  This research was supported by an appointment to the NASA
  Postdoctoral Program at the Goddard Space Flight Center,
  administered by Oak Ridge Associated Universities through a contract
  with NASA.
  E.R. was partially supported by the Spanish MINECO project
  AYA2012-38491-C02-01 and by the 
  Generalitat Valenciana project PROMETEO II/2014/057. We thank J.E.~Davis
  for the development of the \texttt{slxfig} module that has been used
  to prepare the figures in this work.
  We thank T.~Johnson for the \Fermi/LAT SED scripts, which were used to
  calculate the \Fermi/LAT spectra.
 This research has made use of a collection of
  \ISIS scripts provided by the Dr. Karl Remeis-Observatory, Bamberg,
  Germany at \url{http://www.sternwarte.uni-erlangen.de/isis/}. The
  Long Baseline Array and Australia Telescope Compact Array are part
  of the Australia Telescope National Facility, which is funded by the
  Commonwealth of Australia for operation as a National Facility
  managed by CSIRO.

  This paper has made use of up-to-date SMARTS optical/near-infrared
  light curves that are available at
  \url{www.astro.yale.edu/smarts/glast/home.php}. 

  The \textsl{Fermi} LAT Collaboration acknowledges generous ongoing
  support from a number of agencies and institutes that have supported
  both the development and the operation of the LAT as well as
  scientific data analysis. These include the National Aeronautics and
  Space Administration and the Department of Energy in the United
  States, the Commissariat \`a l'Energie Atomique and the Centre
  National de la Recherche Scientifique / Institut National de
  Physique Nucl\'eaire et de Physique des Particules in France, the
  Agenzia Spaziale Italiana and the Istituto Nazionale di Fisica
  Nucleare in Italy, the Ministry of Education, Culture, Sports,
  Science and Technology (MEXT), High Energy Accelerator Research
  Organization (KEK) and Japan Aerospace Exploration Agency (JAXA) in
  Japan, and the K.A.~Wallenberg Foundation, the Swedish Research
  Council and the Swedish National Space Board in Sweden. Additional
  support for science analysis during the operations phase is
  gratefully acknowledged from the Istituto Nazionale di Astrofisica
  in Italy and the Centre National d'\'Etudes Spatiales in France.
 \end{acknowledgements}

\bibliographystyle{jwaabib}
\bibliography{mnemonic,aa_abbrv,tanami,sources,xray,inst_cat,results,stat,agn,sed}

\begin{appendix}
\onecolumn

\section{Results: Light curves and SEDs }
\subsection{\Fermi/LAT light curves}\label{ap-lc}
\begin{figure}[h!]
\label{fig-seds}
\centering
\includegraphics[width=0.48\textwidth]{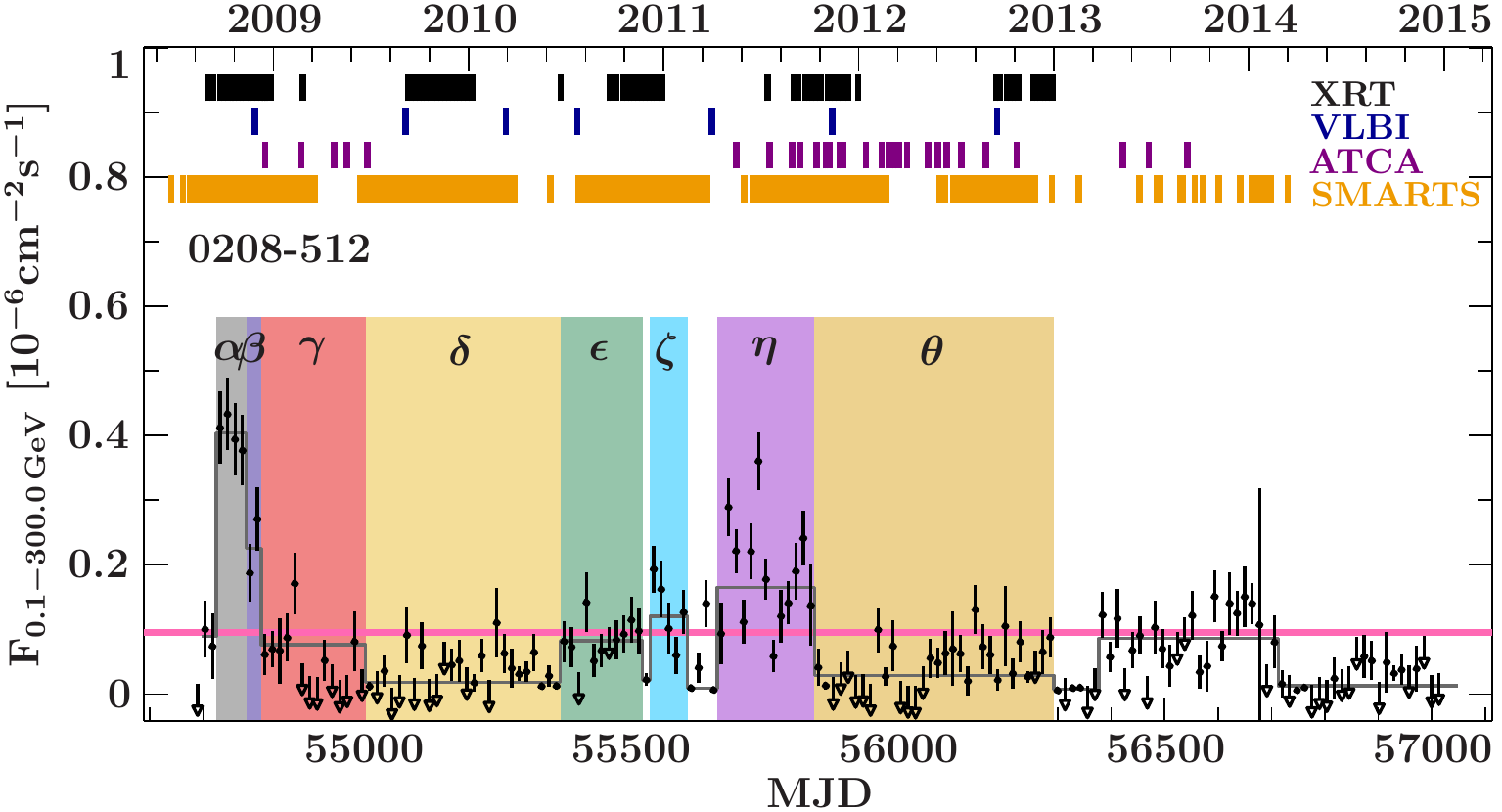}\hfill
\includegraphics[width=0.48\textwidth]{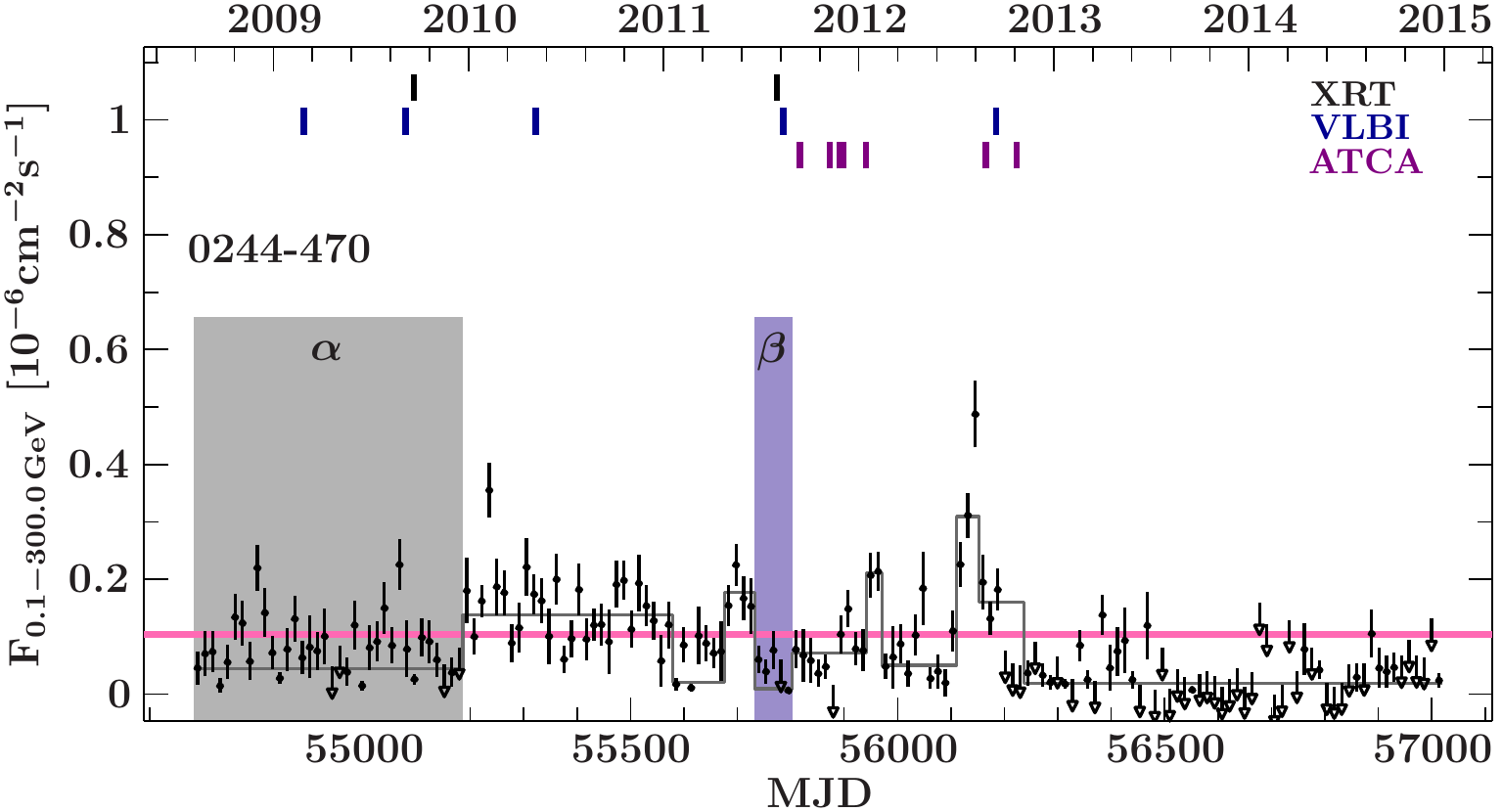}\\[0.5\baselineskip]
\includegraphics[width=0.48\textwidth]{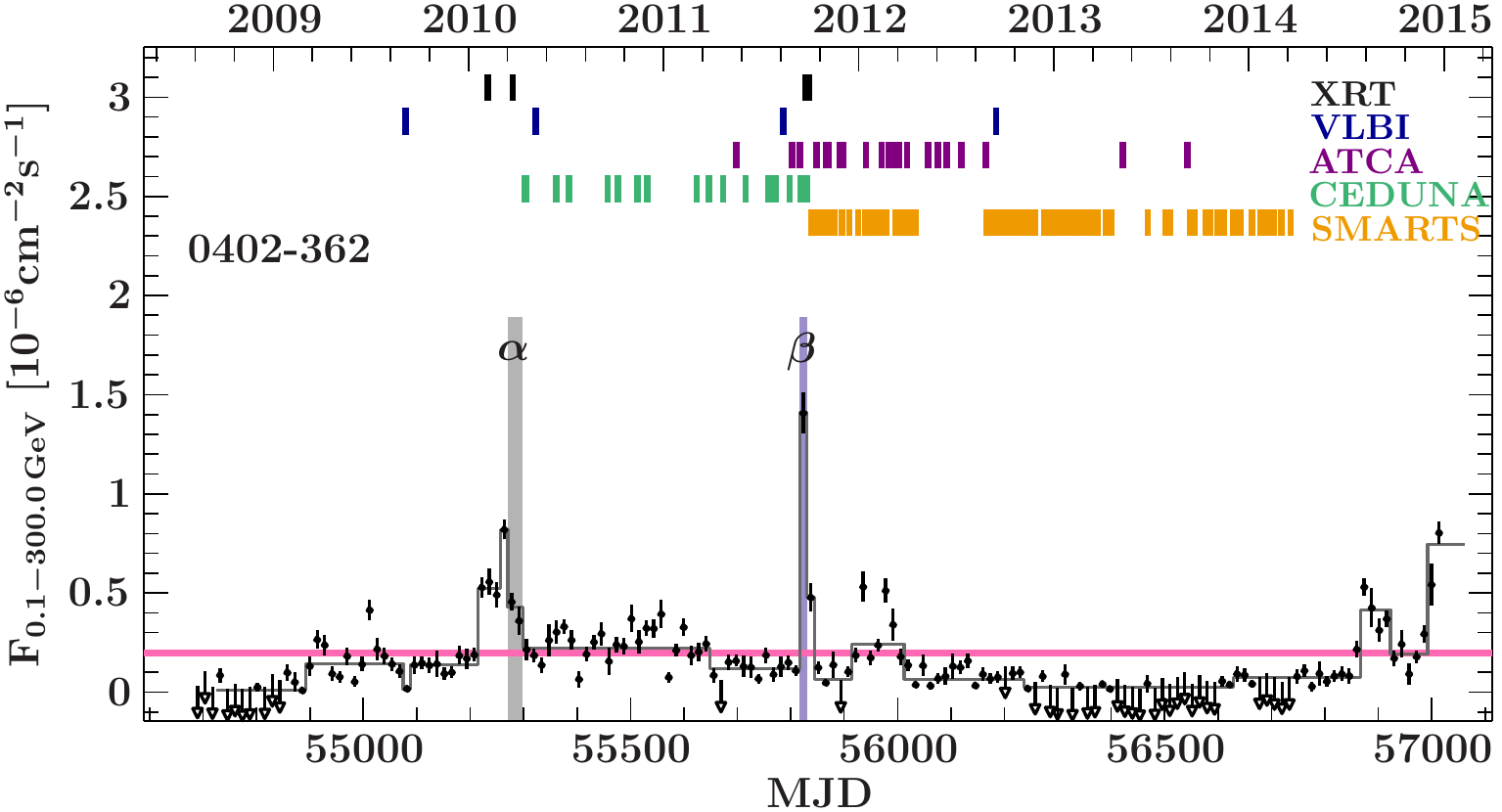}\hfill
\includegraphics[width=0.48\textwidth]{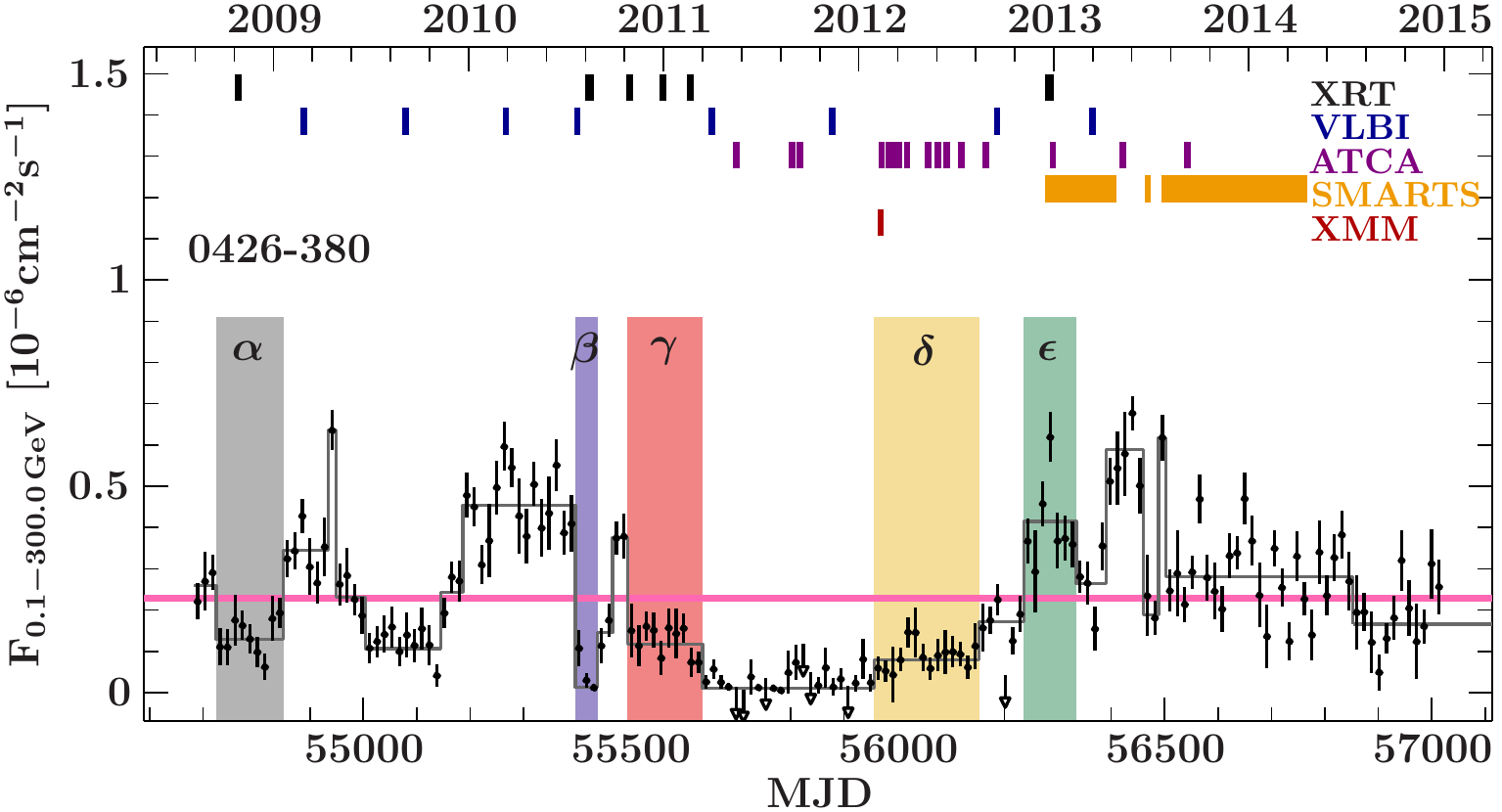}\\[0.5\baselineskip]
\includegraphics[width=0.48\textwidth]{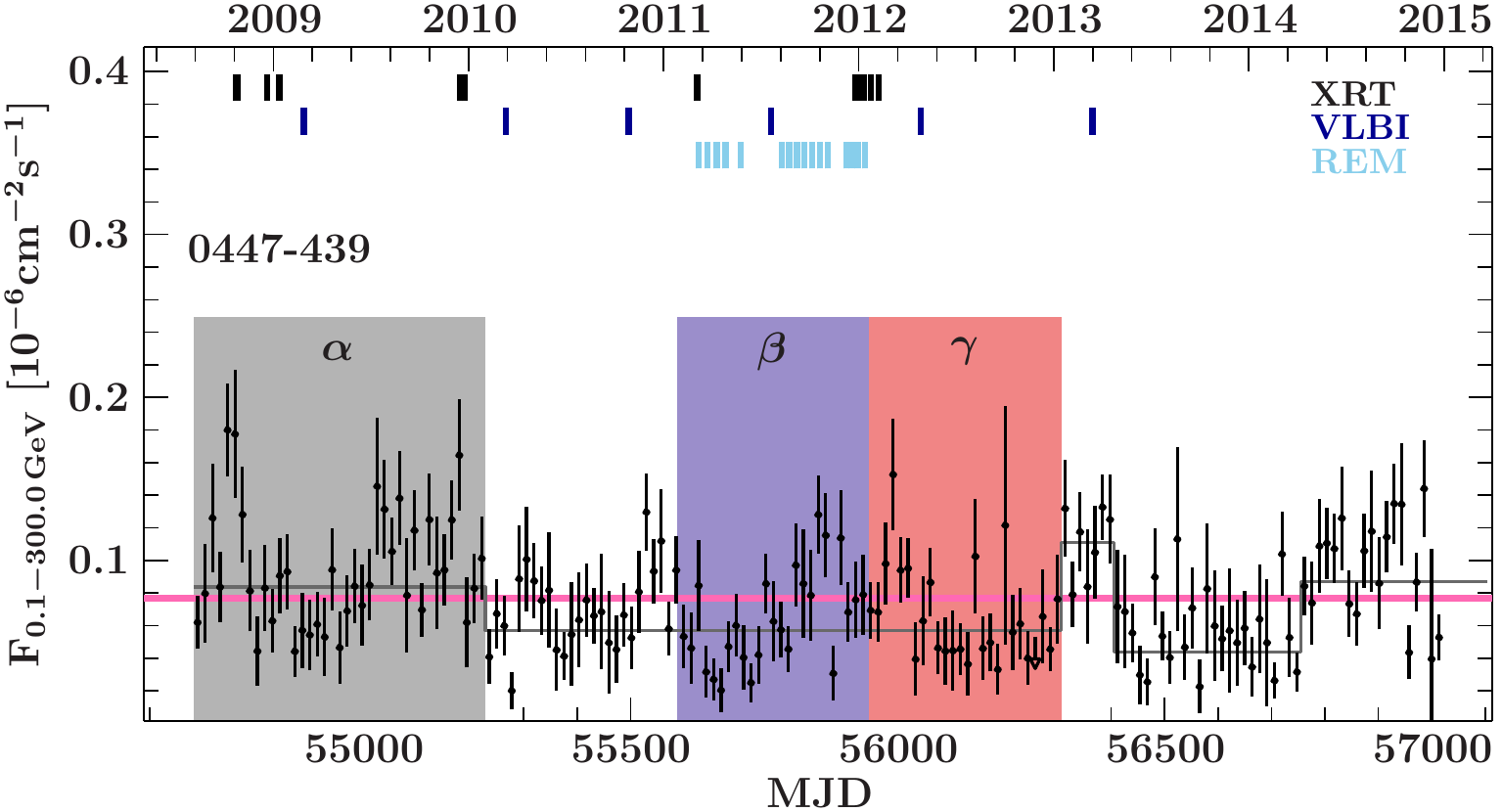}\hfill
\includegraphics[width=0.48\textwidth]{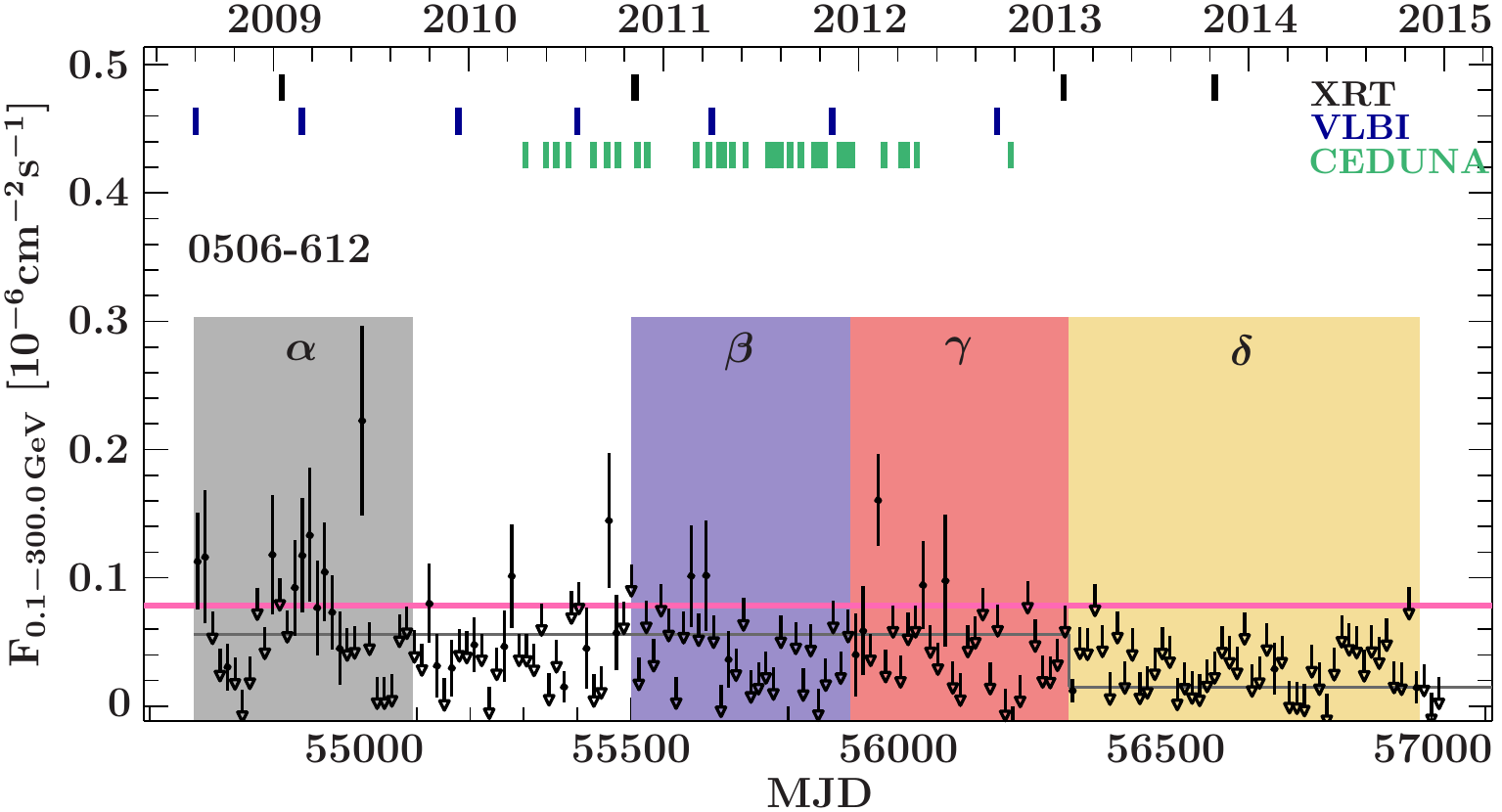}\\[0.5\baselineskip]
\includegraphics[width=0.48\textwidth]{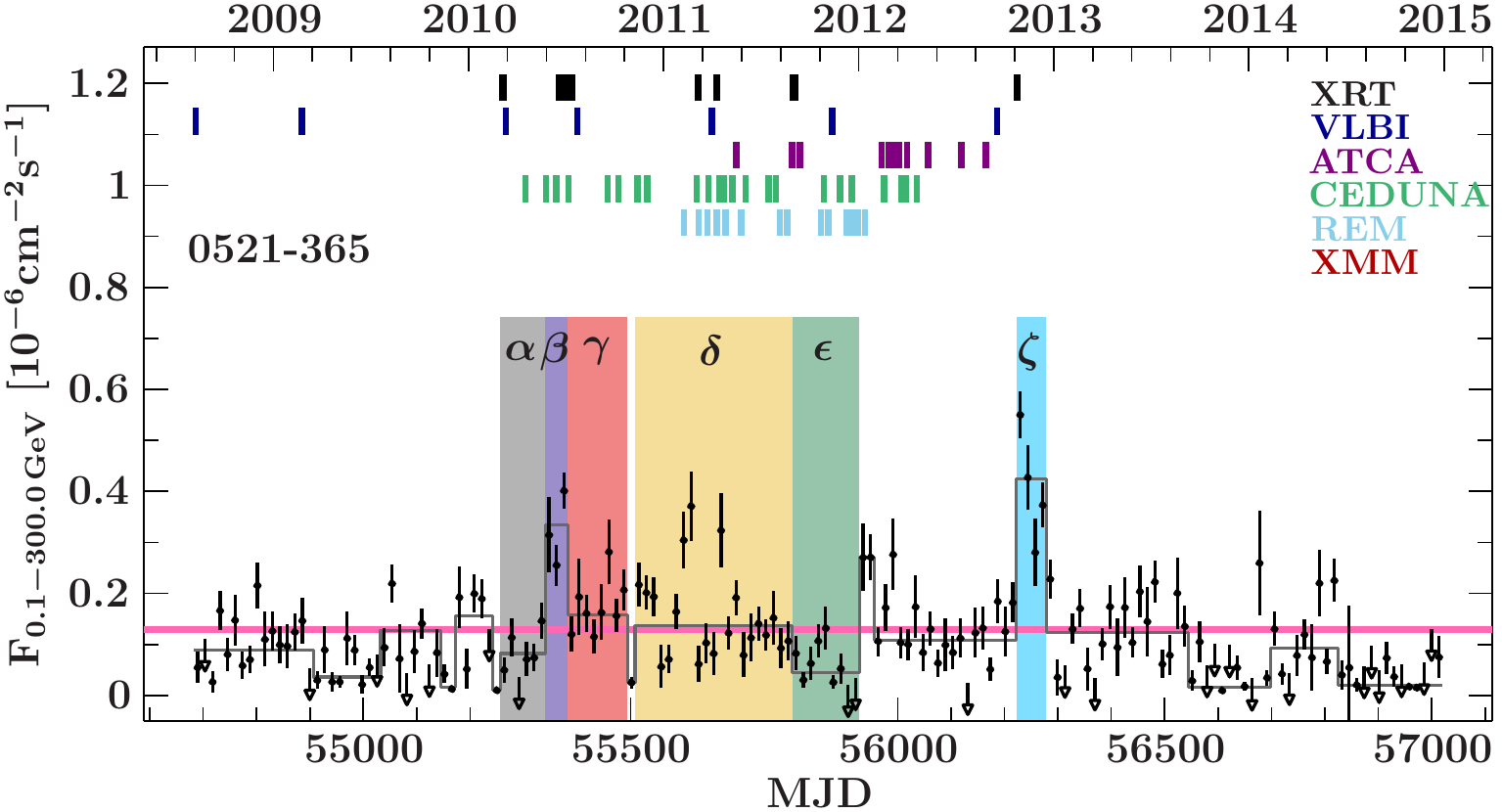}\hfill
\includegraphics[width=0.48\textwidth]{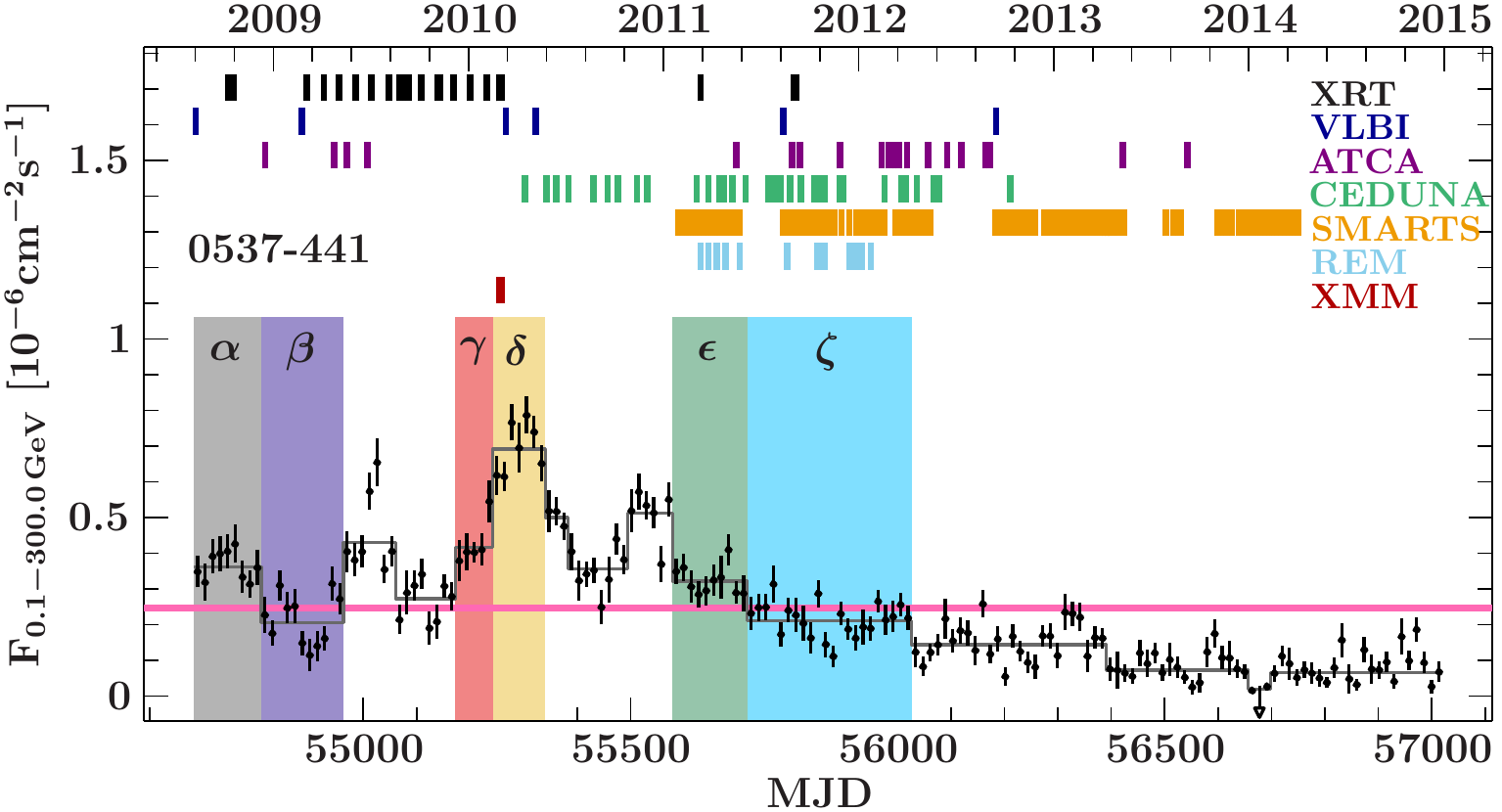}\\[0.5\baselineskip]
\caption{\Fermi/LAT light curves for all sources with a known
  redshift, from August 4, 2008 up to January 1, 2015. A Bayesian blocks
  analysis was performed on the data and is shown in dark gray. The
  horizontal pink line shows the average flux over the full
  light curve. Observations by \Swift, \XMMNewton, REM, SMARTS,
  Ceduna, ATCA, or VLBI are marked with a line at the corresponding
  time. Blocks with sufficient data for a broadband SED are marked in
  color, and labeled with Greek letters. The colors correspond to the
  colors used in the broadband spectra. }
\end{figure}
\clearpage

\begin{figure}
\addtocounter{figure}{-1}
\includegraphics[width=0.48\textwidth]{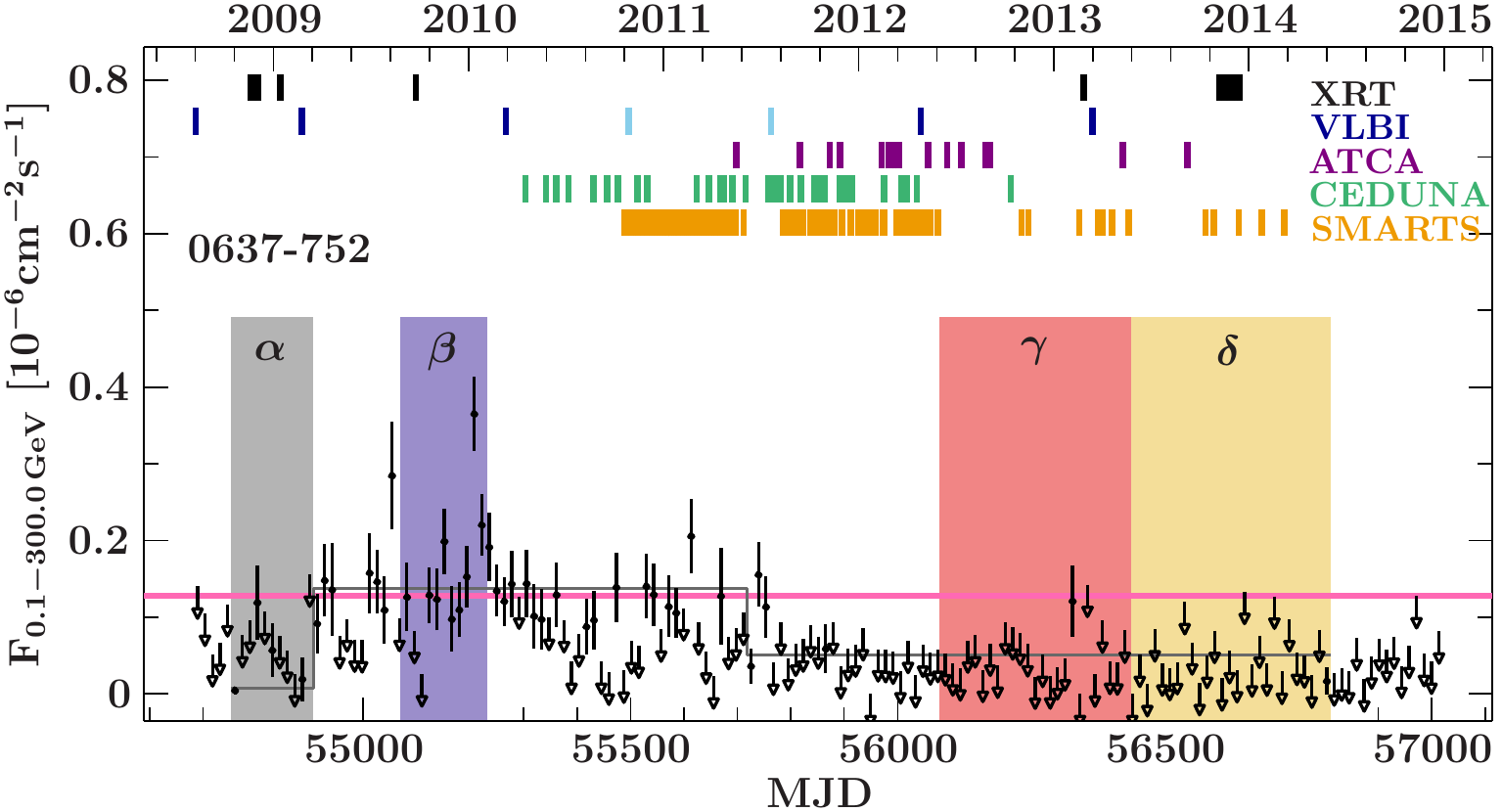}\hfill
\includegraphics[width=0.48\textwidth]{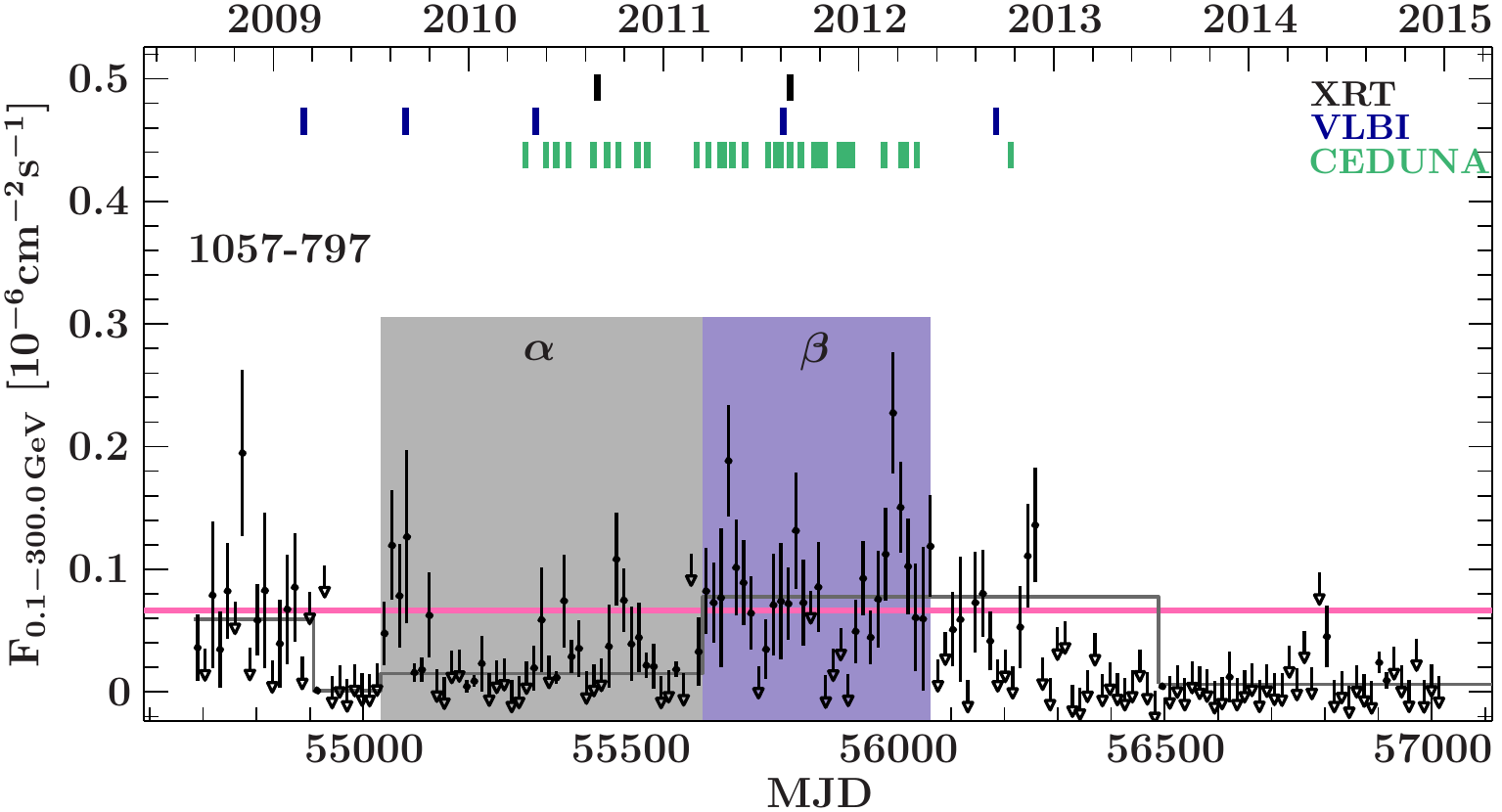}\\[0.5\baselineskip]
\includegraphics[width=0.48\textwidth]{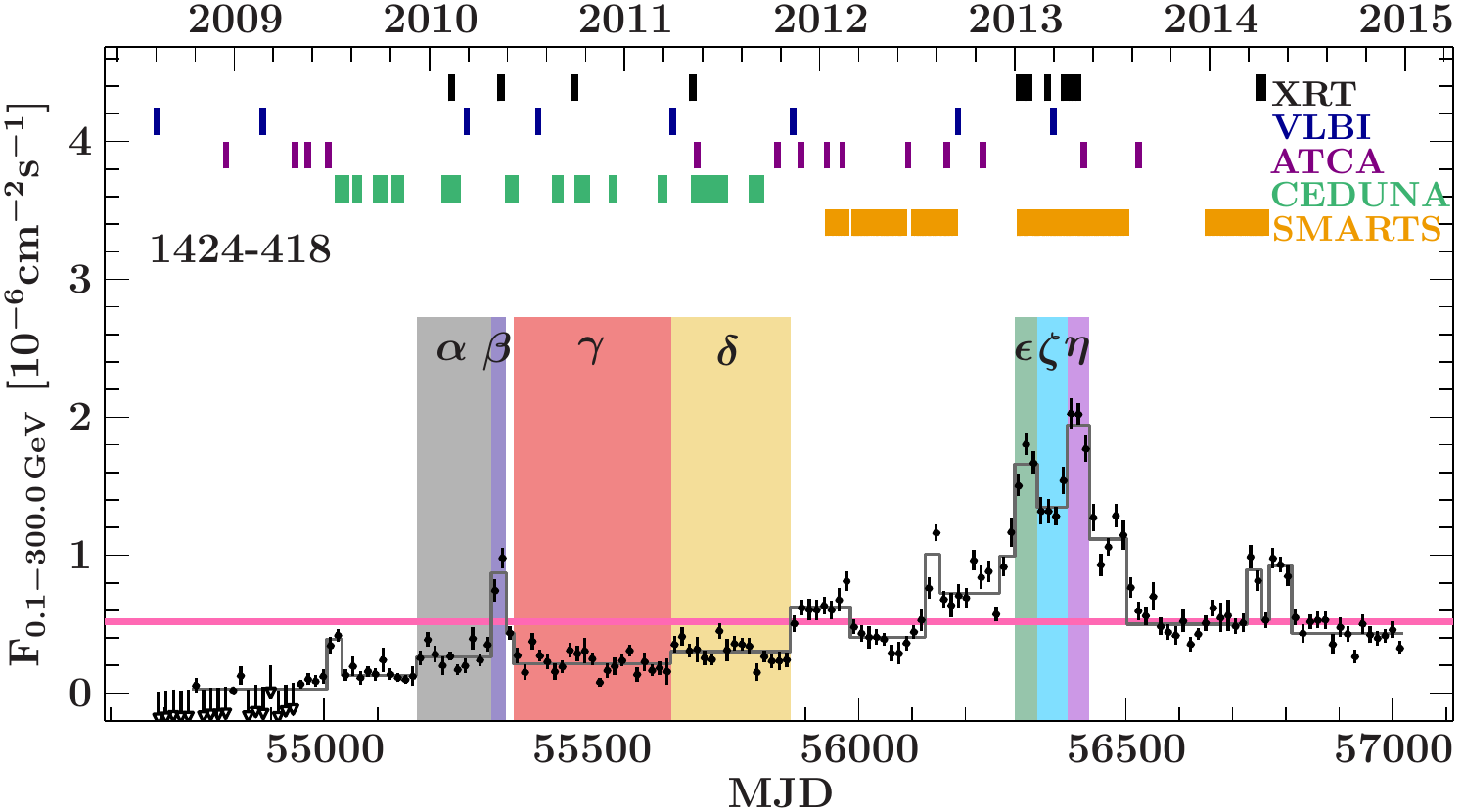}\hfill
\includegraphics[width=0.48\textwidth]{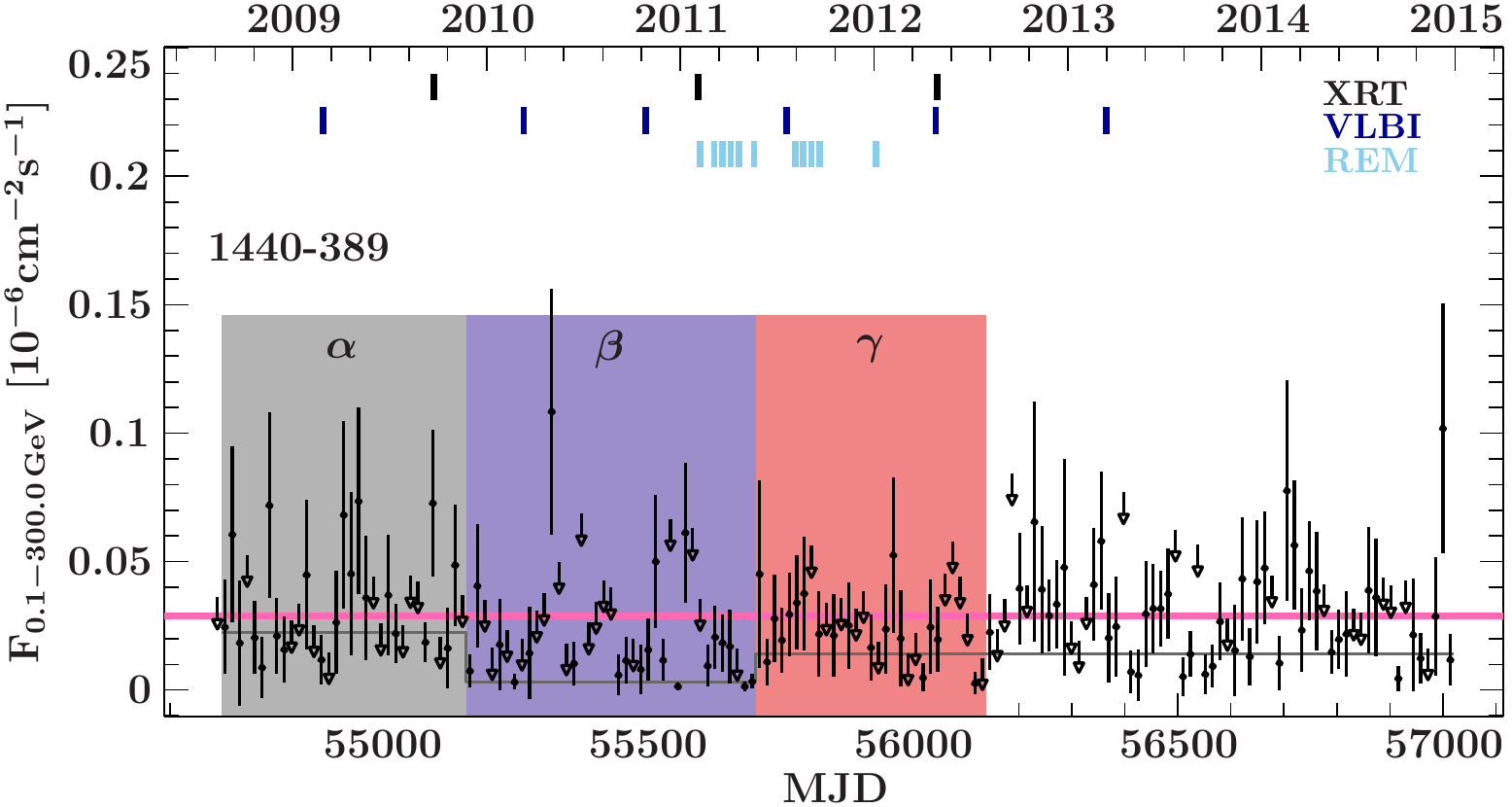}\\[0.5\baselineskip]
\includegraphics[width=0.48\textwidth]{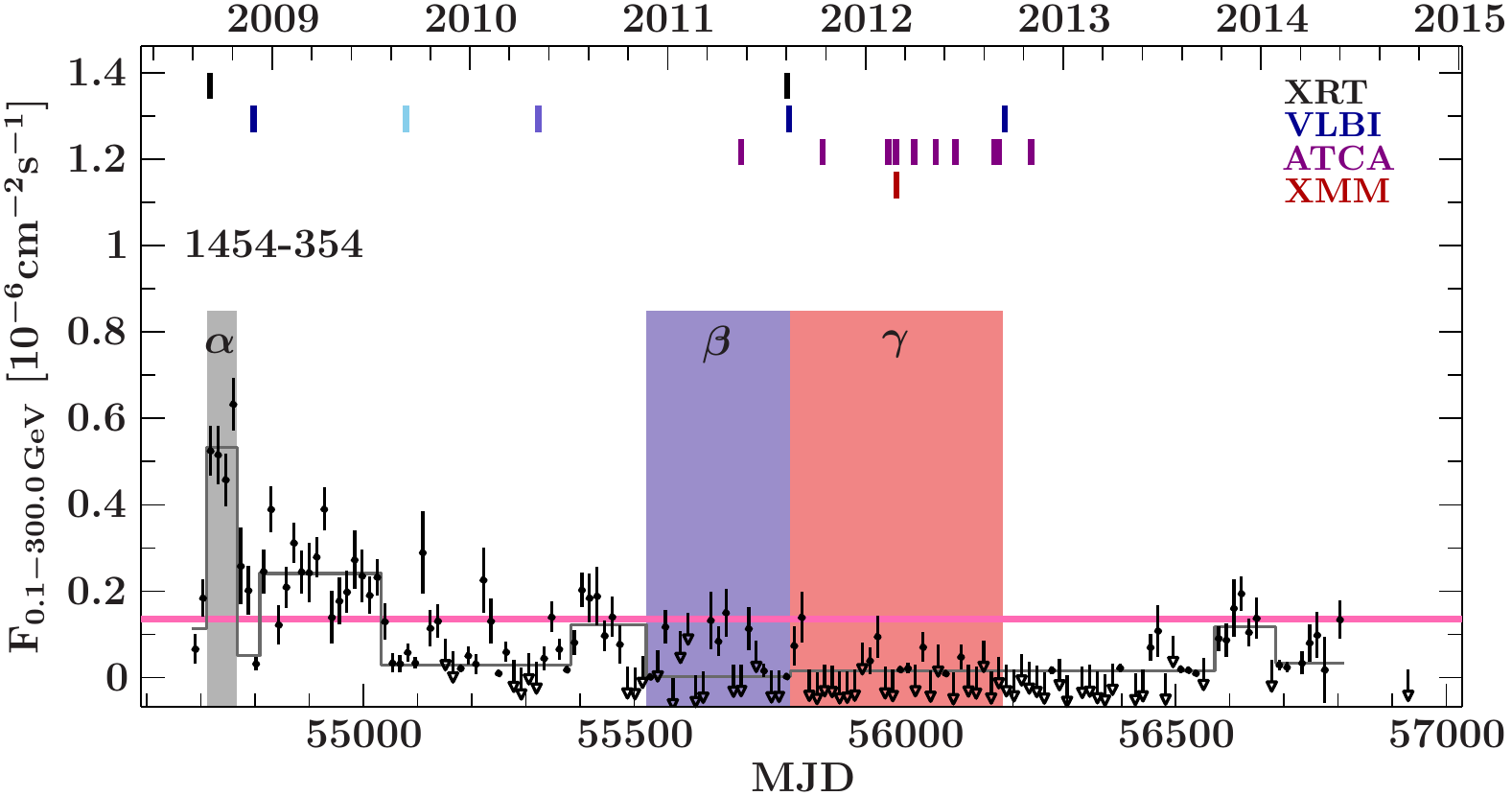}\hfill
\includegraphics[width=0.48\textwidth]{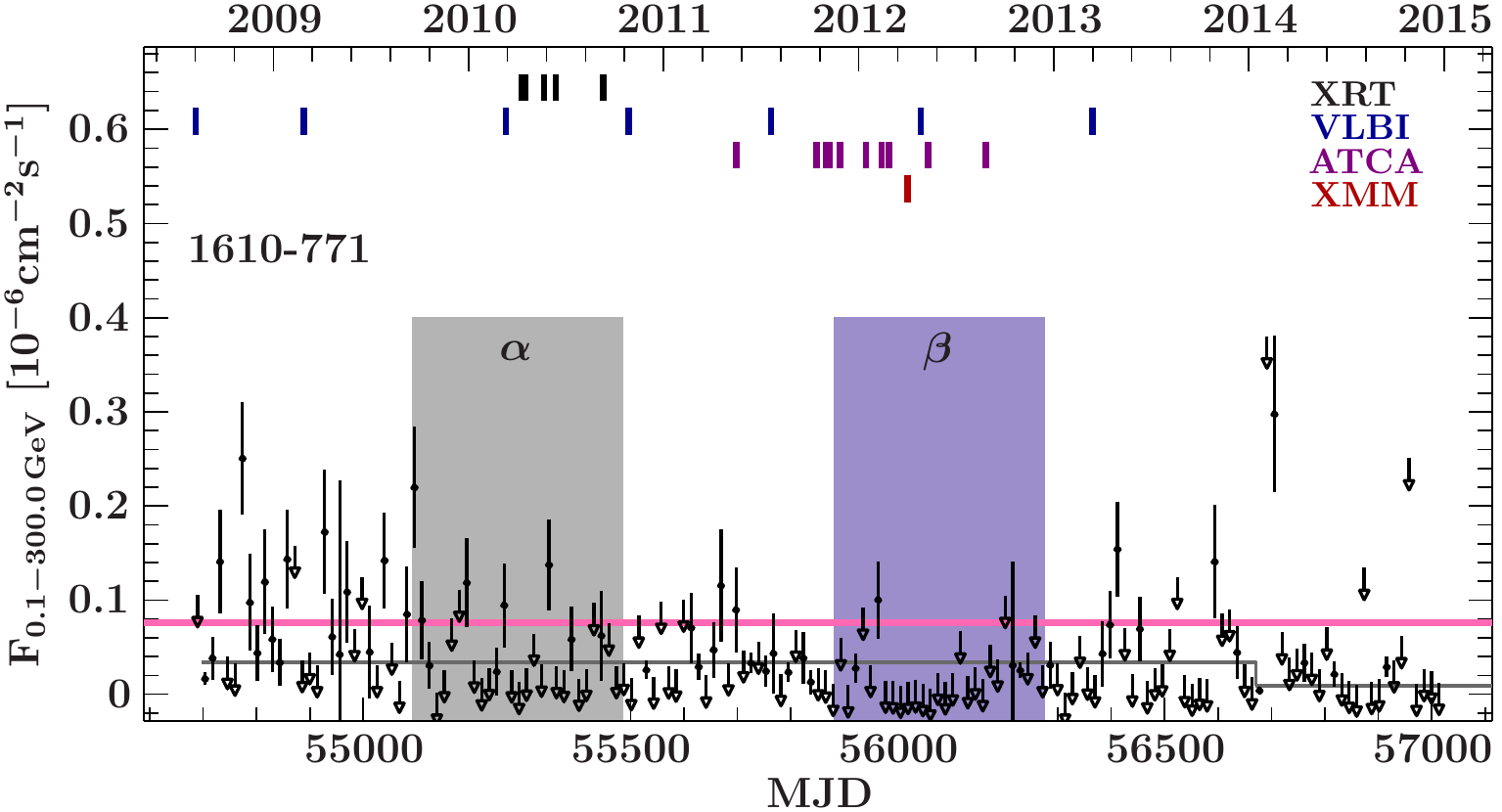}\\[0.5\baselineskip]
\includegraphics[width=0.48\textwidth]{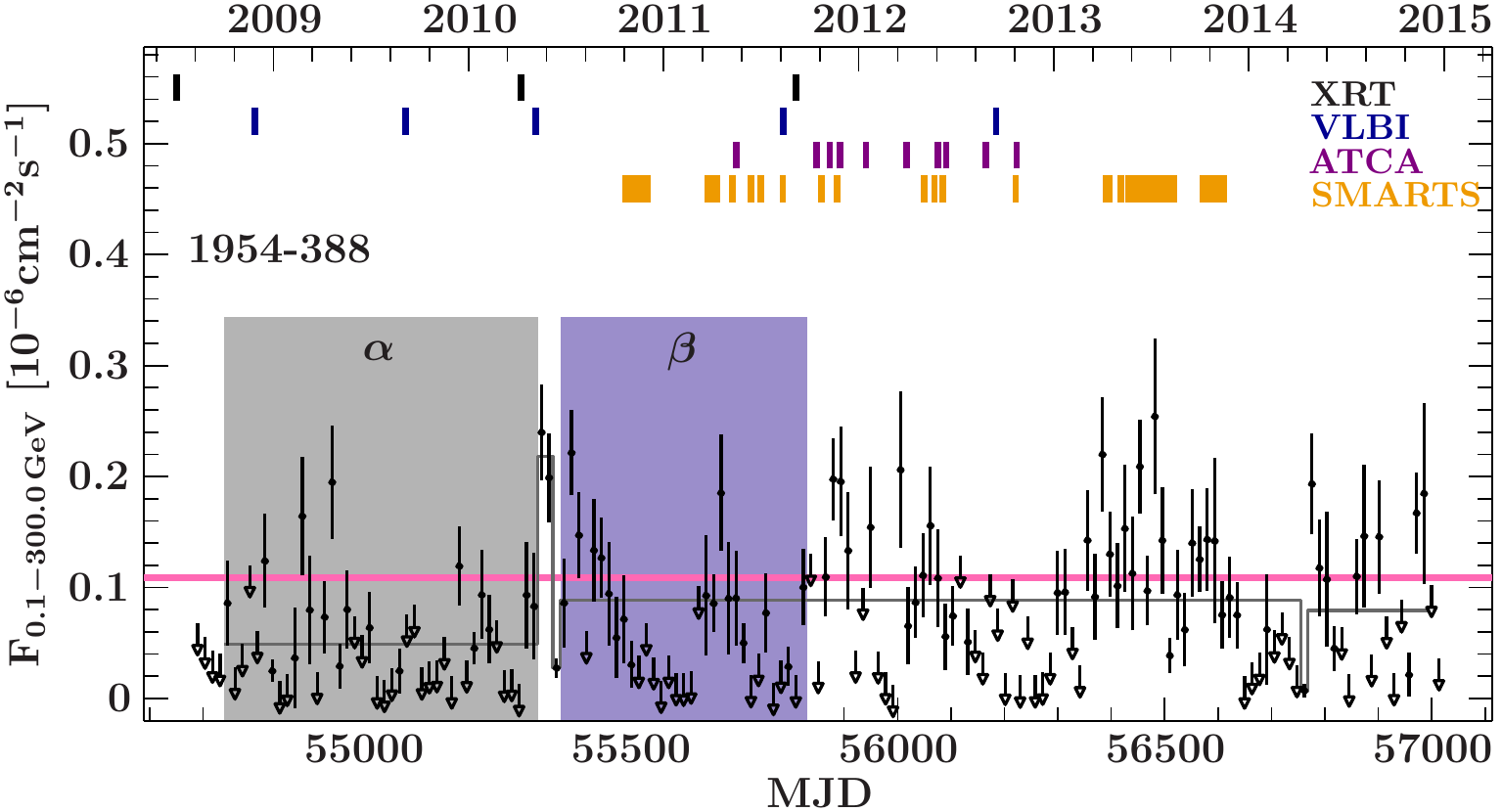}\hfill
\includegraphics[width=0.48\textwidth]{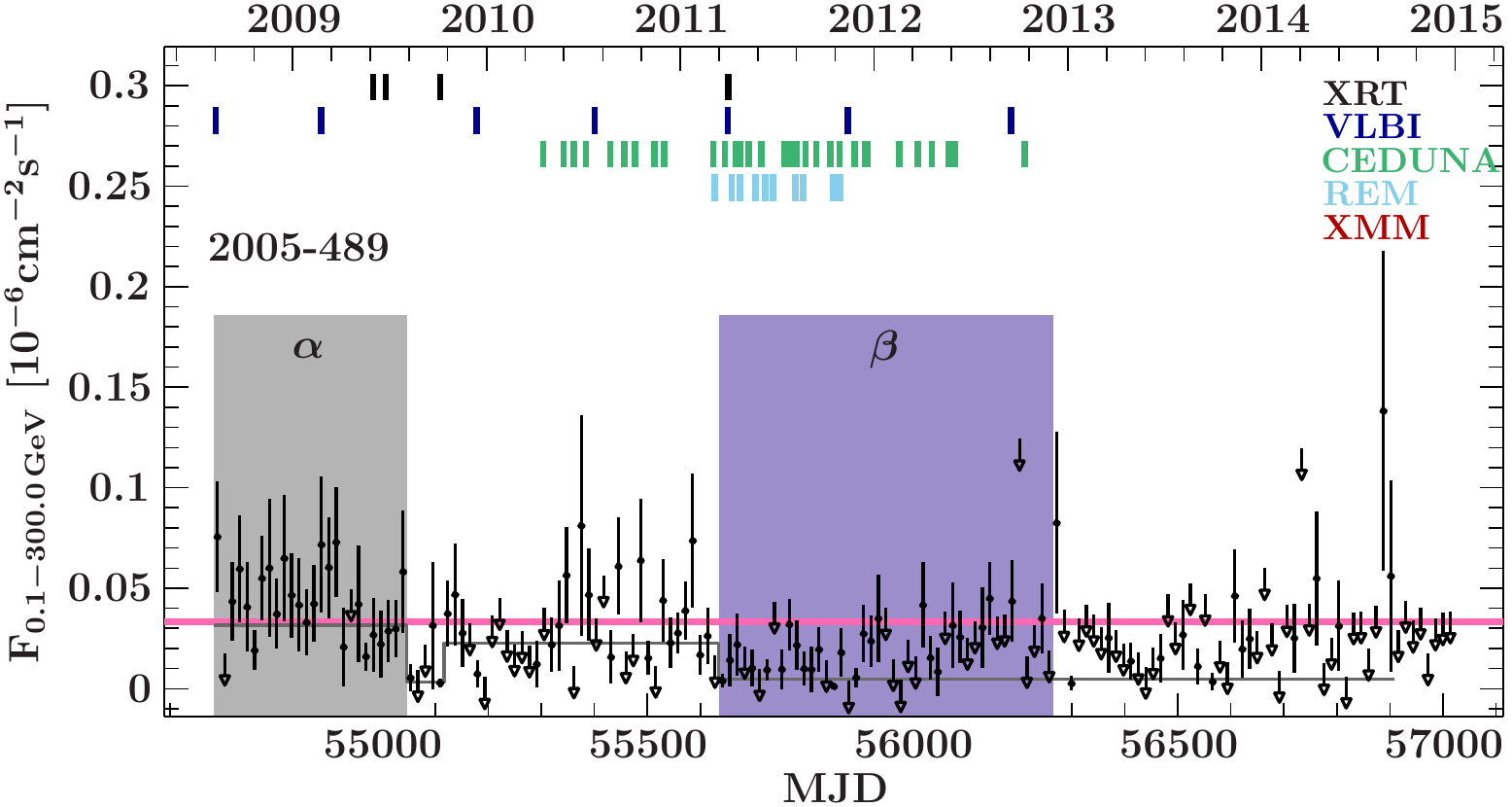}\\[0.5\baselineskip]
\includegraphics[width=0.48\textwidth]{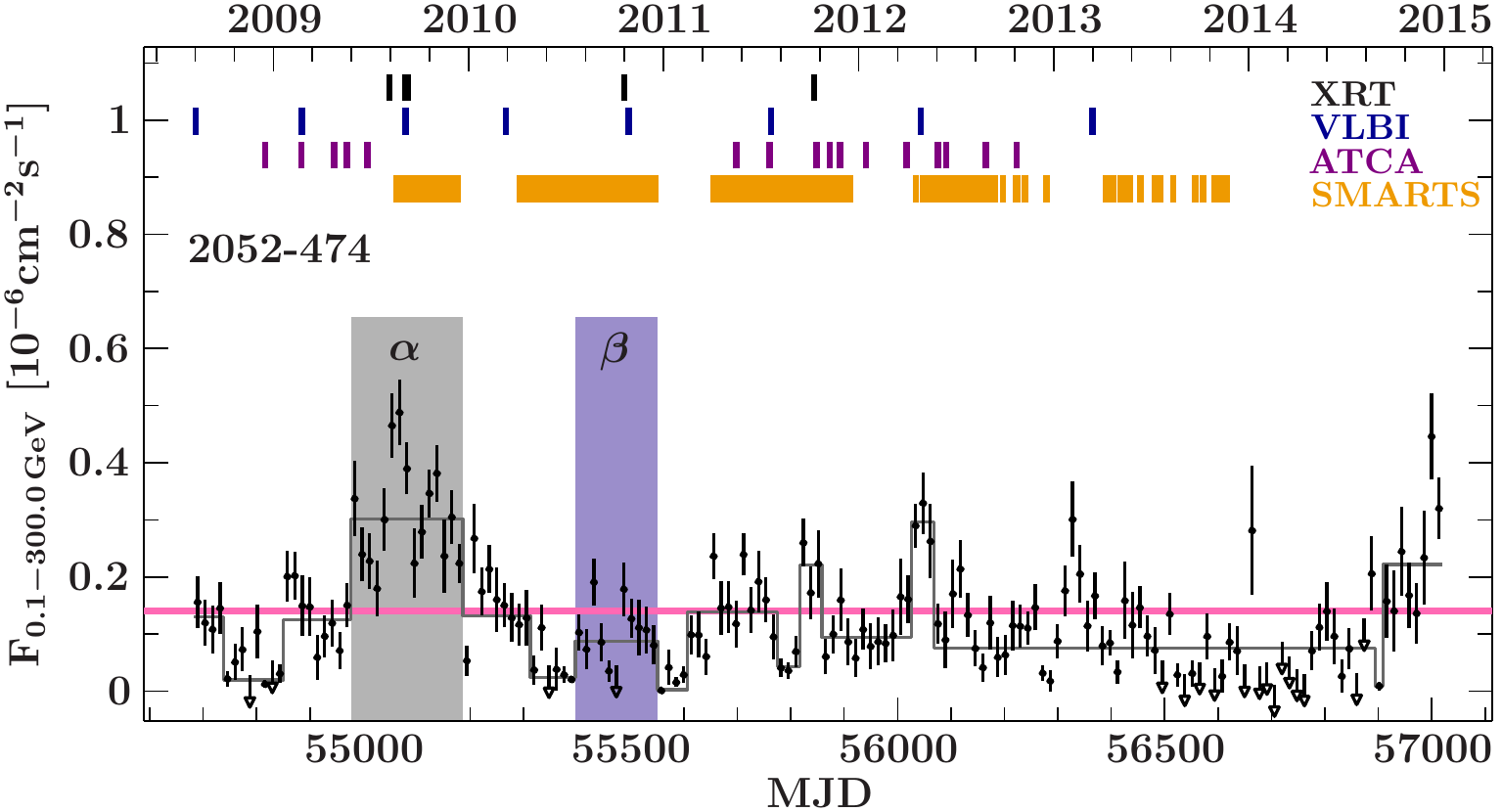}\hfill
\includegraphics[width=0.48\textwidth]{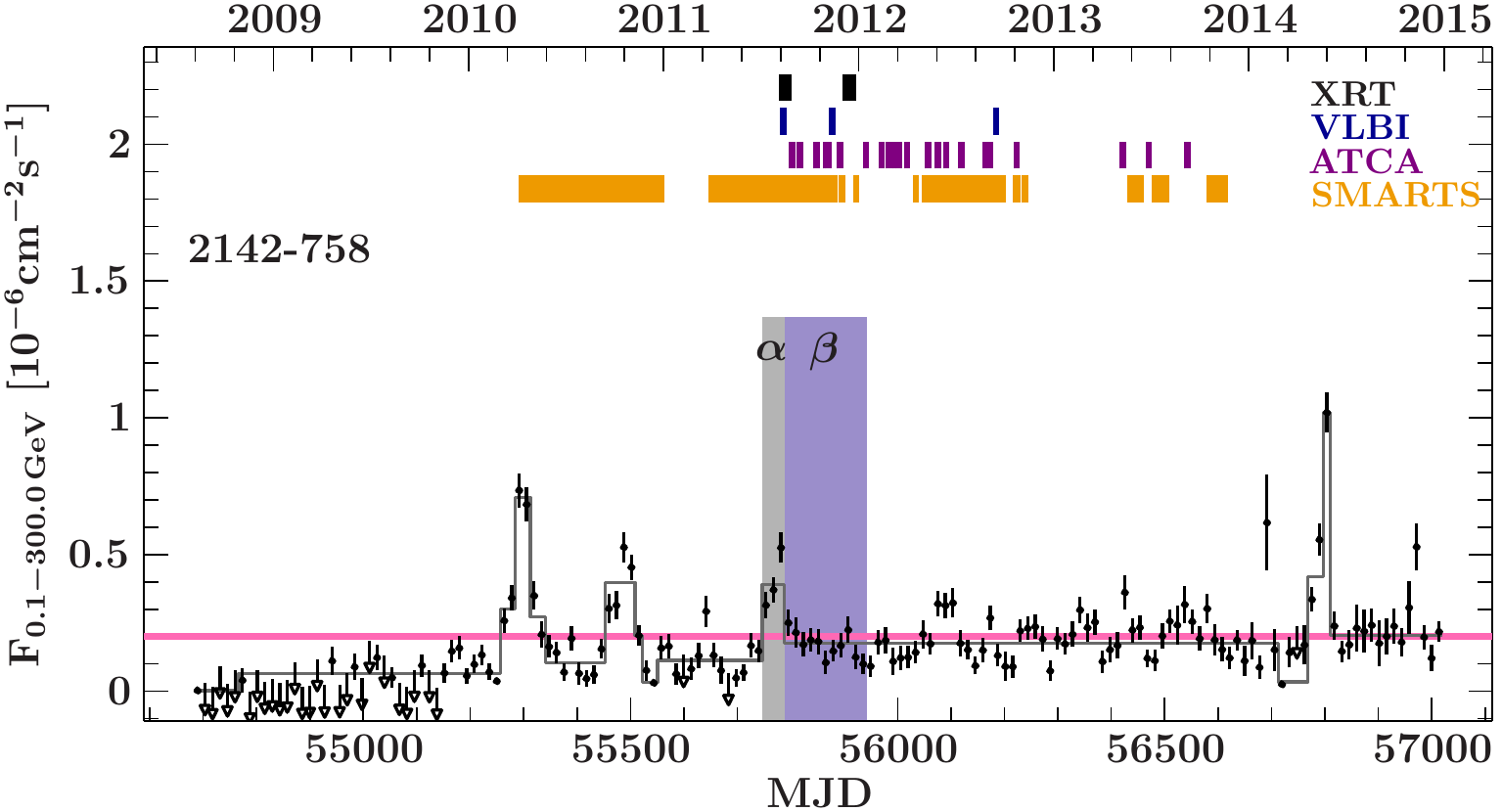}\\[0.5\baselineskip]
\caption{(contd.)}
\end{figure}
\clearpage

\begin{figure}
\addtocounter{figure}{-1}
\includegraphics[width=0.48\textwidth]{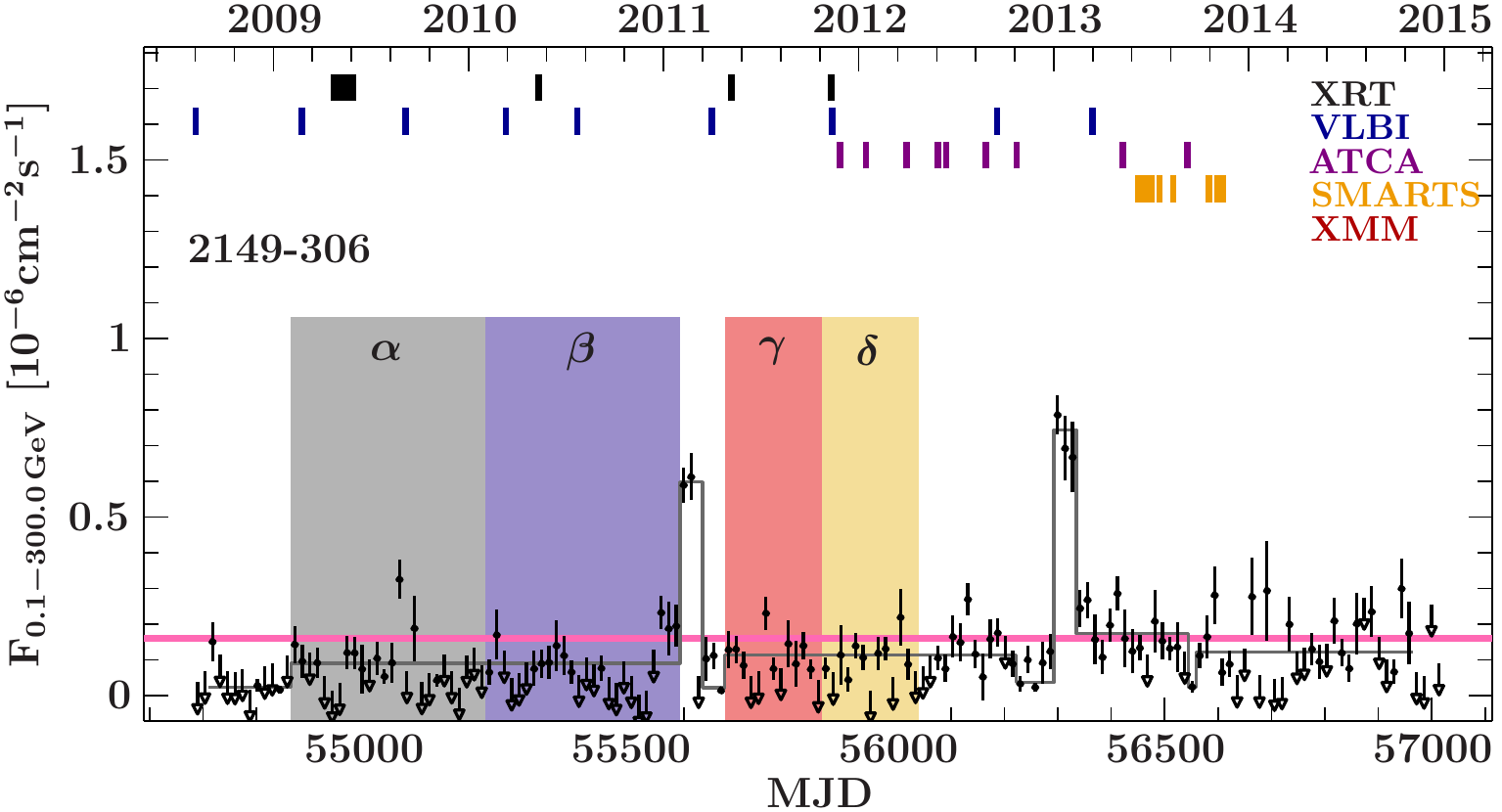}\hfill
\includegraphics[width=0.48\textwidth]{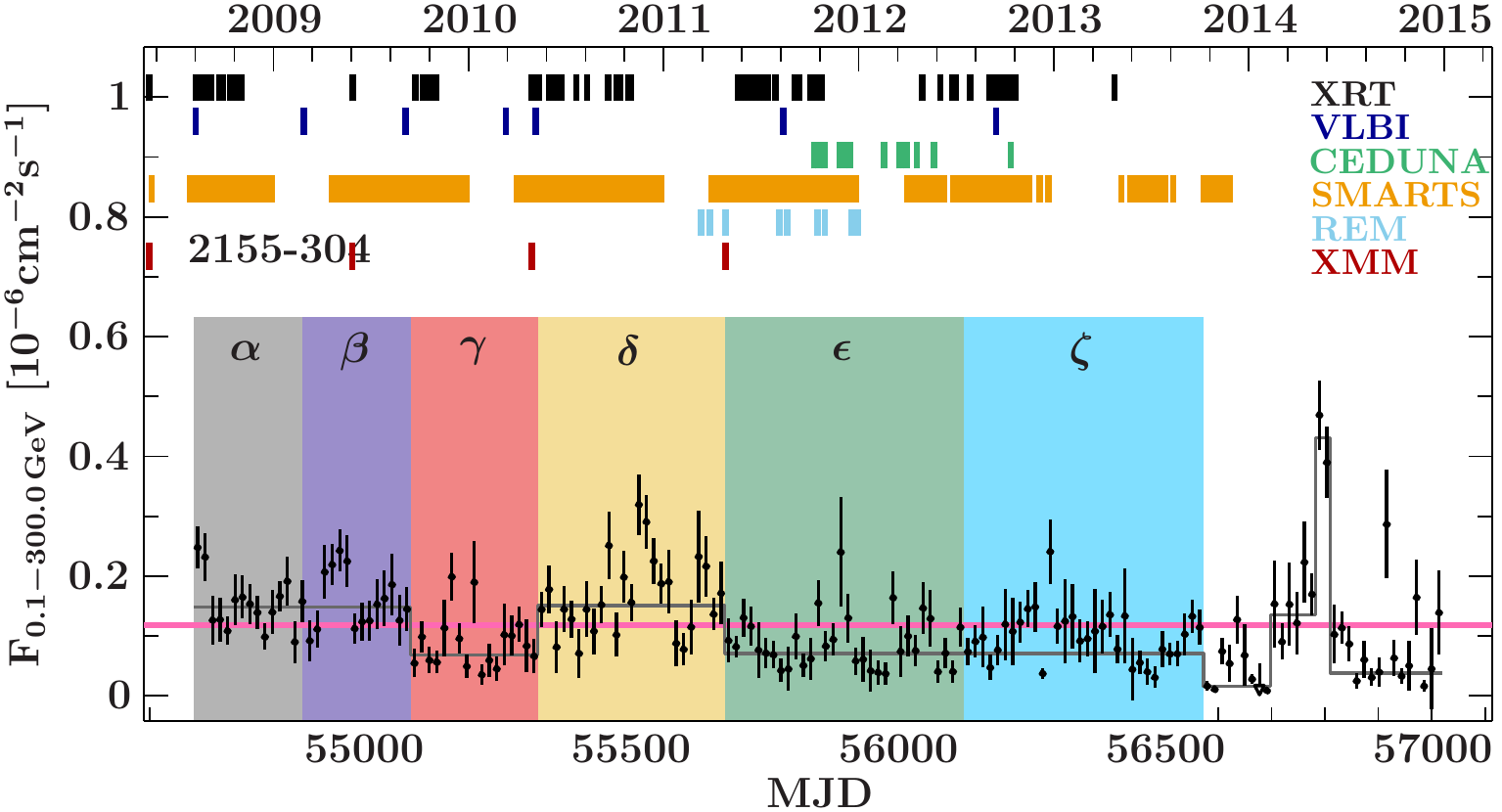}\\[0.5\baselineskip]
\caption{(contd.)}
\end{figure}

\clearpage

\begin{figure}
\includegraphics[width=0.48\textwidth]{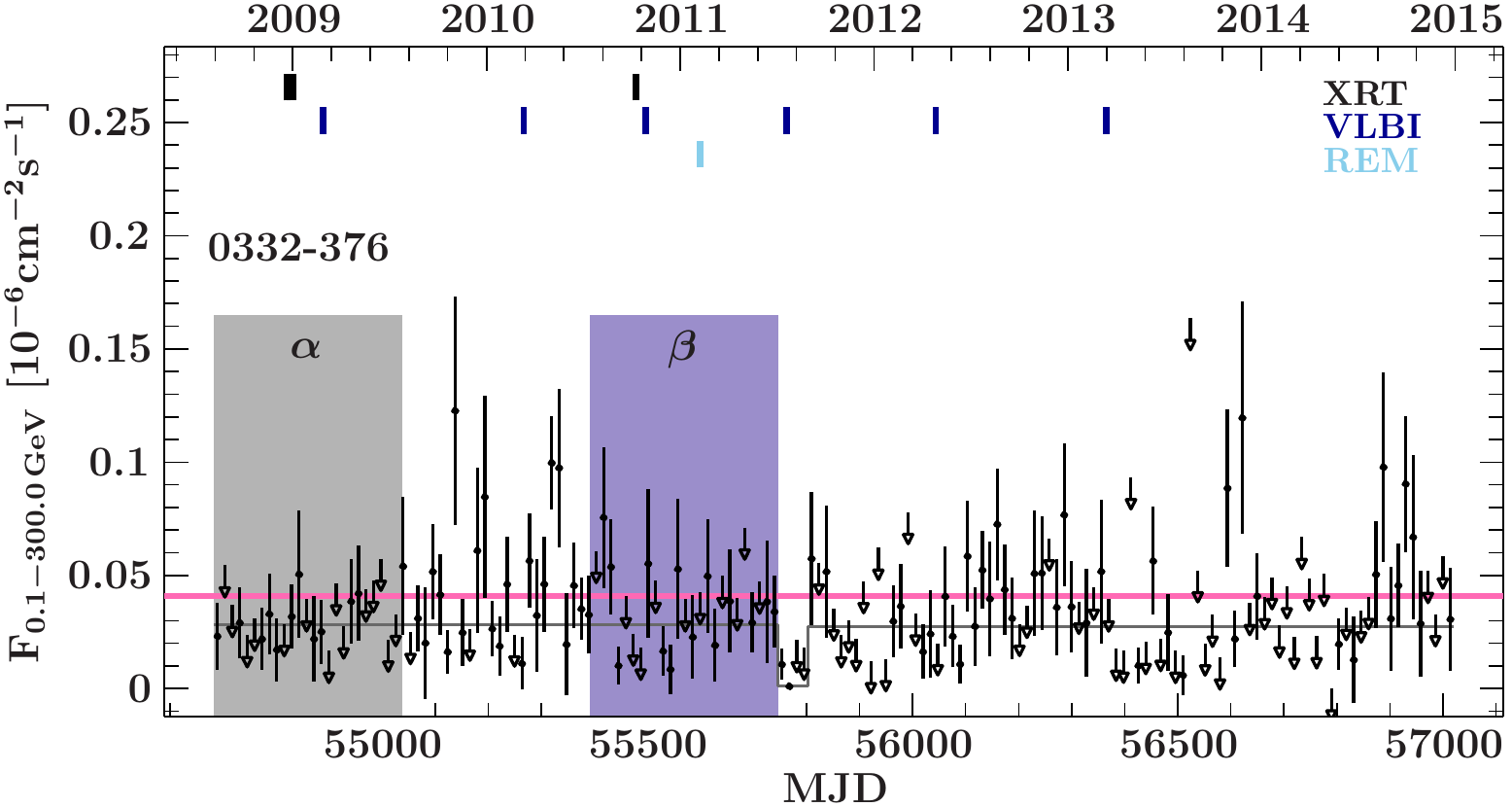}\hfill
\includegraphics[width=0.48\textwidth]{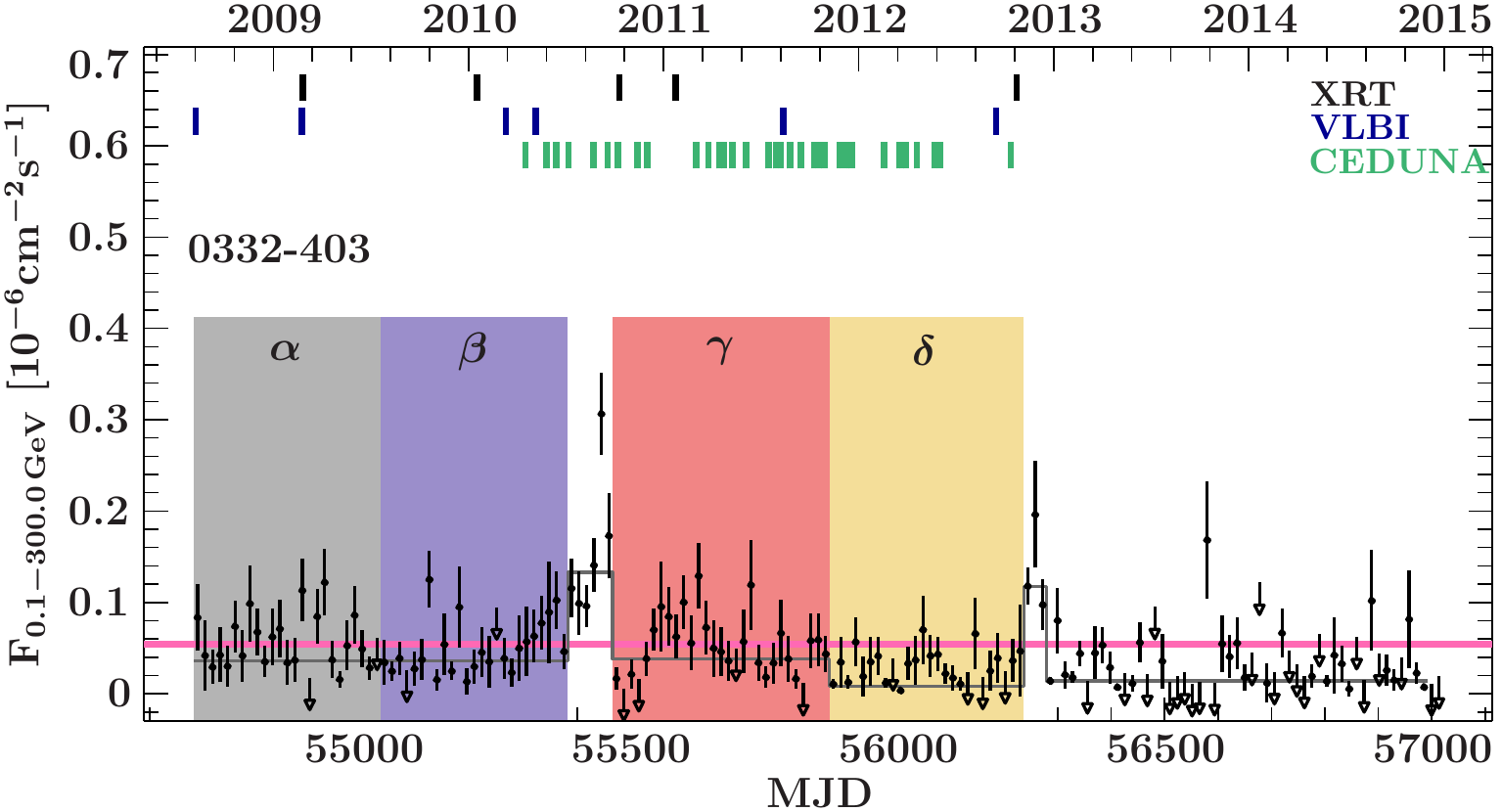}\\[0.5\baselineskip]
\caption{\Fermi/LAT light curves for both sources without a known
  redshift, from August 4, 2008 up to 2015 January 1. A Bayesian blocks
  analysis was performed on the data and is shown in dark gray. The
  horizontal pink line shows the average flux over the full
  light curve. Observations by \Swift, REM, Ceduna, and VLBI are marked with a
  line at the corresponding time. Blocks with sufficient data for a
  broadband SED are marked in color, and labeled with Greek letters.
  The colors correspond to the colors used in the broadband spectra.}
\end{figure}

\clearpage
\subsection{Broadband spectral energy distributions}
\label{ap-sed}

\begin{figure}[h!]
\includegraphics[height=0.24\textheight]{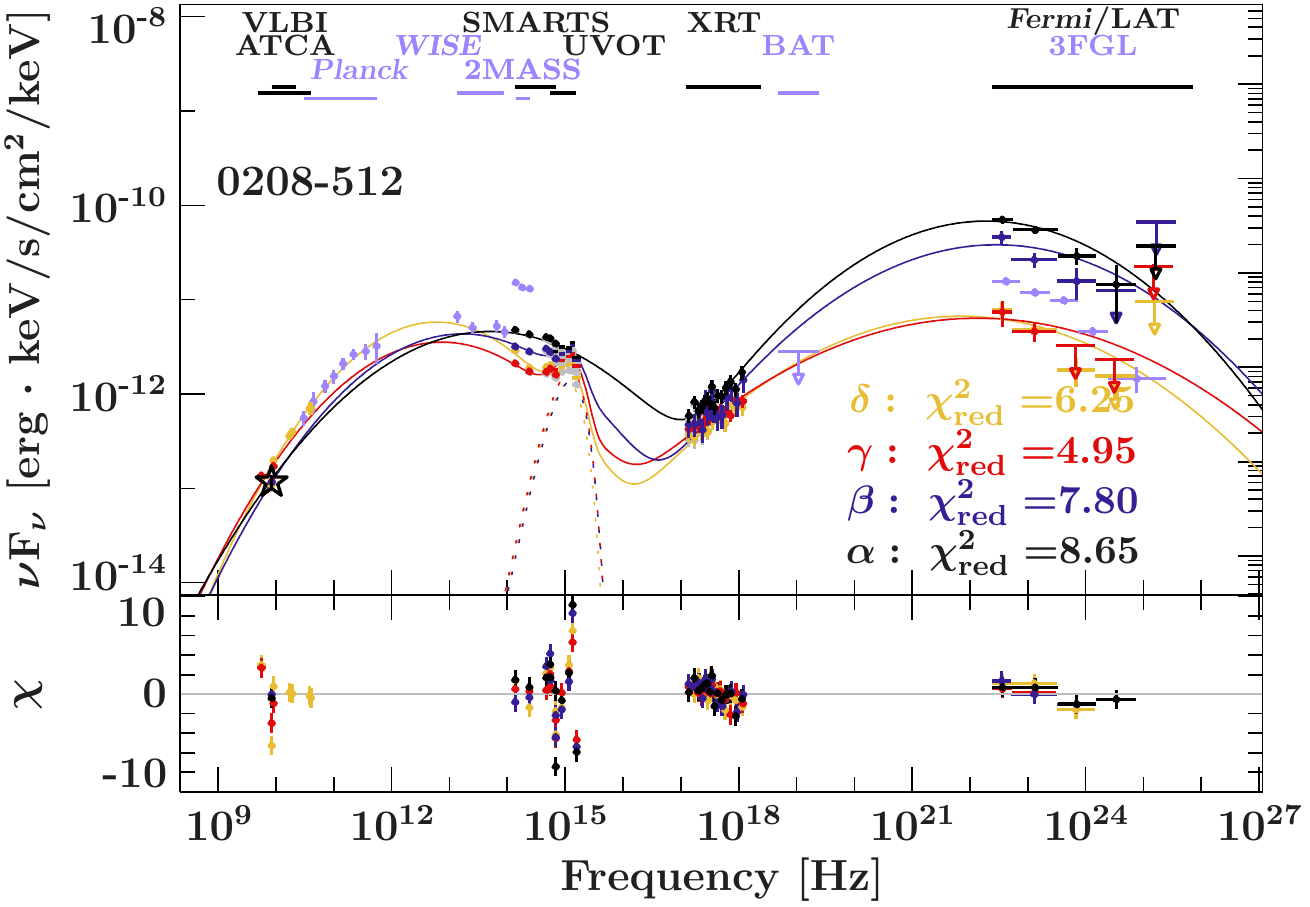}\hfill
\includegraphics[height=0.24\textheight]{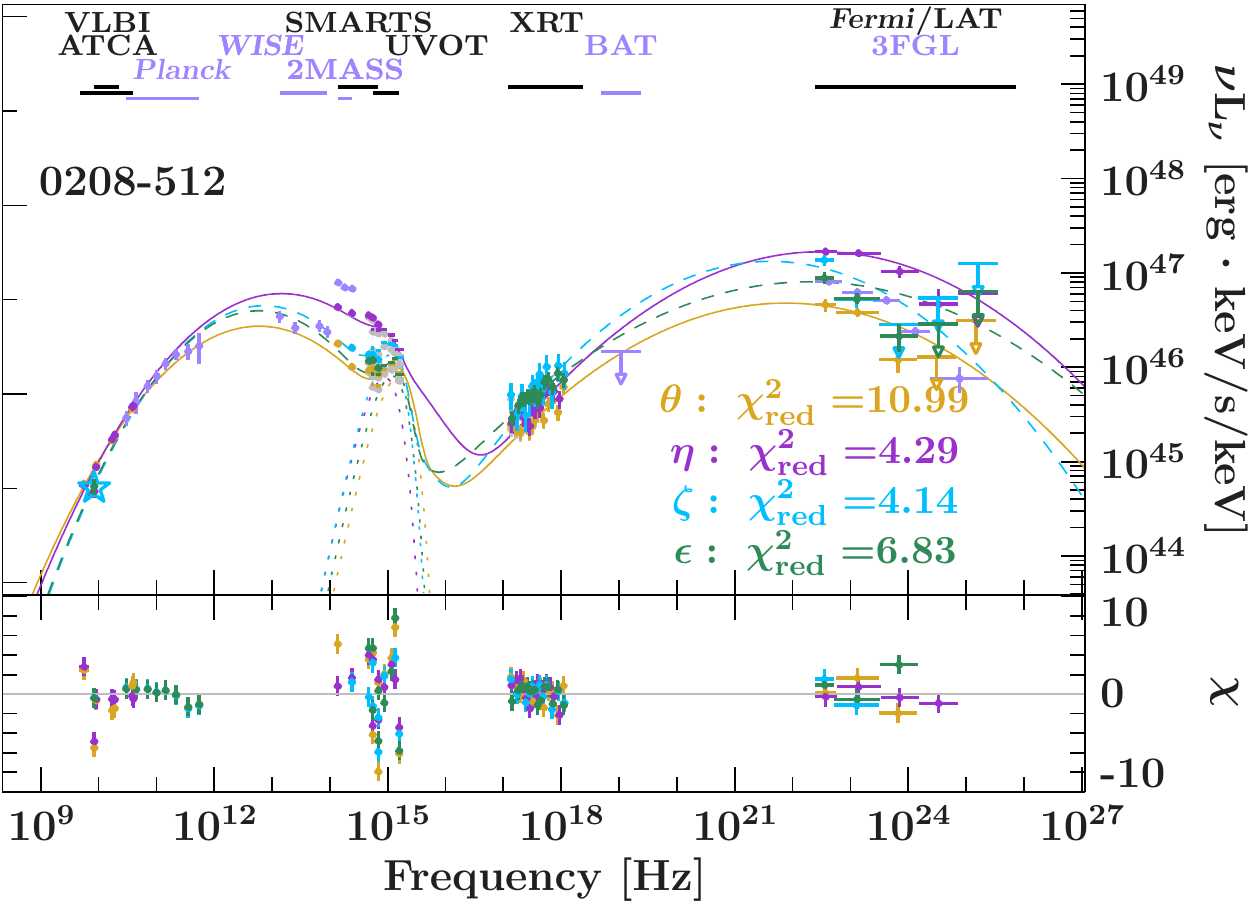}\\[0.5\baselineskip]
\includegraphics[height=0.24\textheight]{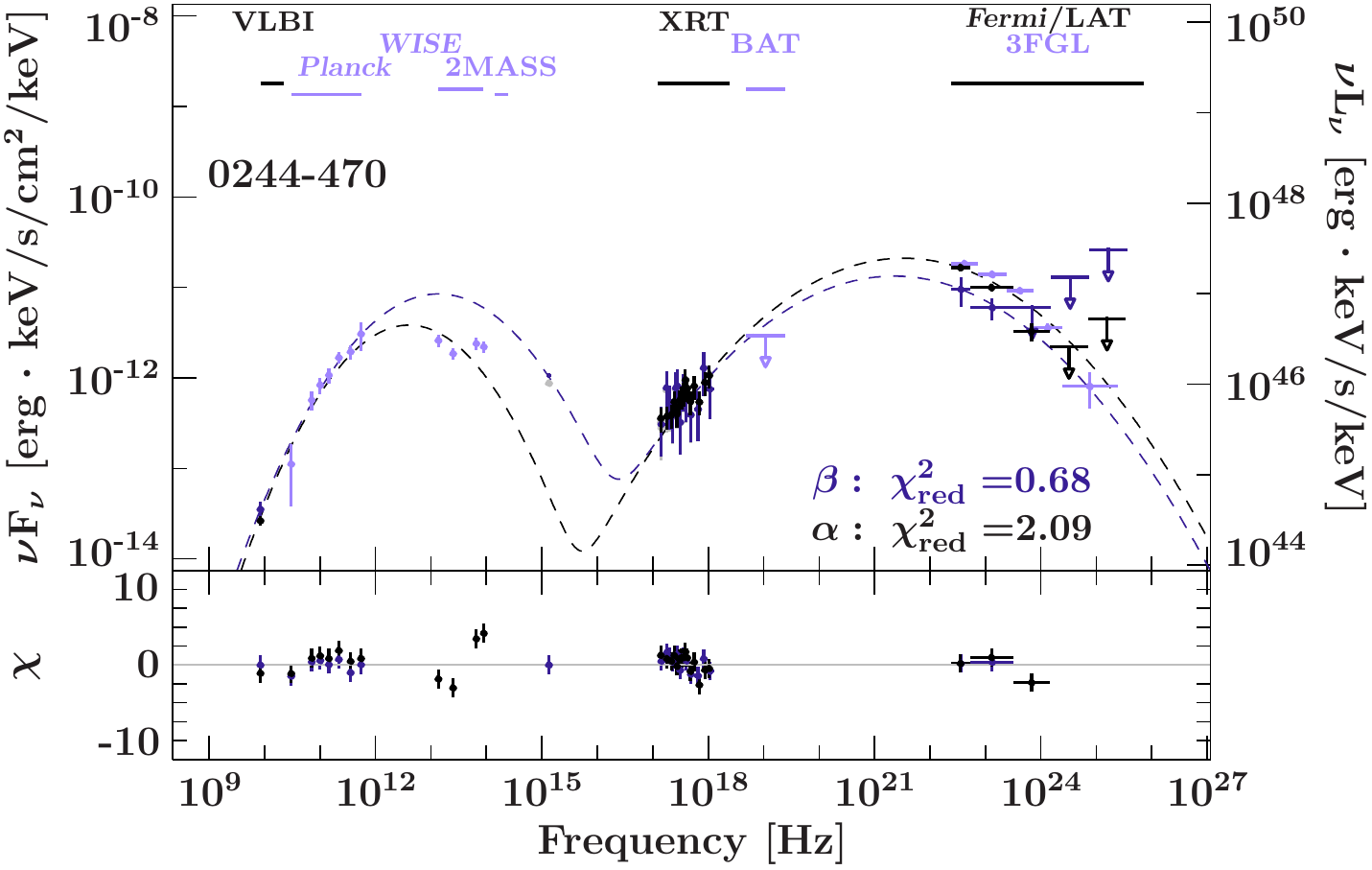}\\[0.5\baselineskip]
\includegraphics[height=0.25\textheight]{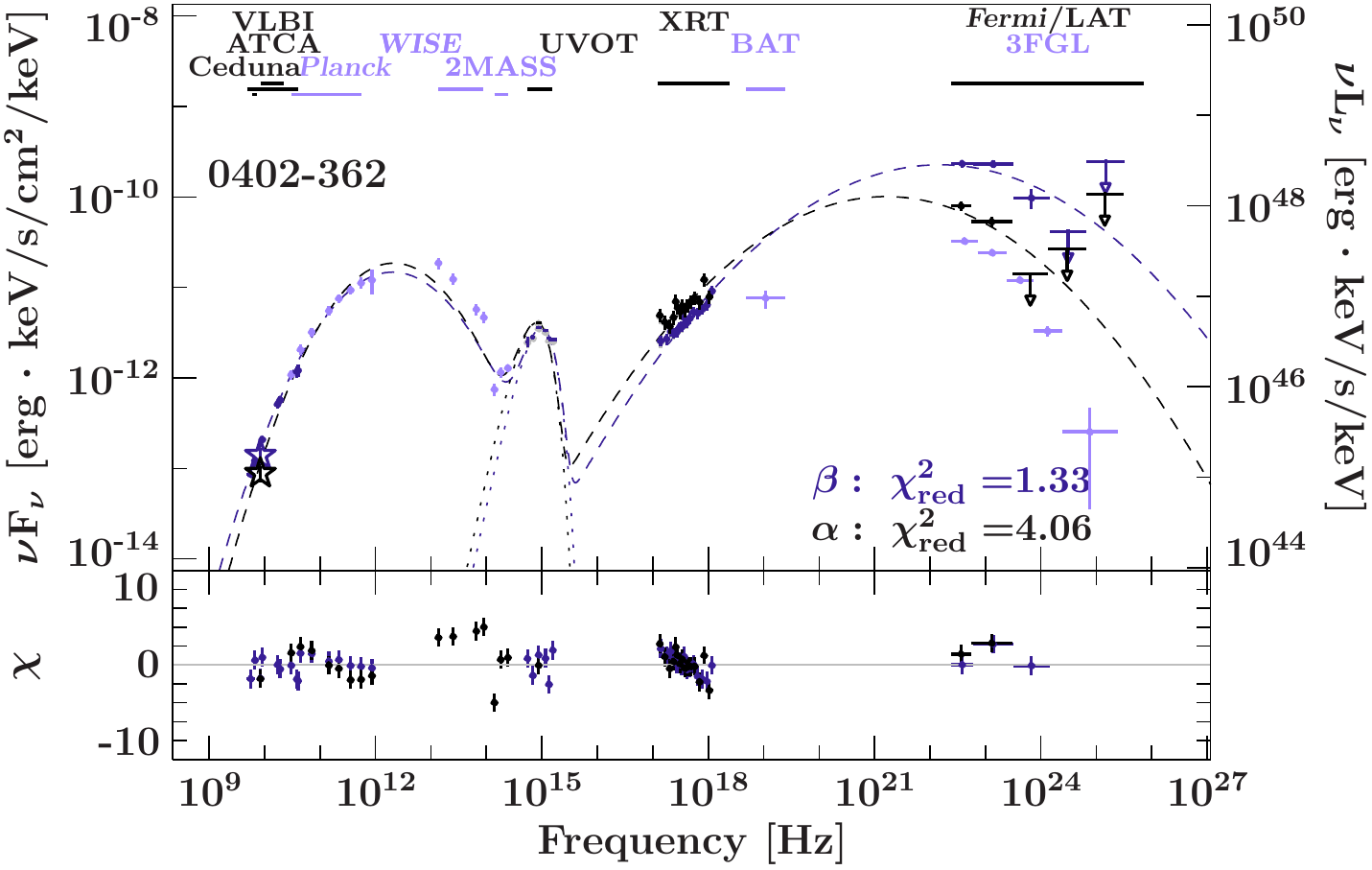}\\[0.5\baselineskip]
\caption{Broadband spectral energy distributions of all sources  with
  a redshift in the loglog $\nu F_\nu$ representation.
  For sources with more than 3 states with sufficient data, the plots
  were split into two parts, to ensure that the SEDs are easily
  visible. Fit models are shown in dashed if archival data had to be
  included in the fit. For sources with a thermal excess in the
  optical/UV, a blackbody was included (dotted). The instruments
  (including their spectral range) are shown above the spectrum. The
  colors correspond to the colors used in the light curve. The best
  fit reduced $\chi^2$ value is shown at the bottom right for
  every state. Residuals are shown in the lower panel. The spectra
  have not been k-corrected.}
\end{figure}
\clearpage

\begin{figure}
\addtocounter{figure}{-1}
\includegraphics[height=0.25\textheight]{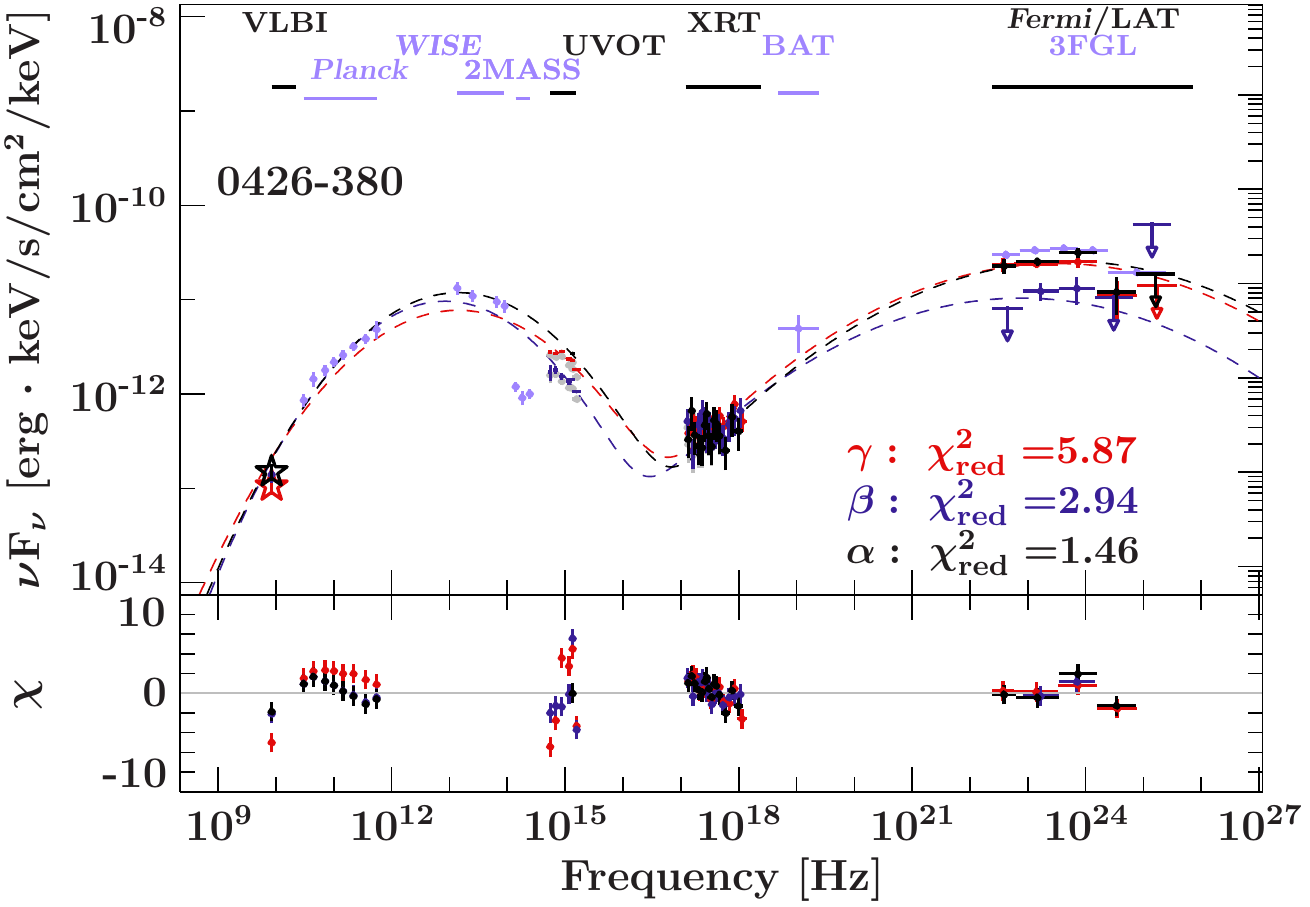}\hfill
\includegraphics[height=0.25\textheight]{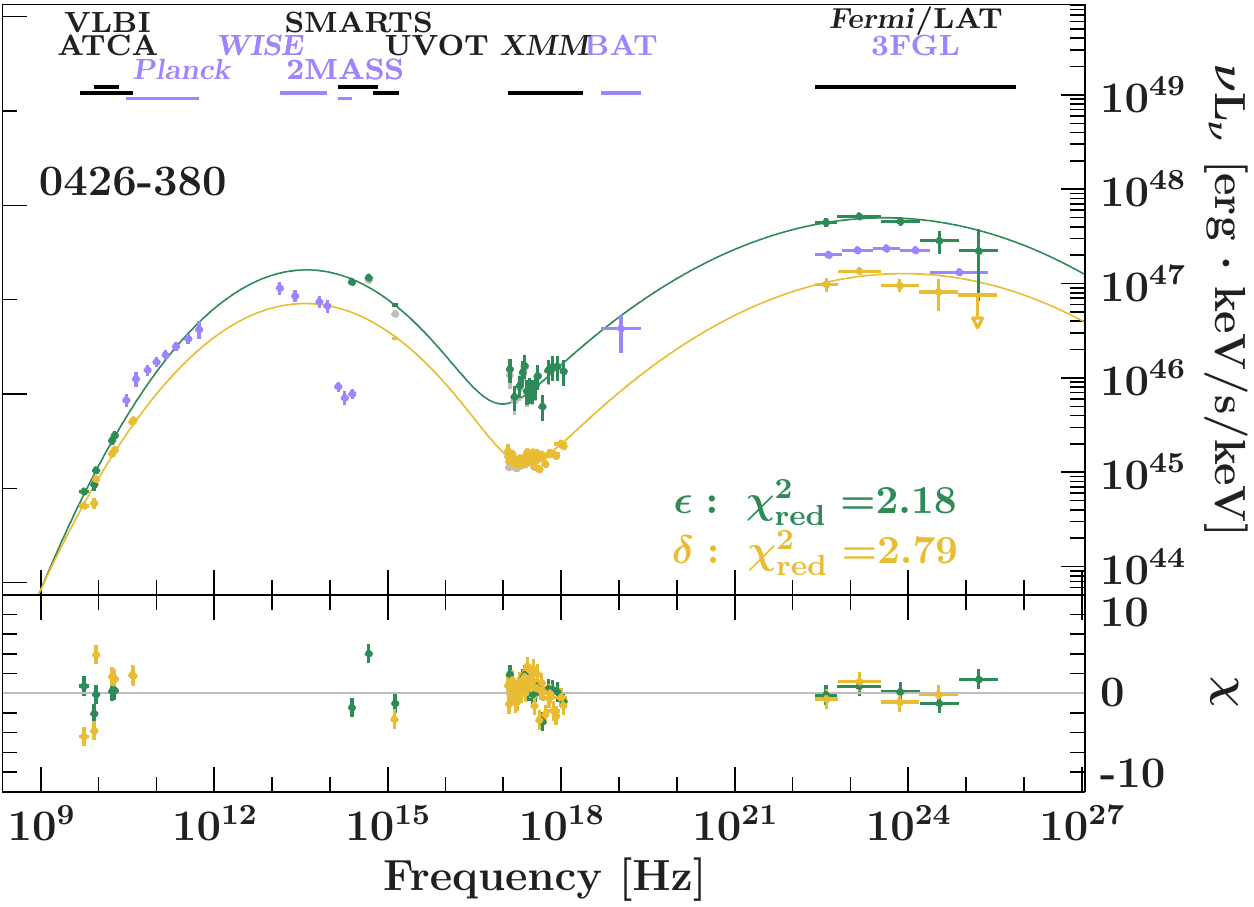}\\[0.5\baselineskip]
\includegraphics[height=0.25\textheight]{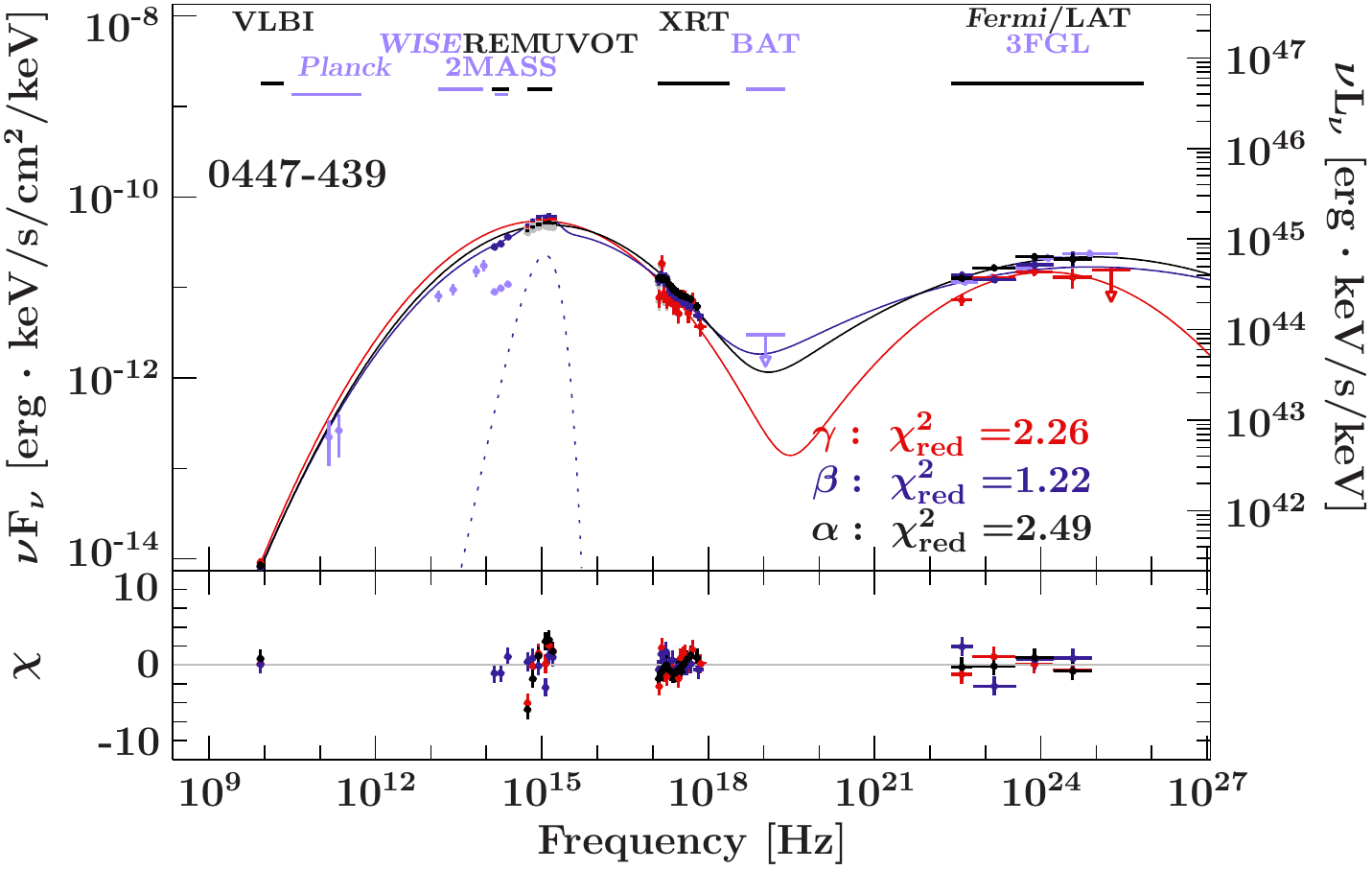}\\[0.5\baselineskip]
\includegraphics[height=0.25\textheight]{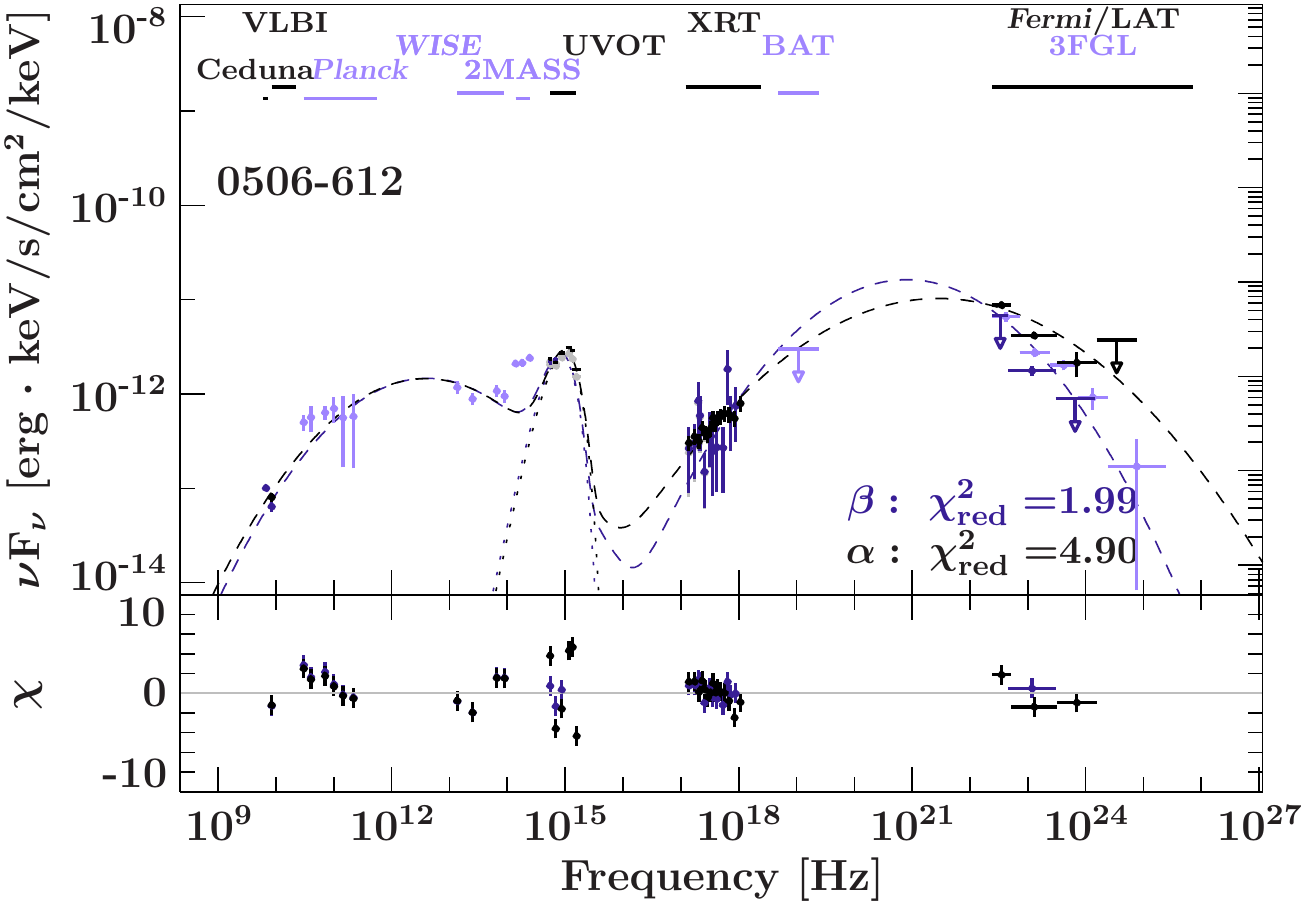}\hfill
\includegraphics[height=0.25\textheight]{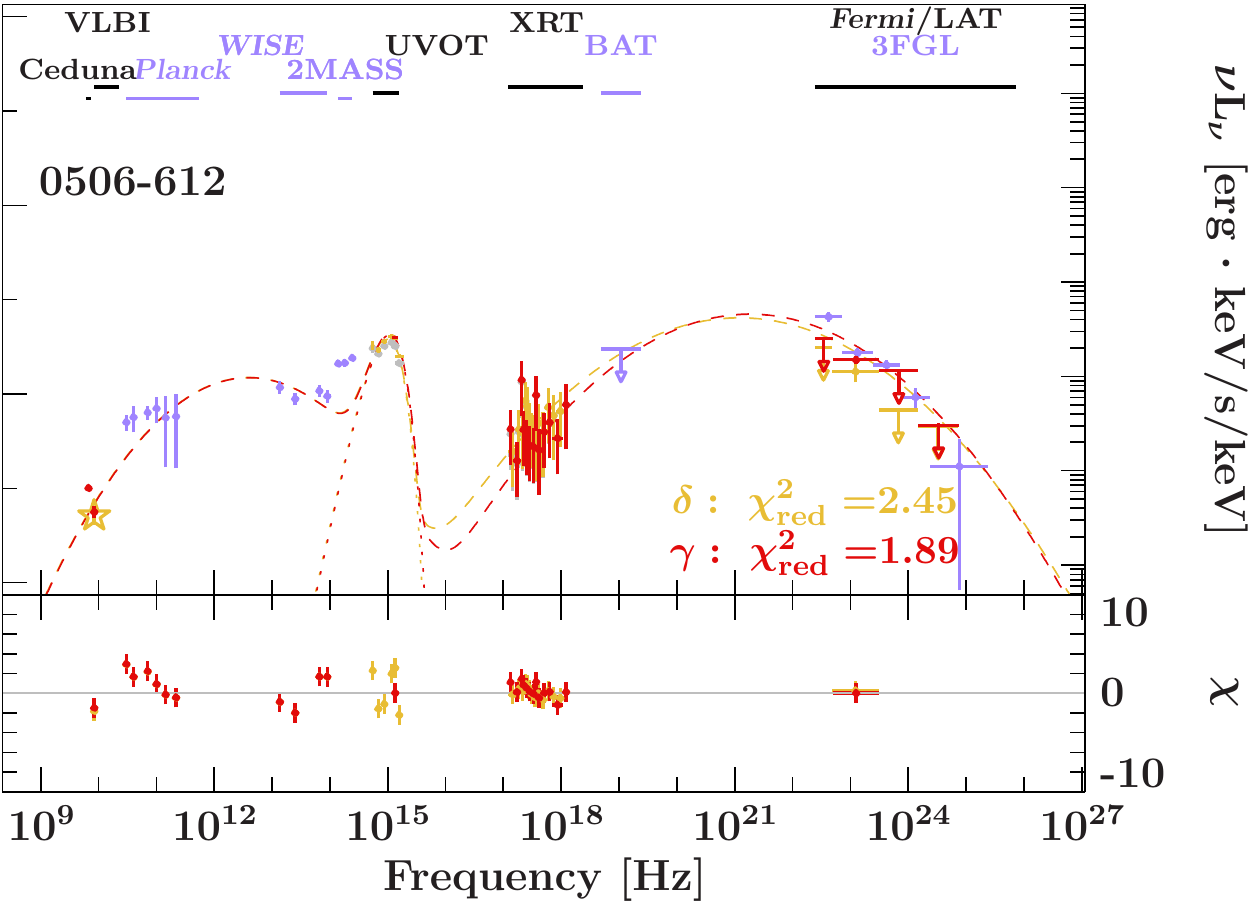}\\
\caption{(contd.)}
\end{figure}
\clearpage

\begin{figure}
\addtocounter{figure}{-1}
\includegraphics[height=0.25\textheight]{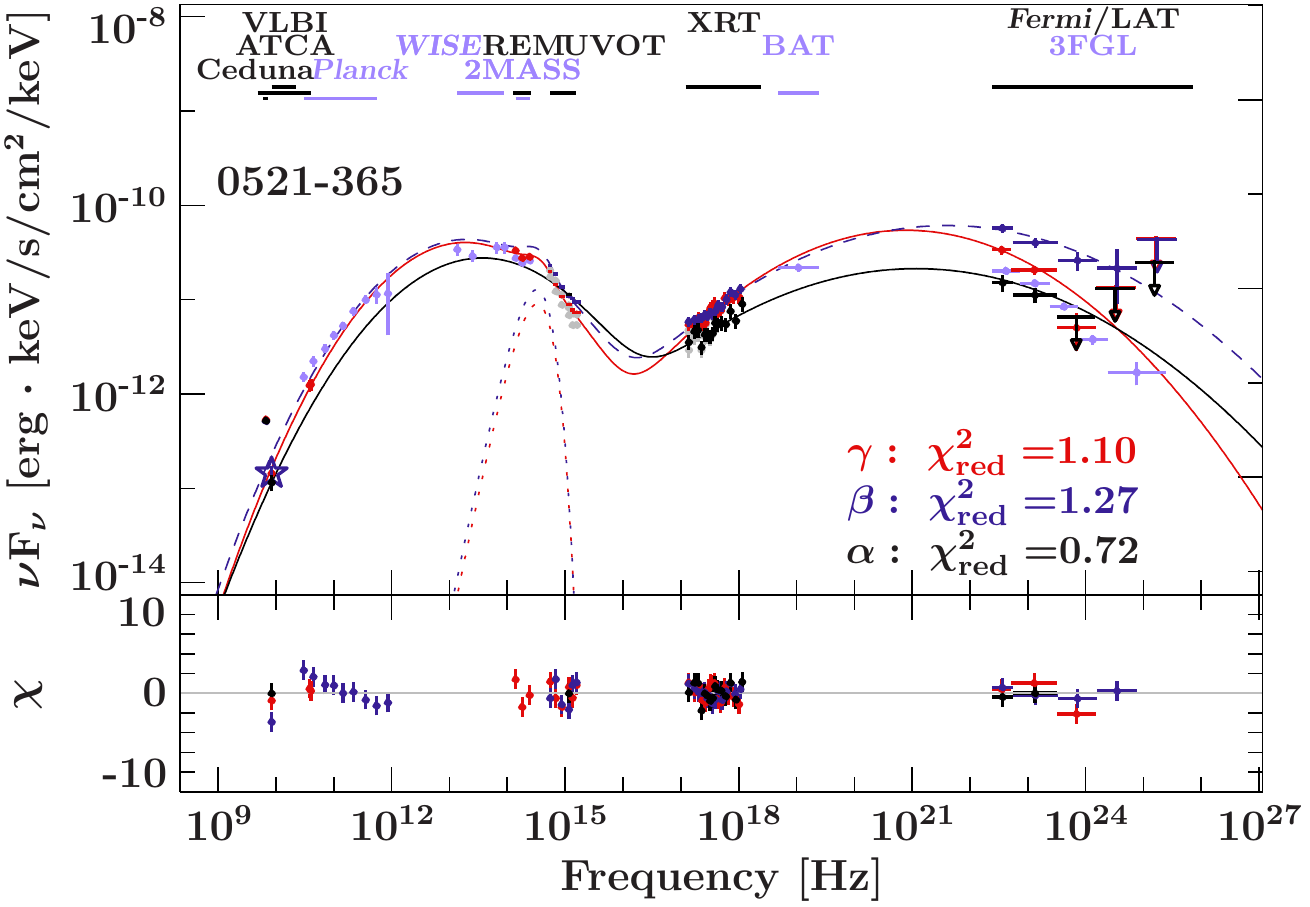}\hfill
\includegraphics[height=0.25\textheight]{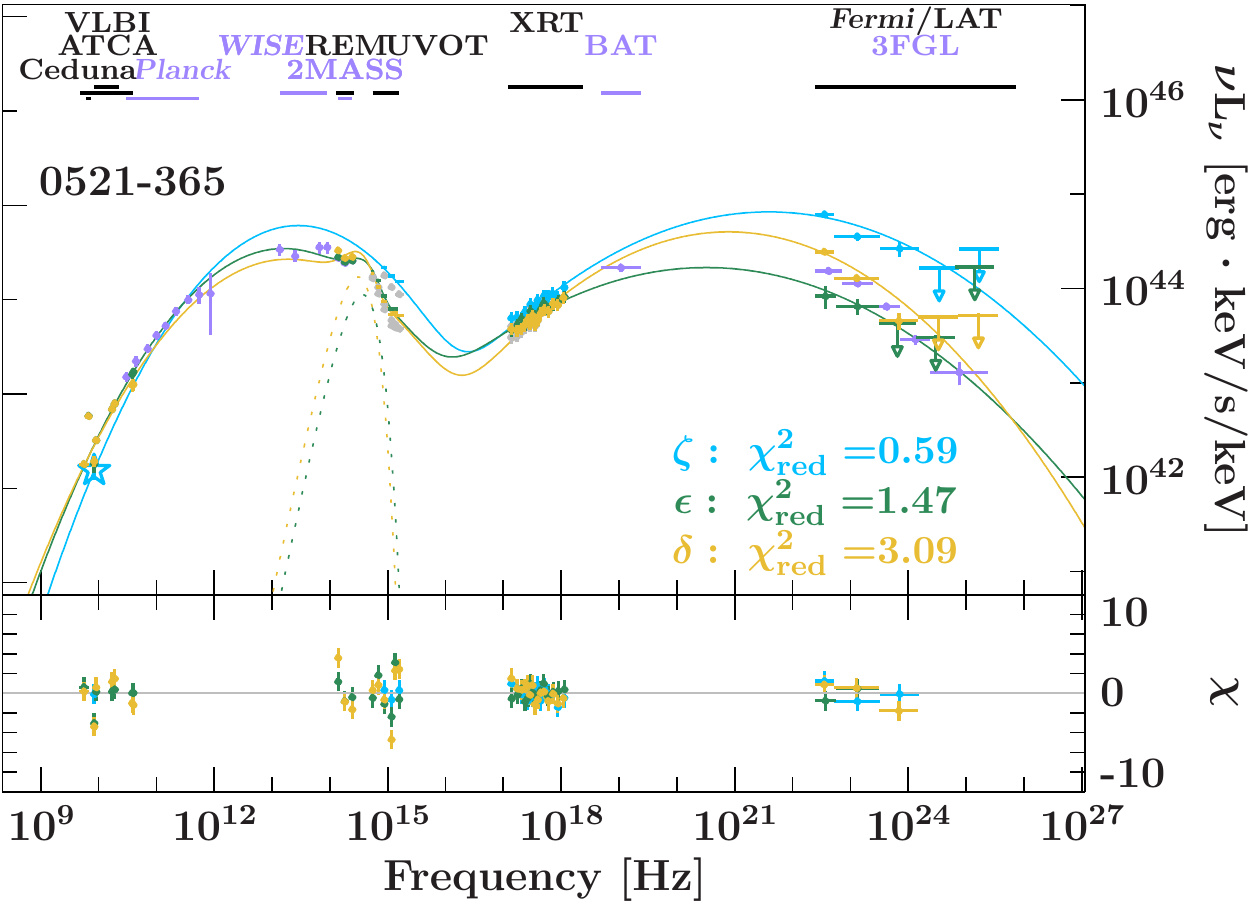}\\[0.5\baselineskip]
\includegraphics[height=0.25\textheight]{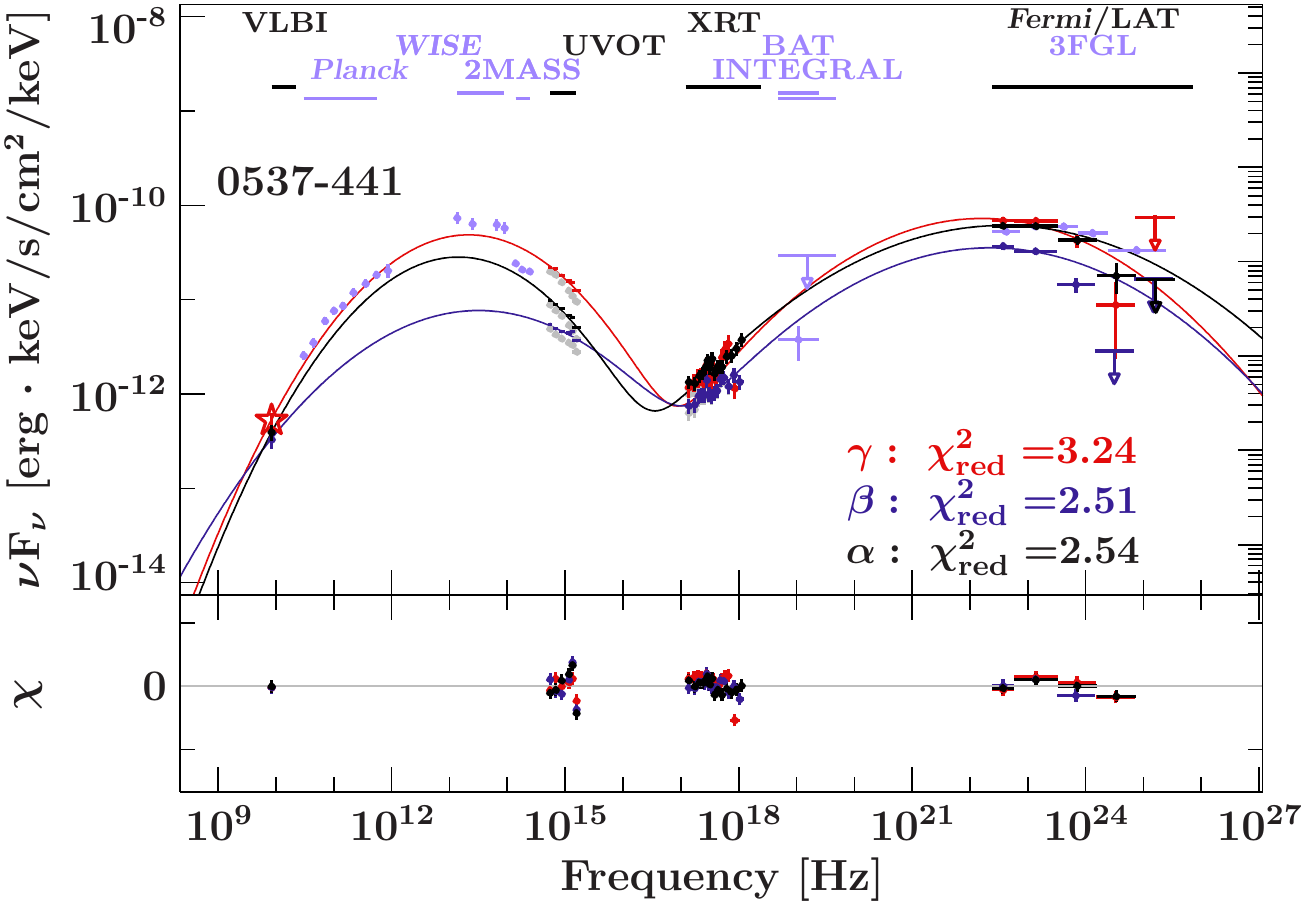}\hfill
\includegraphics[height=0.25\textheight]{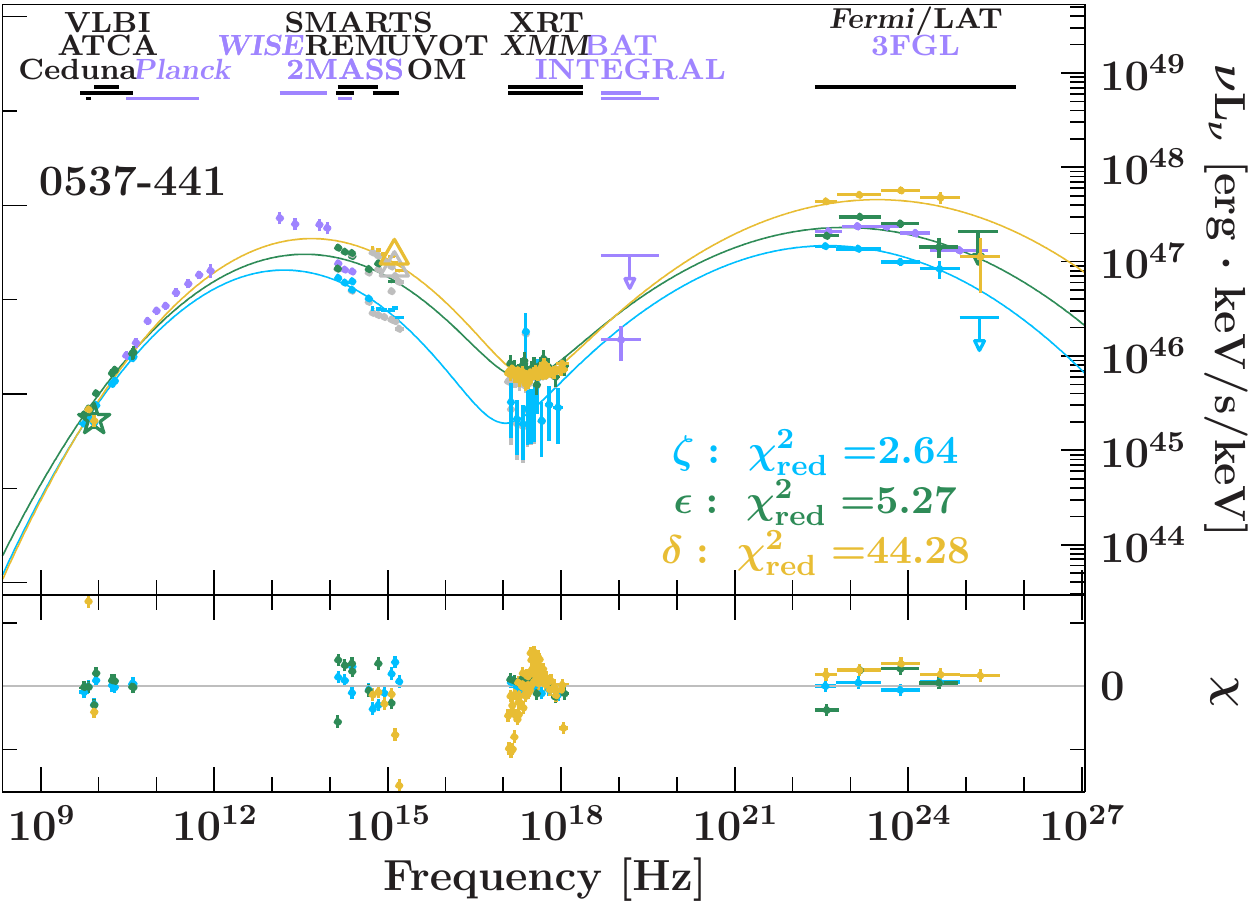}\\[0.5\baselineskip]
\includegraphics[height=0.25\textheight]{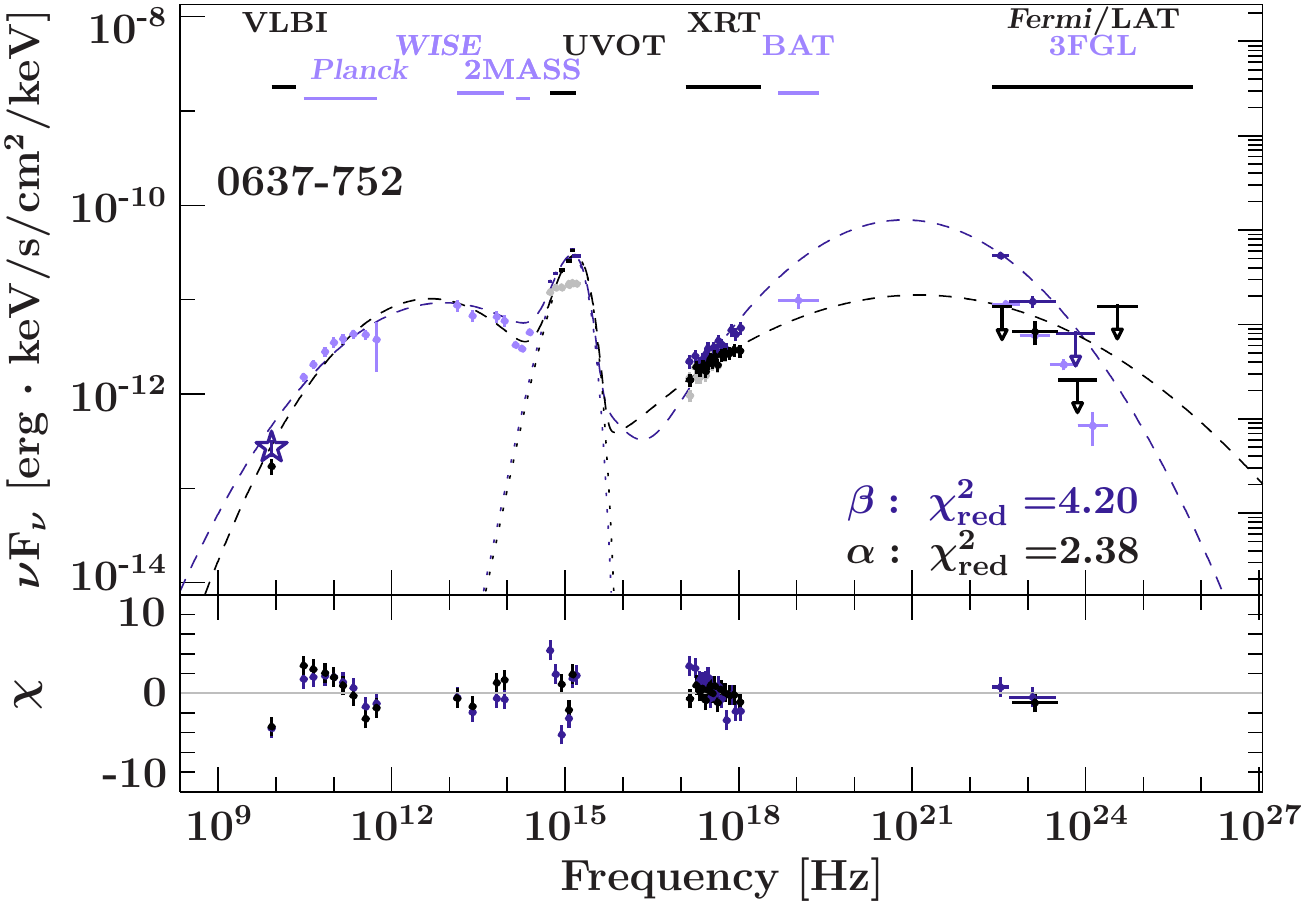}\hfill
\includegraphics[height=0.25\textheight]{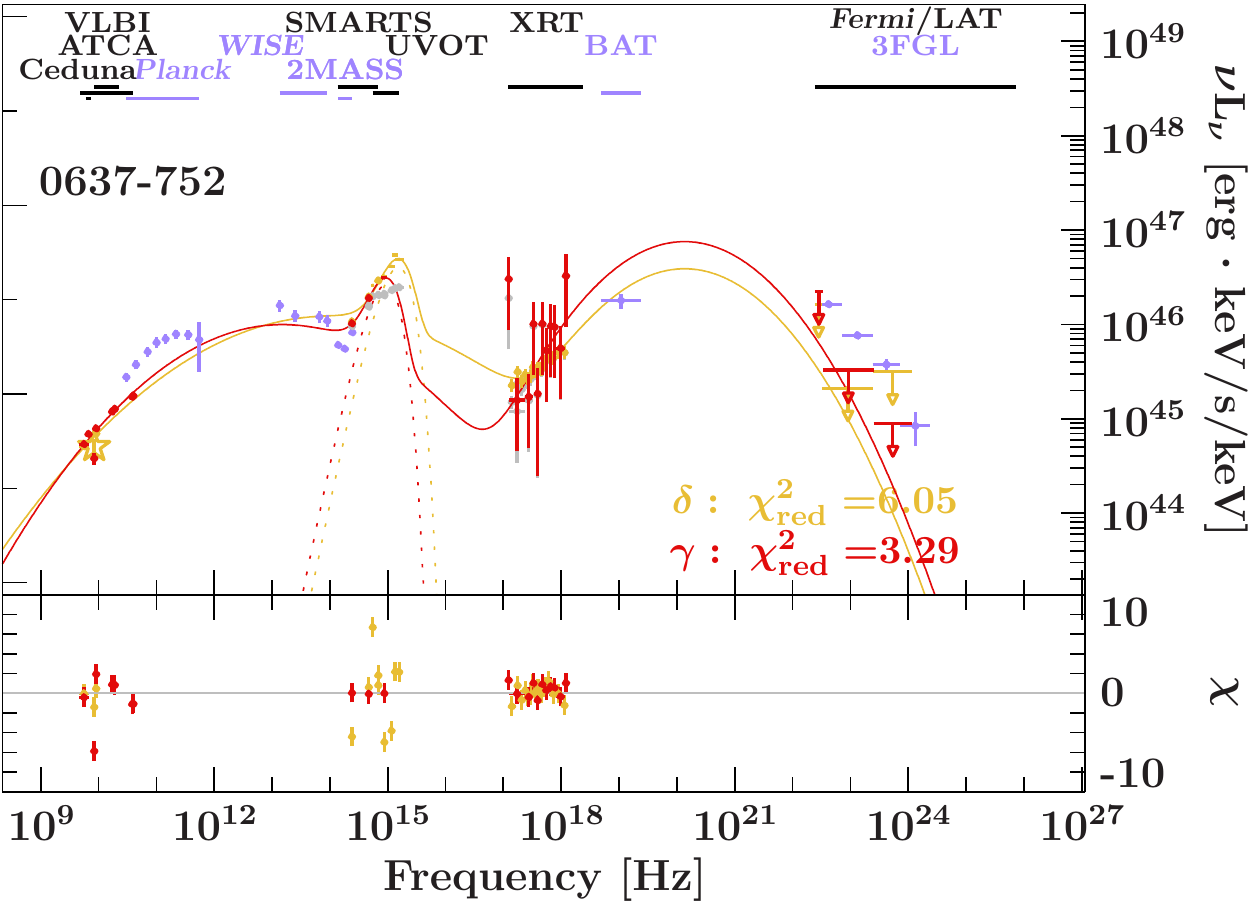}\\
\caption{(contd.)}
\end{figure}
\clearpage

\begin{figure}
  \addtocounter{figure}{-1}
\includegraphics[height=0.25\textheight]{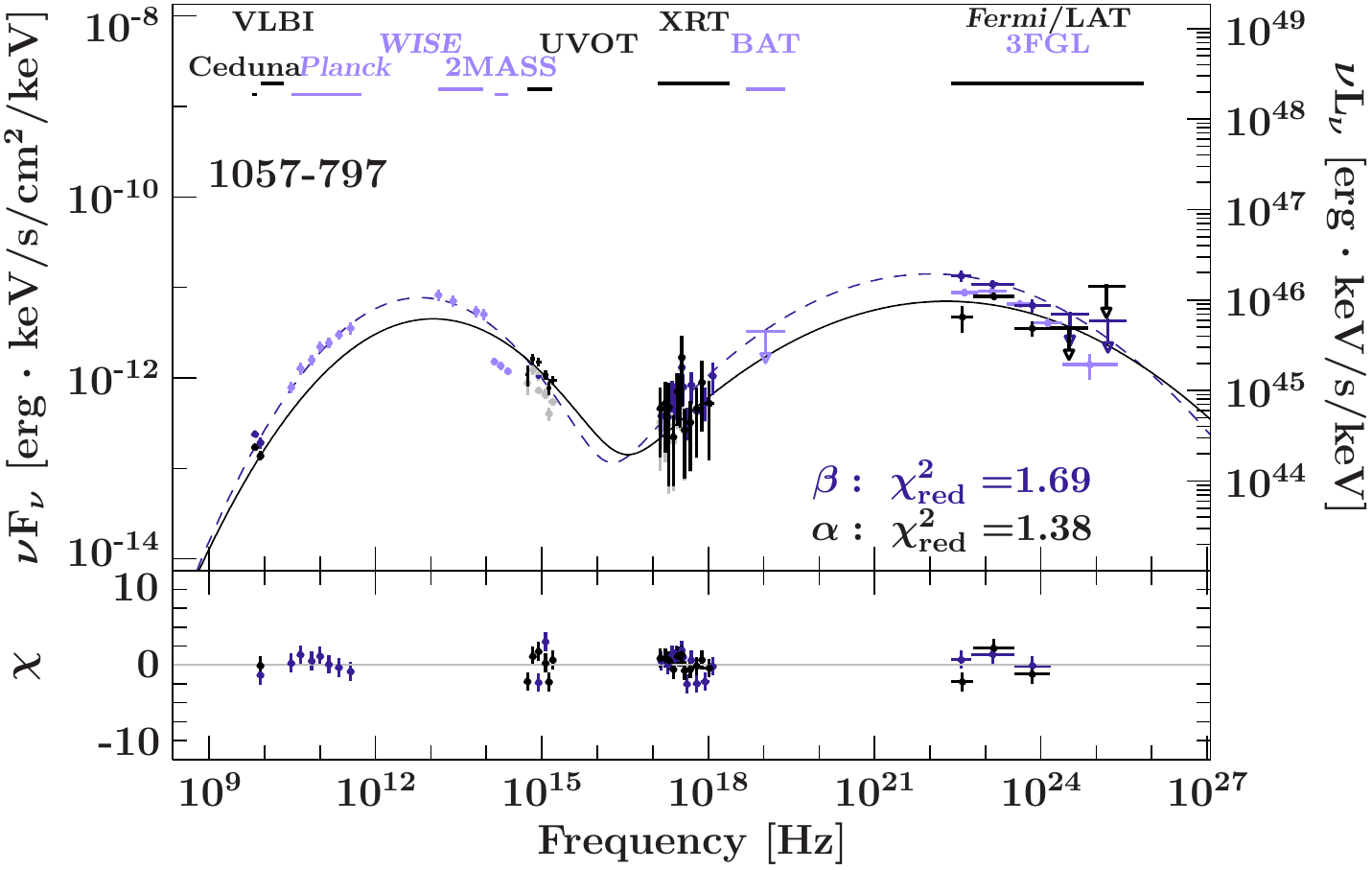}\\[0.5\baselineskip]
\includegraphics[height=0.25\textheight]{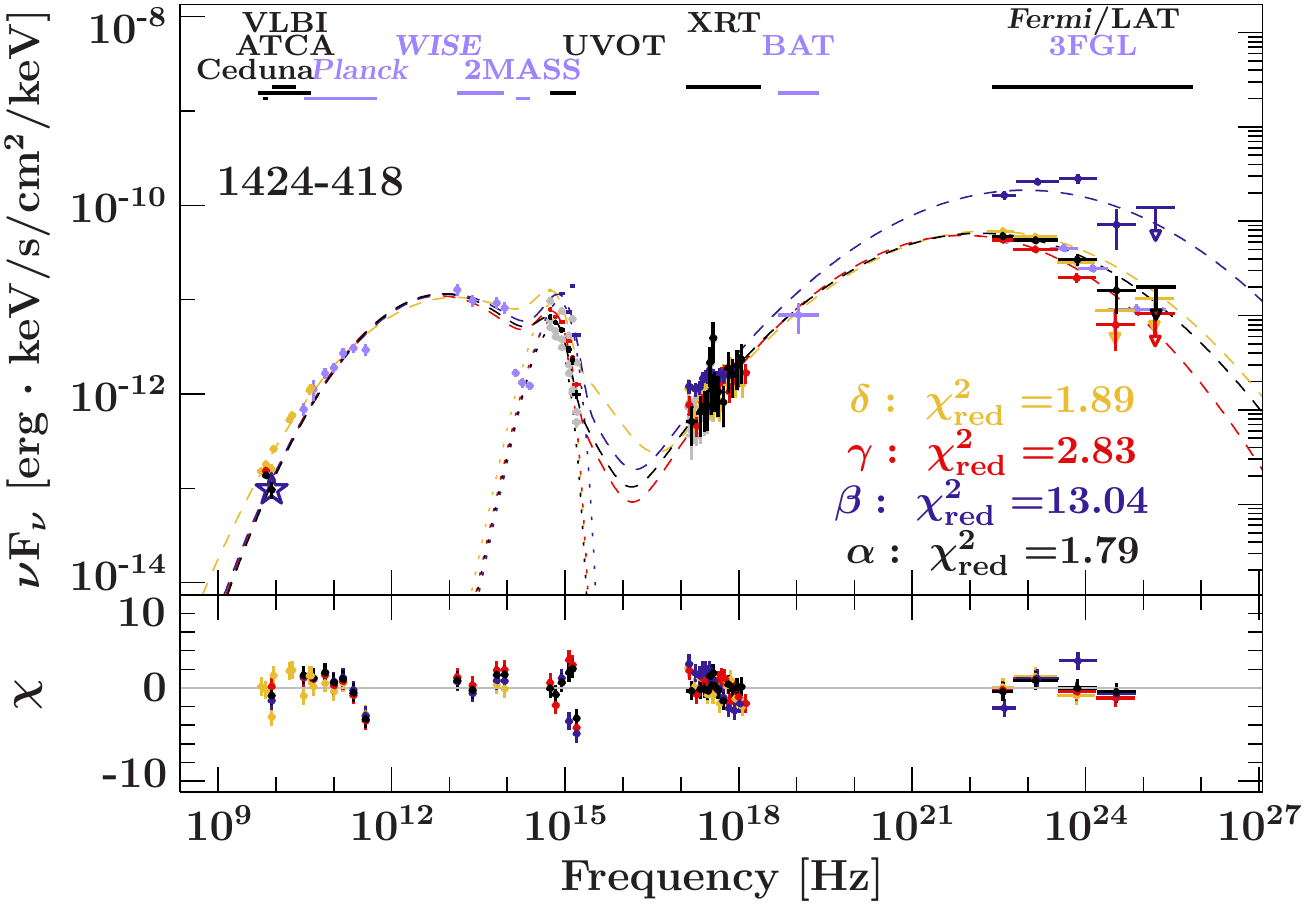}\hfill
\includegraphics[height=0.25\textheight]{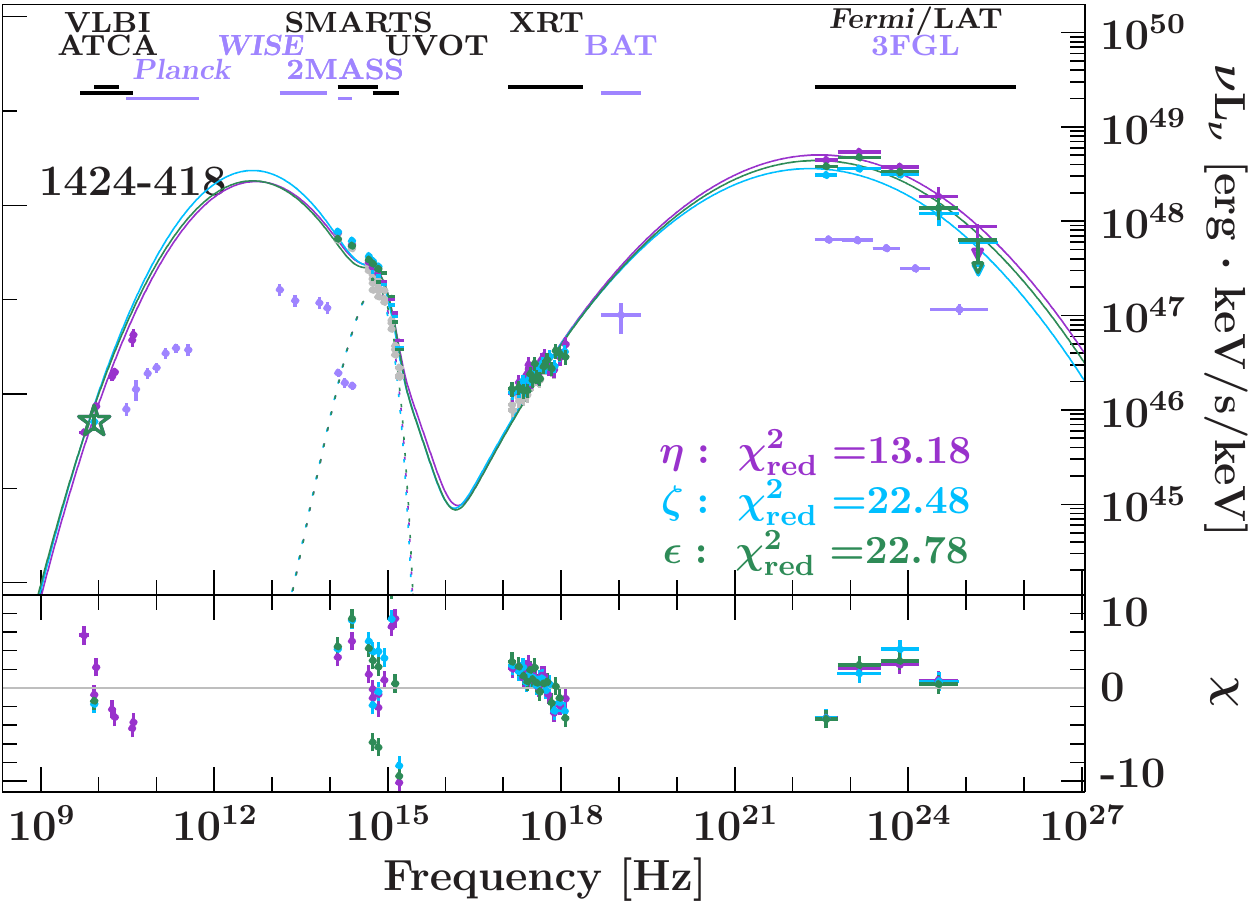}\\[0.5\baselineskip]
\includegraphics[height=0.25\textheight]{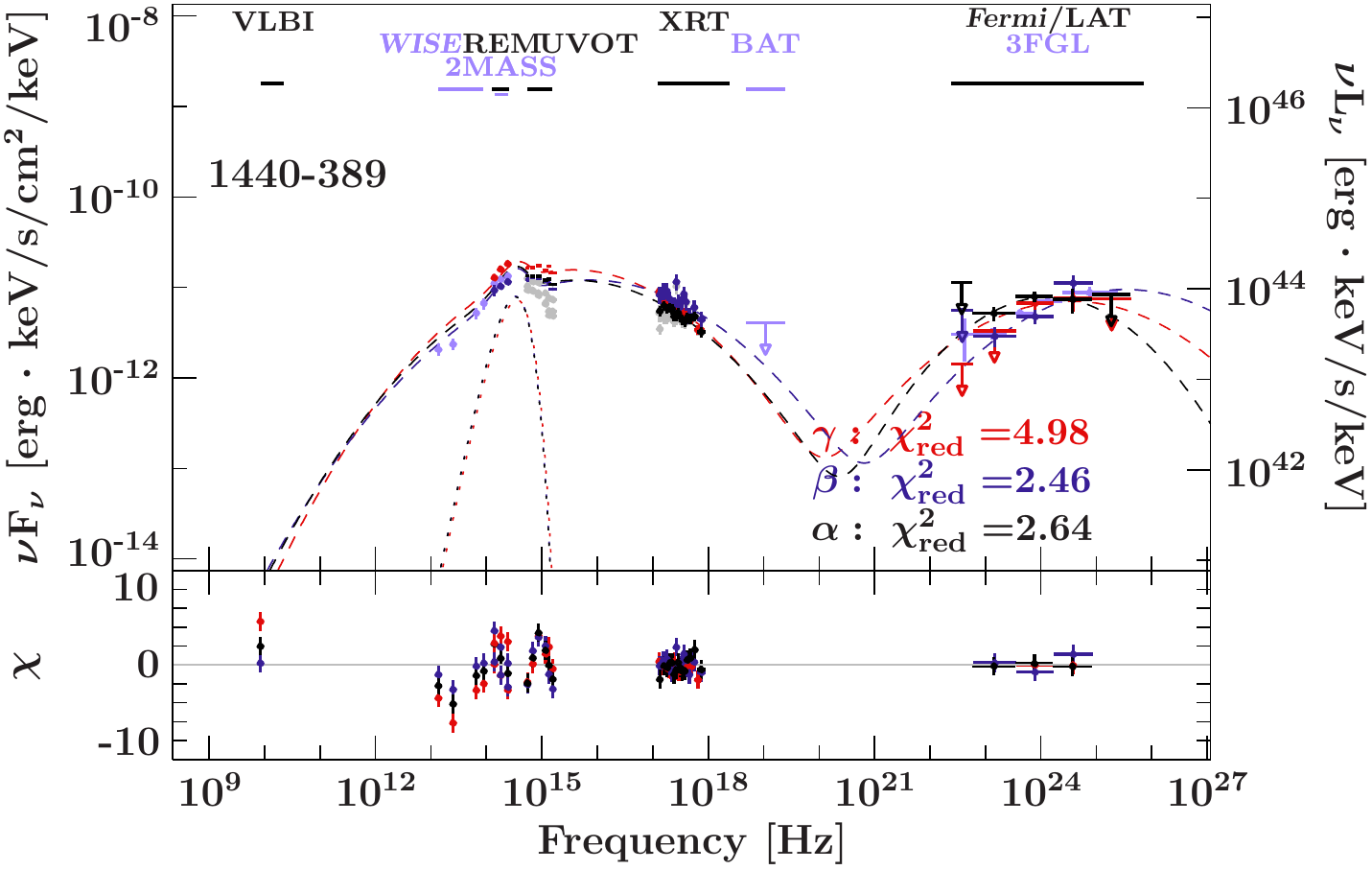}\\[0.5\baselineskip]

\caption{(contd.)}
\end{figure}
\clearpage
\begin{figure}
  \addtocounter{figure}{-1}
\includegraphics[height=0.25\textheight]{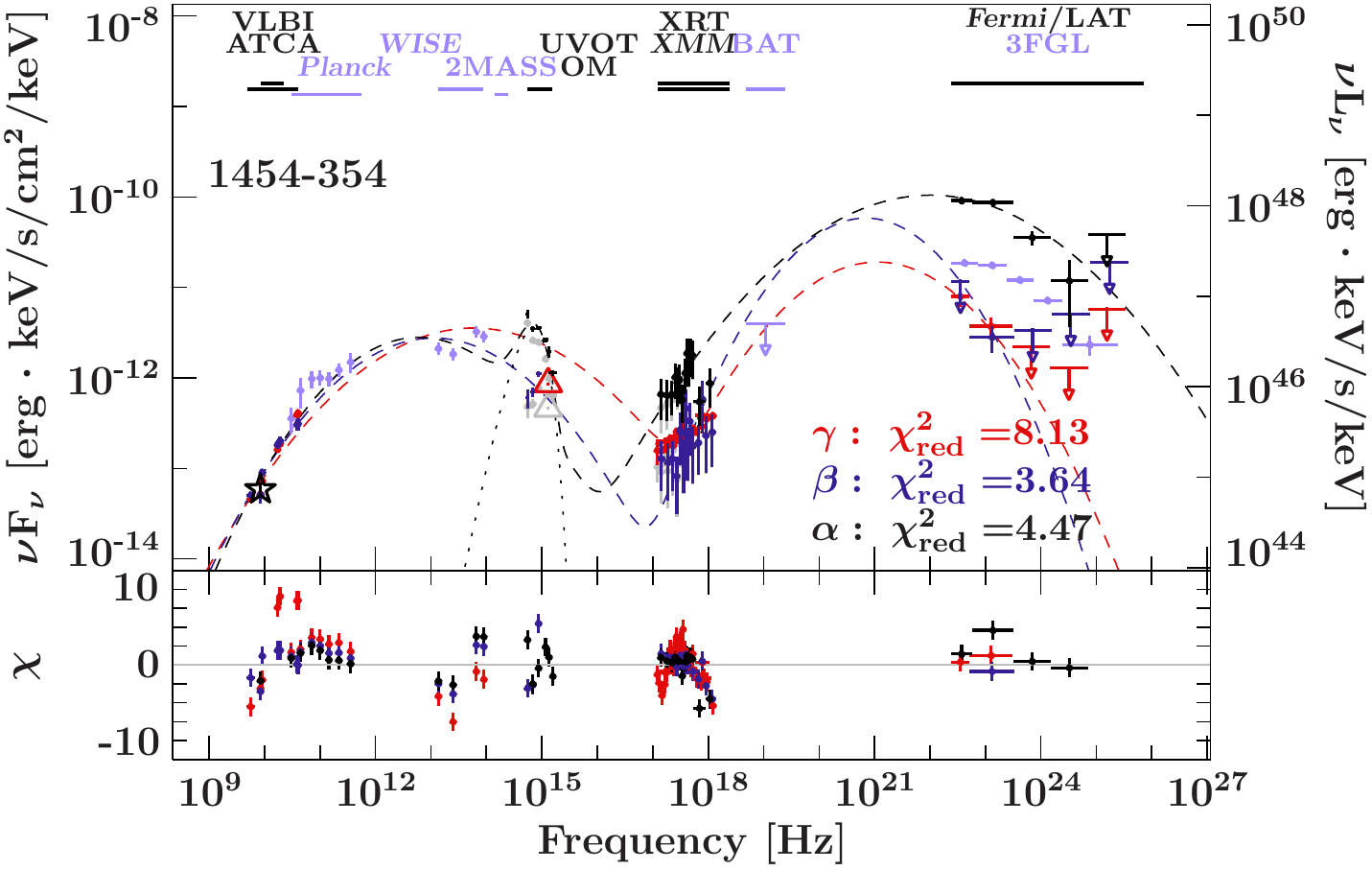}\\[0.5\baselineskip]
\includegraphics[height=0.25\textheight]{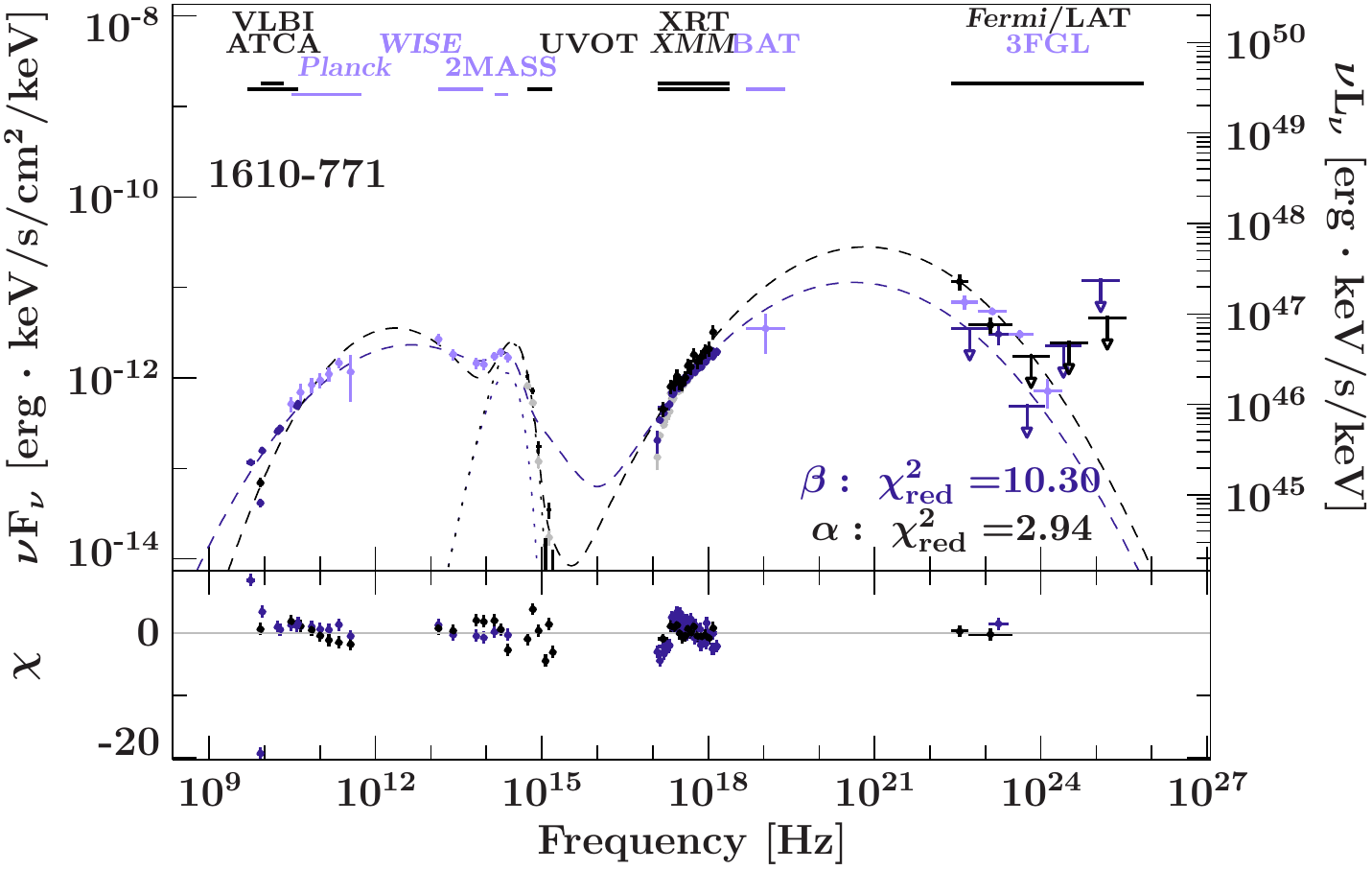}\\[0.5\baselineskip]
\includegraphics[height=0.25\textheight]{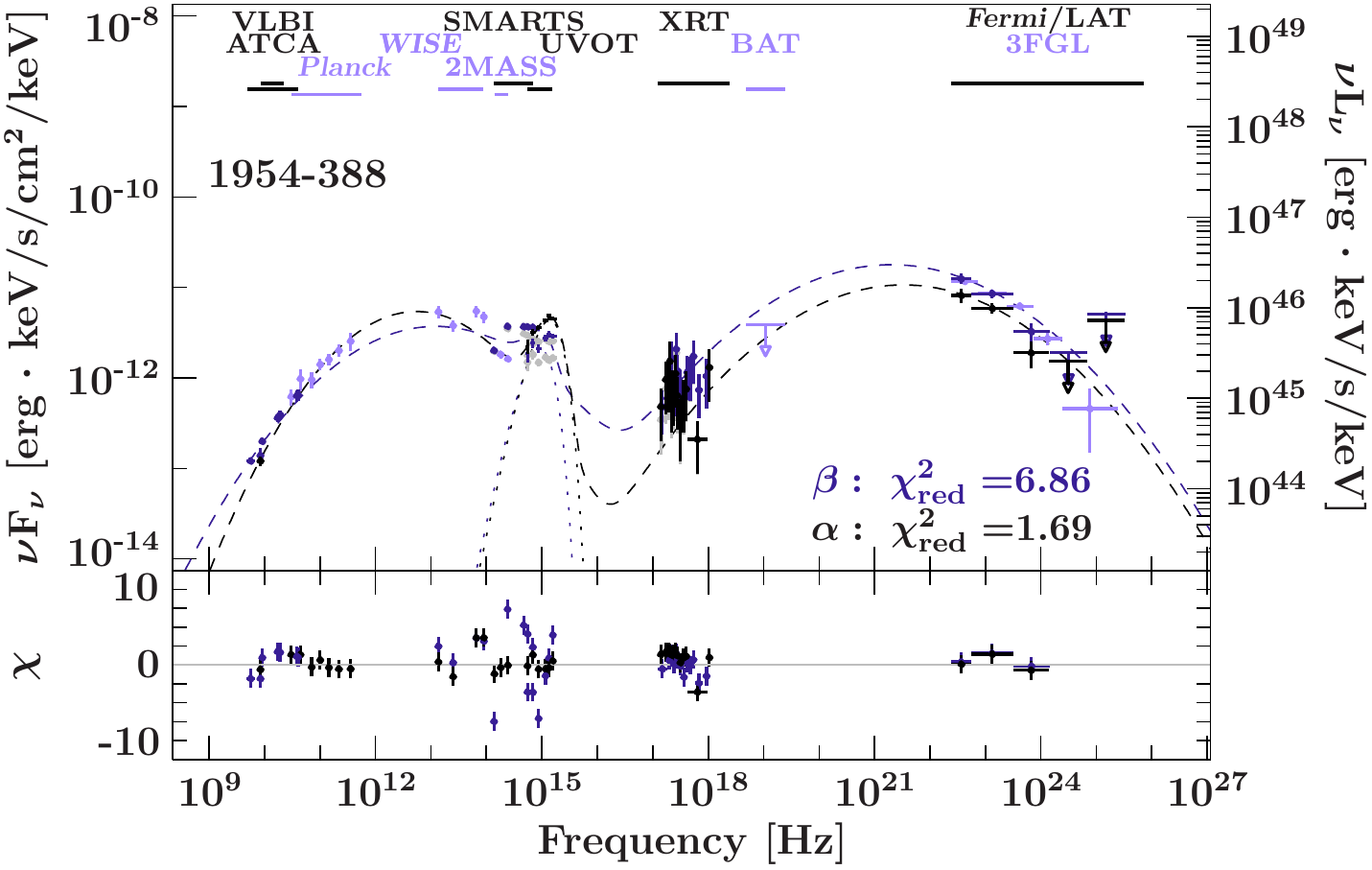}\\[0.5\baselineskip]
\caption{(contd.)}
\end{figure}
\clearpage
\begin{figure}
  \addtocounter{figure}{-1}
\includegraphics[height=0.25\textheight]{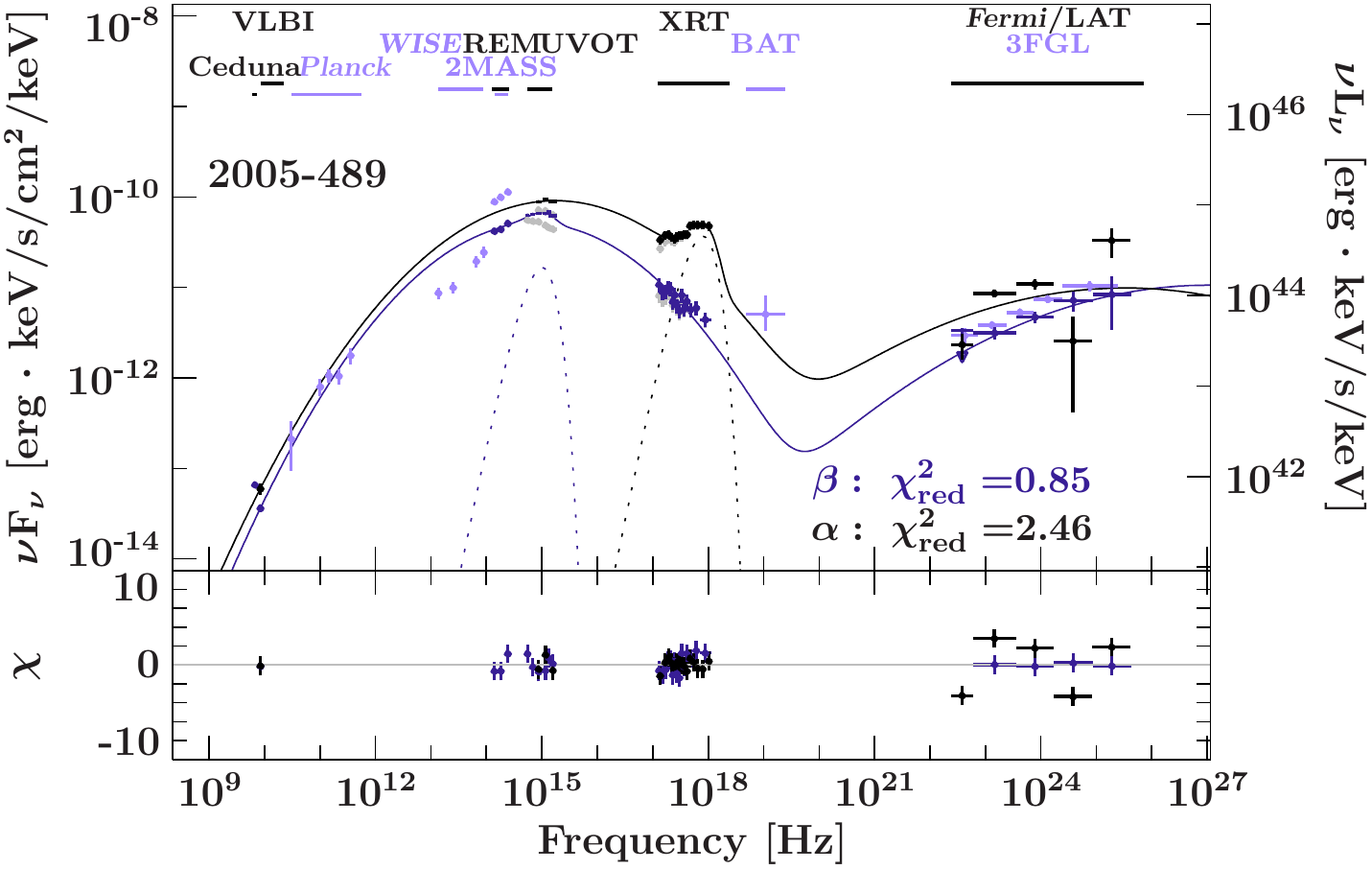}\\[0.5\baselineskip]
\includegraphics[height=0.25\textheight]{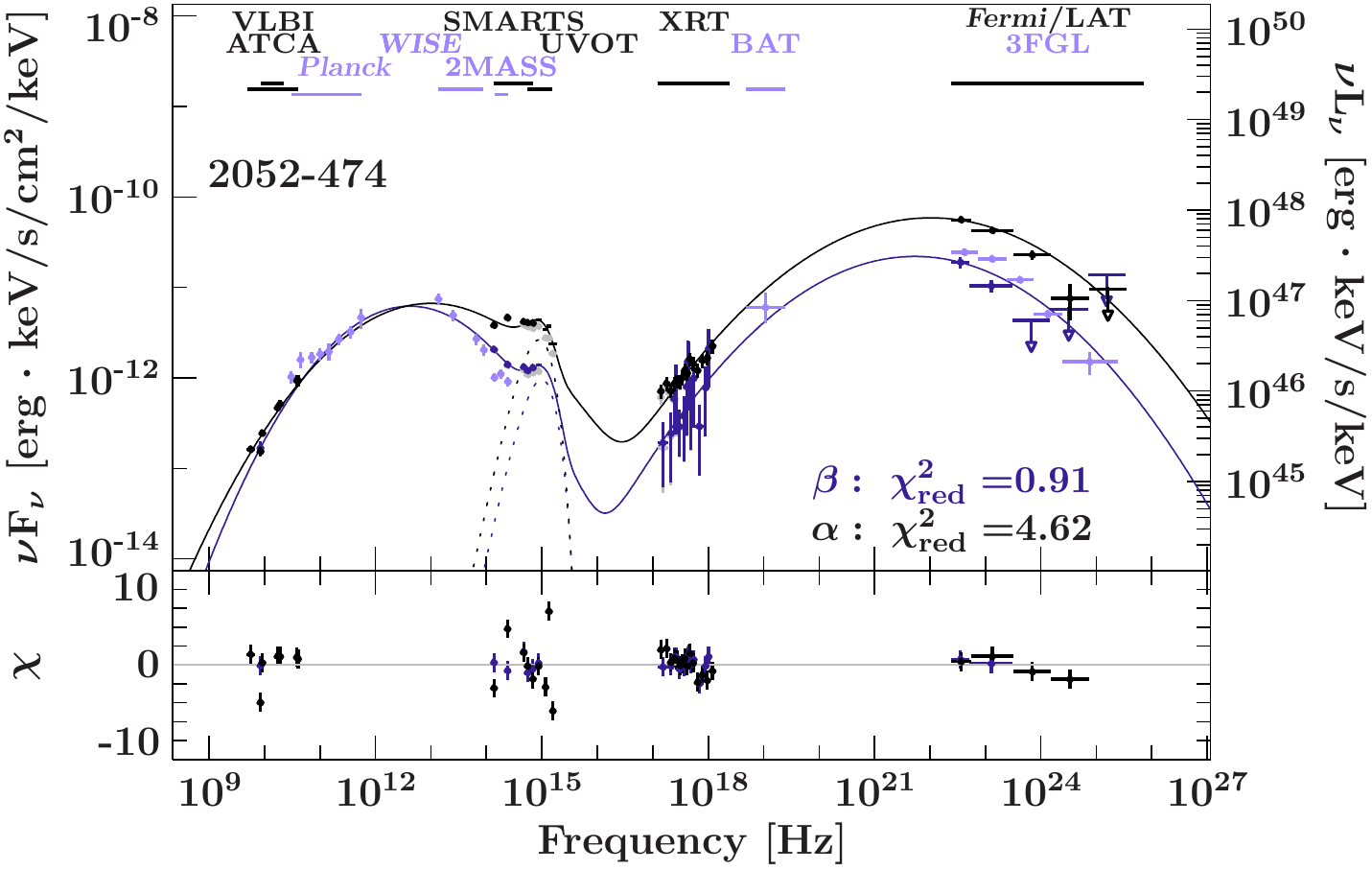}\\[0.5\baselineskip]
\includegraphics[height=0.25\textheight]{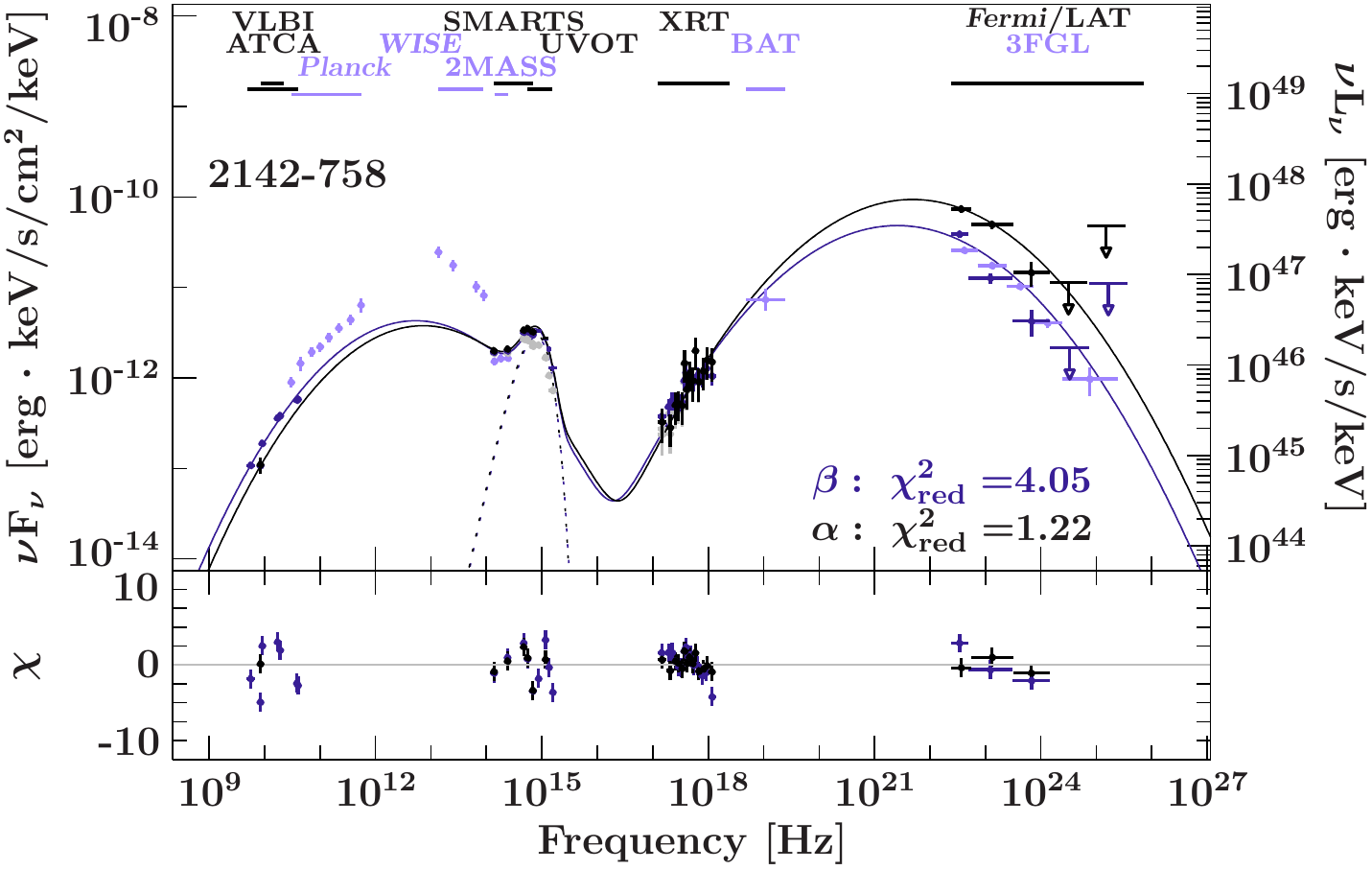}\\[0.5\baselineskip]
\caption{(contd.)}
\end{figure}
\clearpage
\begin{figure}
  \addtocounter{figure}{-1}
\includegraphics[height=0.25\textheight]{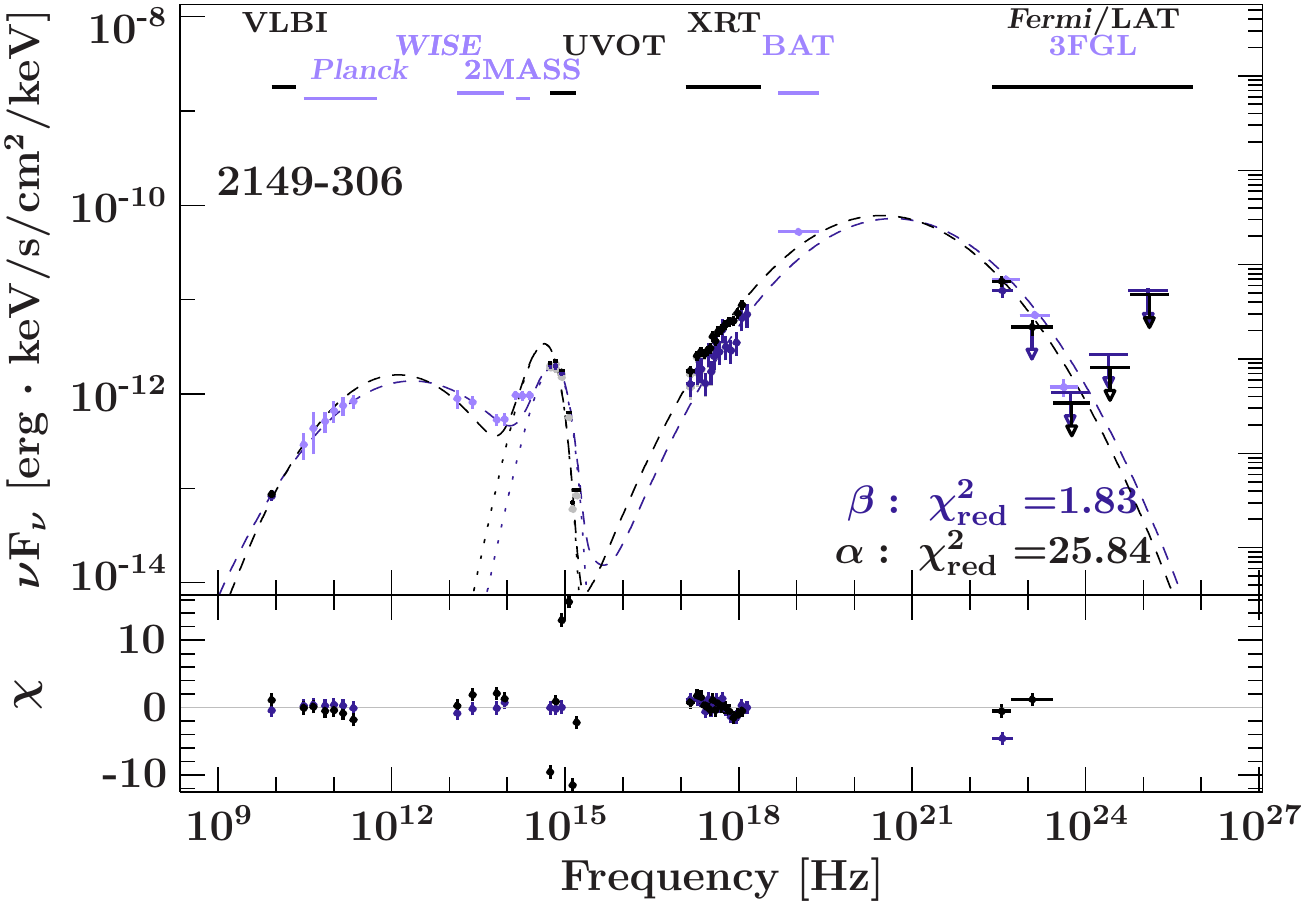}\hfill
\includegraphics[height=0.25\textheight]{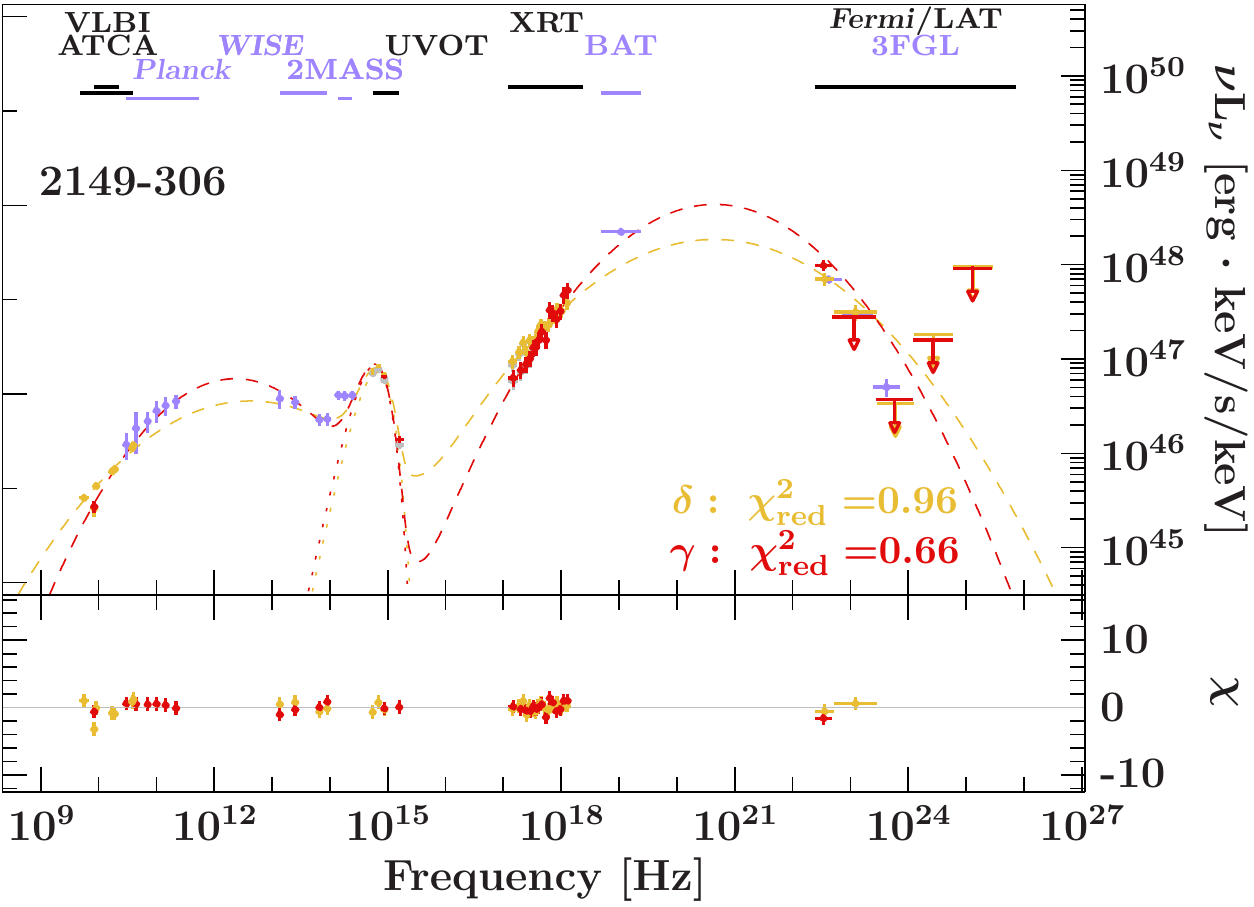}\\[0.5\baselineskip]
\includegraphics[height=0.25\textheight]{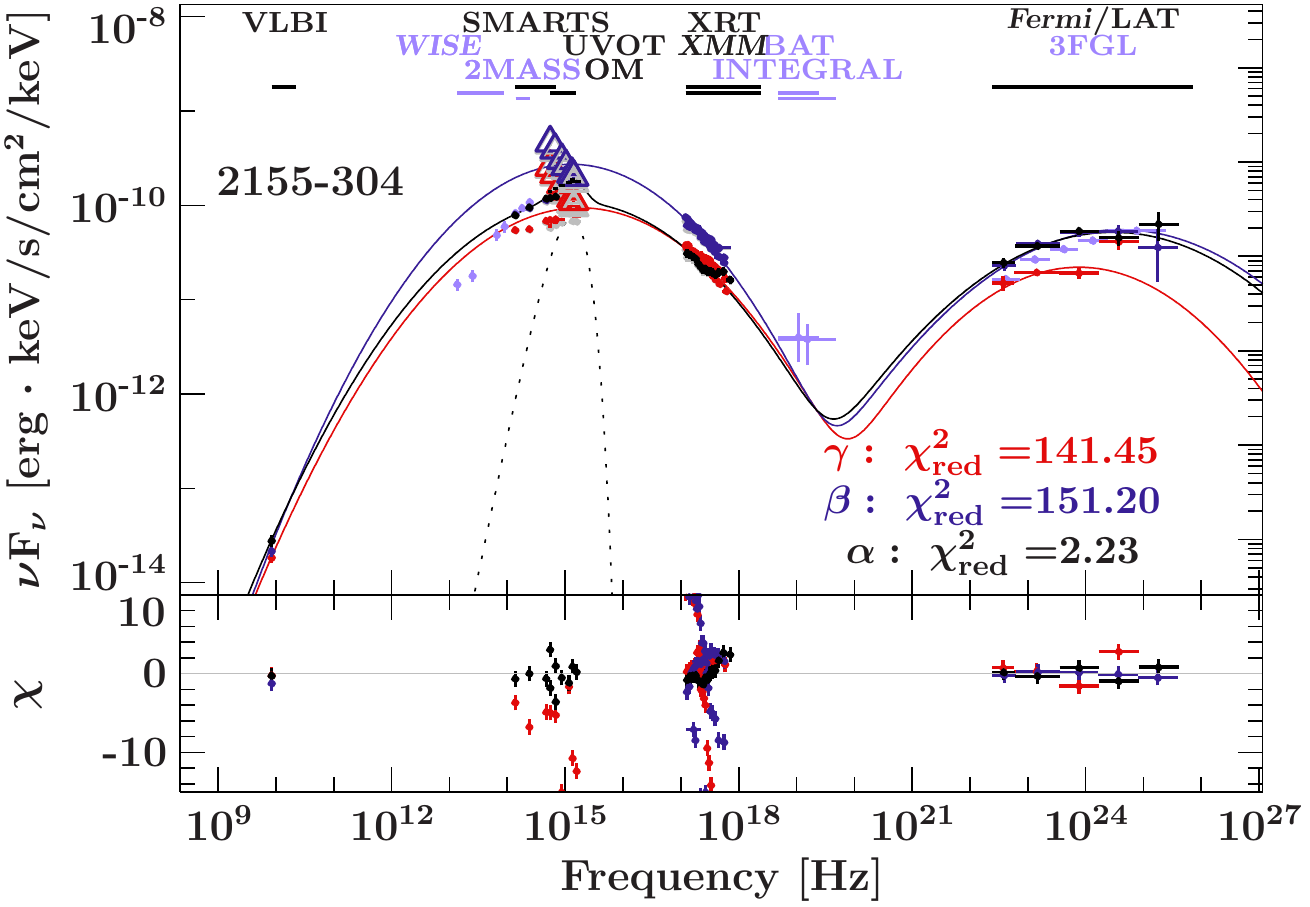}\hfill
\includegraphics[height=0.25\textheight]{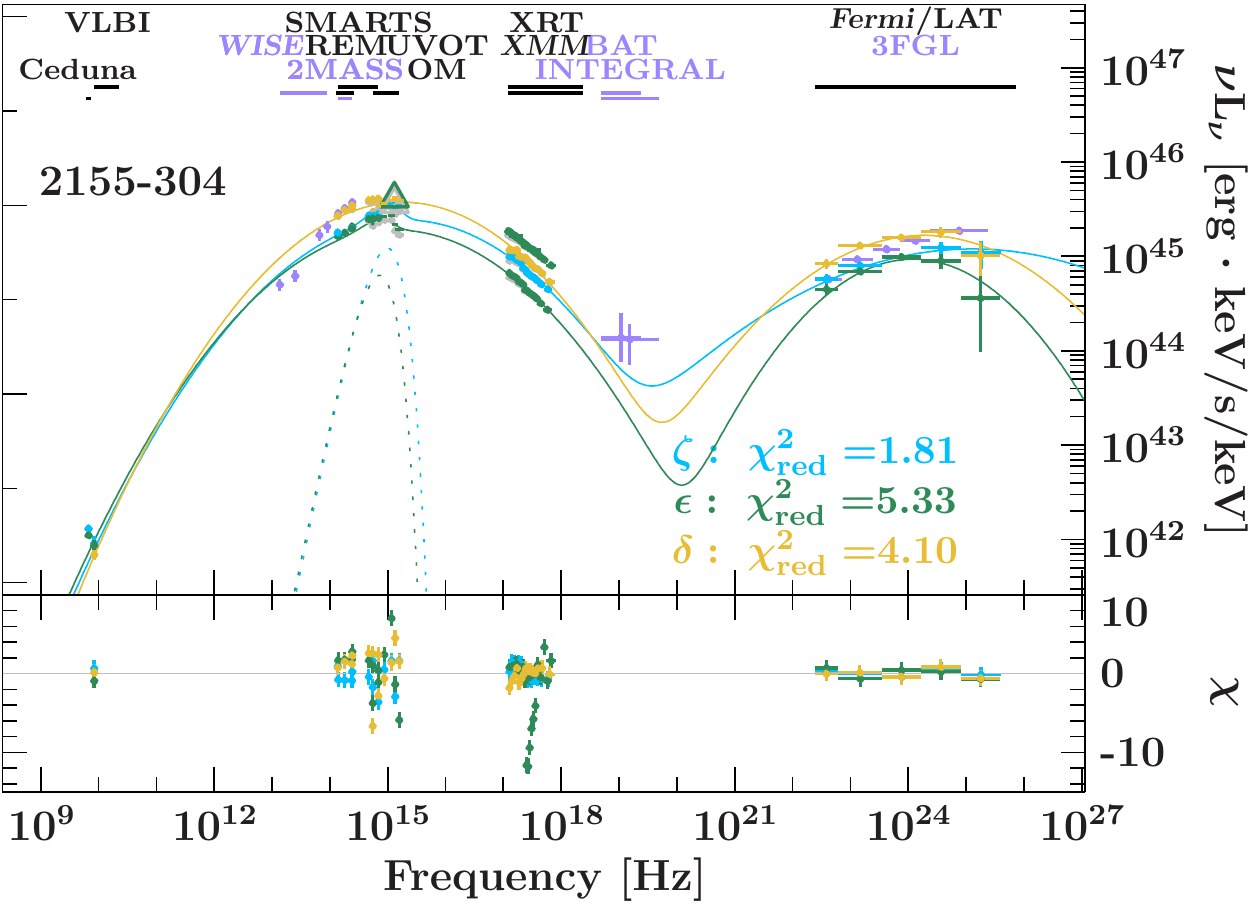}\\
\caption{(contd.)}
\end{figure}
\clearpage

\begin{figure}[h!]
\includegraphics[height=0.25\textheight]{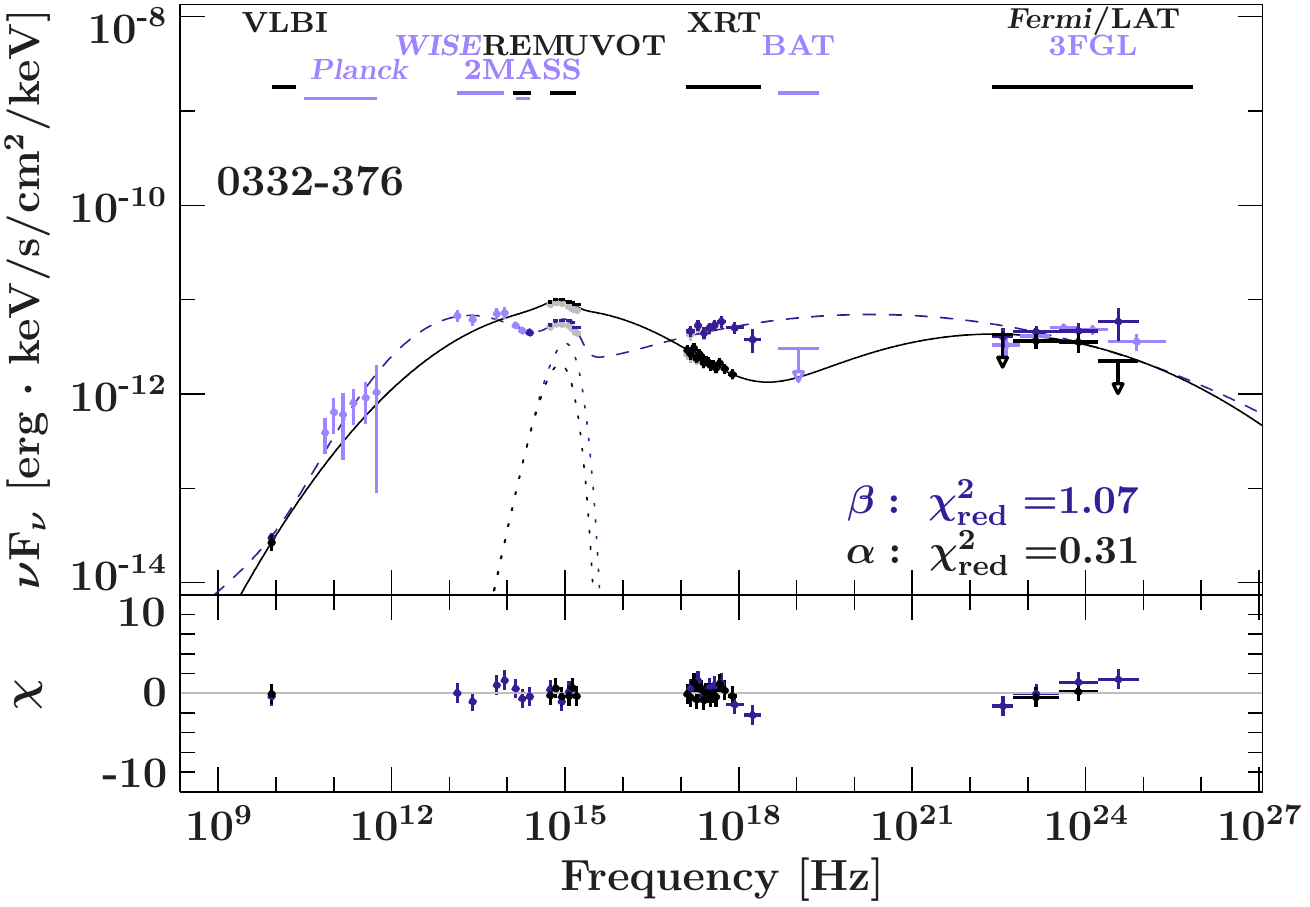}\\[0.5\baselineskip]
\includegraphics[height=0.24\textheight]{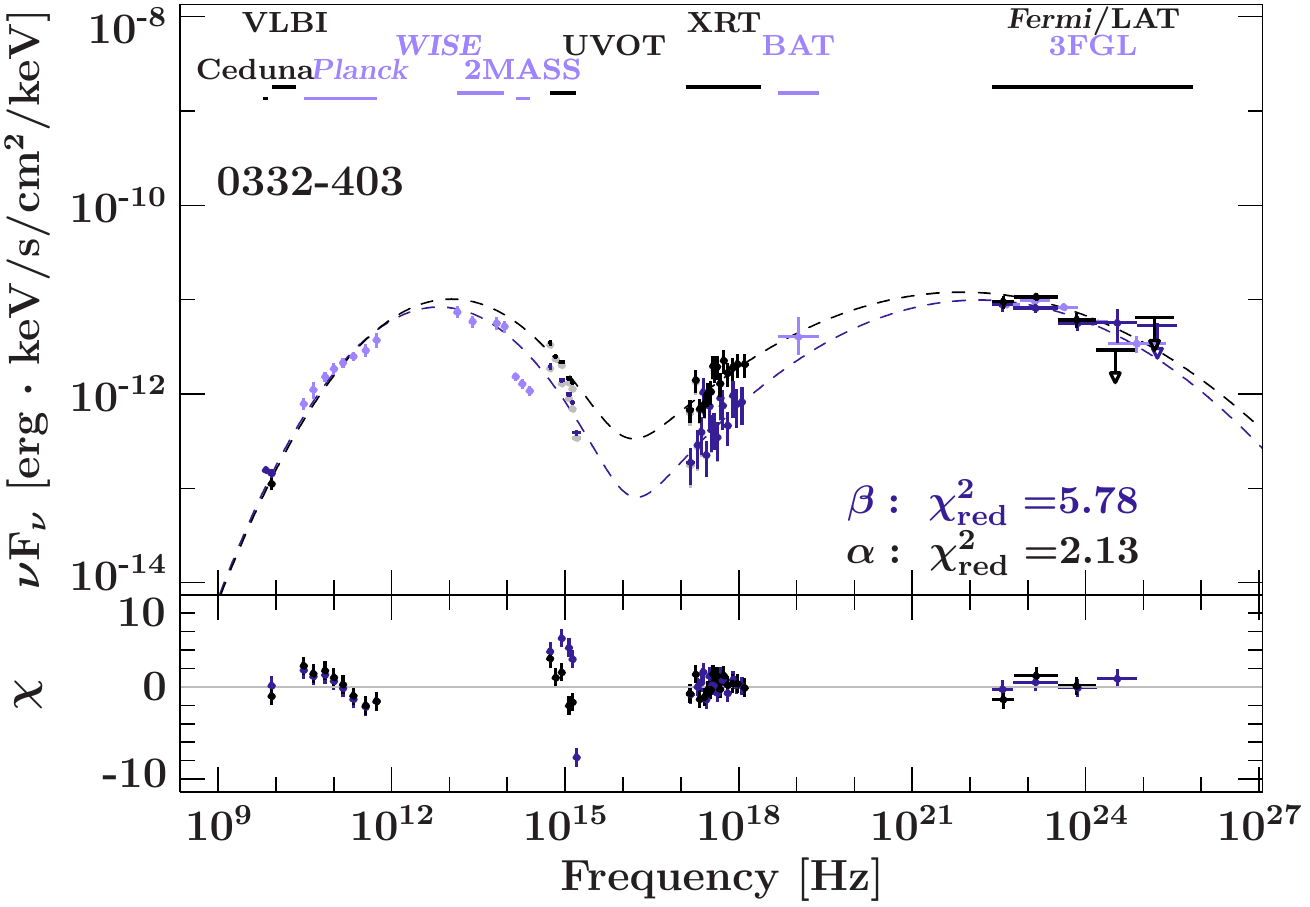}\hfill
\includegraphics[height=0.24\textheight]{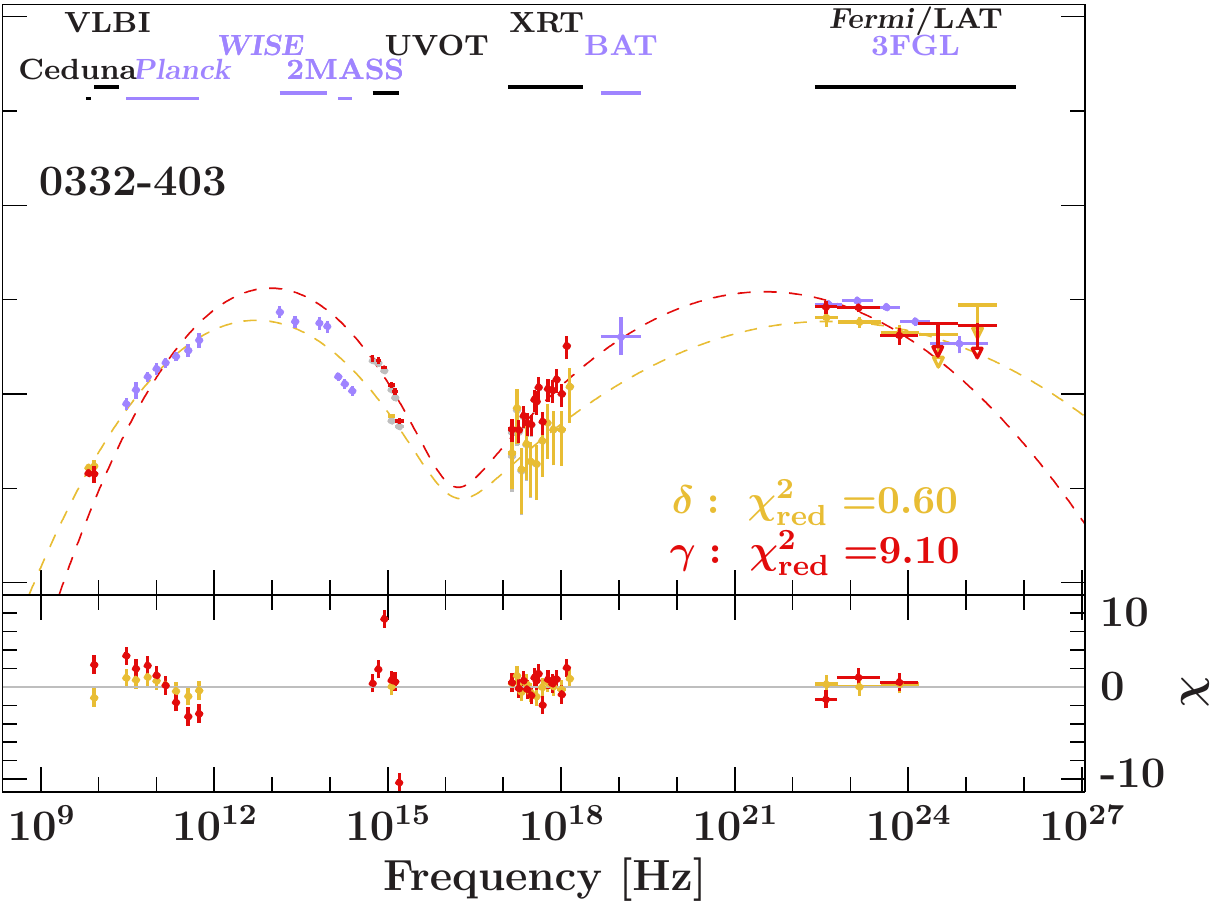}\\
\caption{Broadband spectral energy distributions for both sources
  without a redshift in the loglog $\nu F_\nu$ representation.
  For sources with more than 3 states with sufficient data, the plots
  were split into two parts, to ensure that the SEDs are easily
  visible. Fit models are shown in dashed if archival data had to be
  included in the fit. For sources with a thermal excess in the
  optical/UV, a blackbody was included (dotted). The instruments
  (including their spectral range) are shown above the spectrum. The
  colors correspond to the colors used in the light curve. The best
  fit reduced $\chi^2$ value is shown at the bottom right for
  every state. Residuals are shown in the lower panel. The spectra
  have not been k-corrected.}
\end{figure}

\clearpage
\subsection{Fit results}
\label{ap-fit}

\begin{figure}
\centering
\includegraphics[height=\textheight]{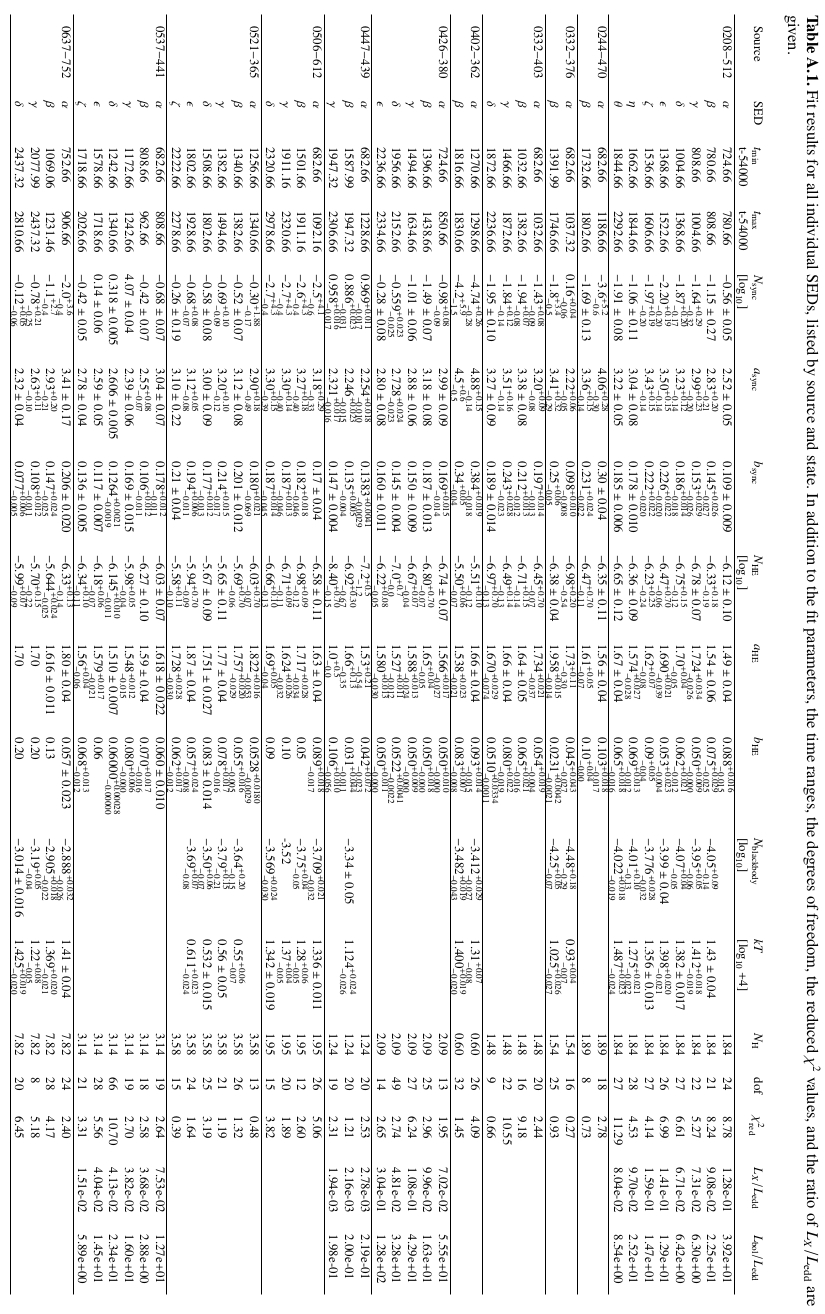} 
\end{figure}

\begin{figure}
\centering
\includegraphics[height=\textheight]{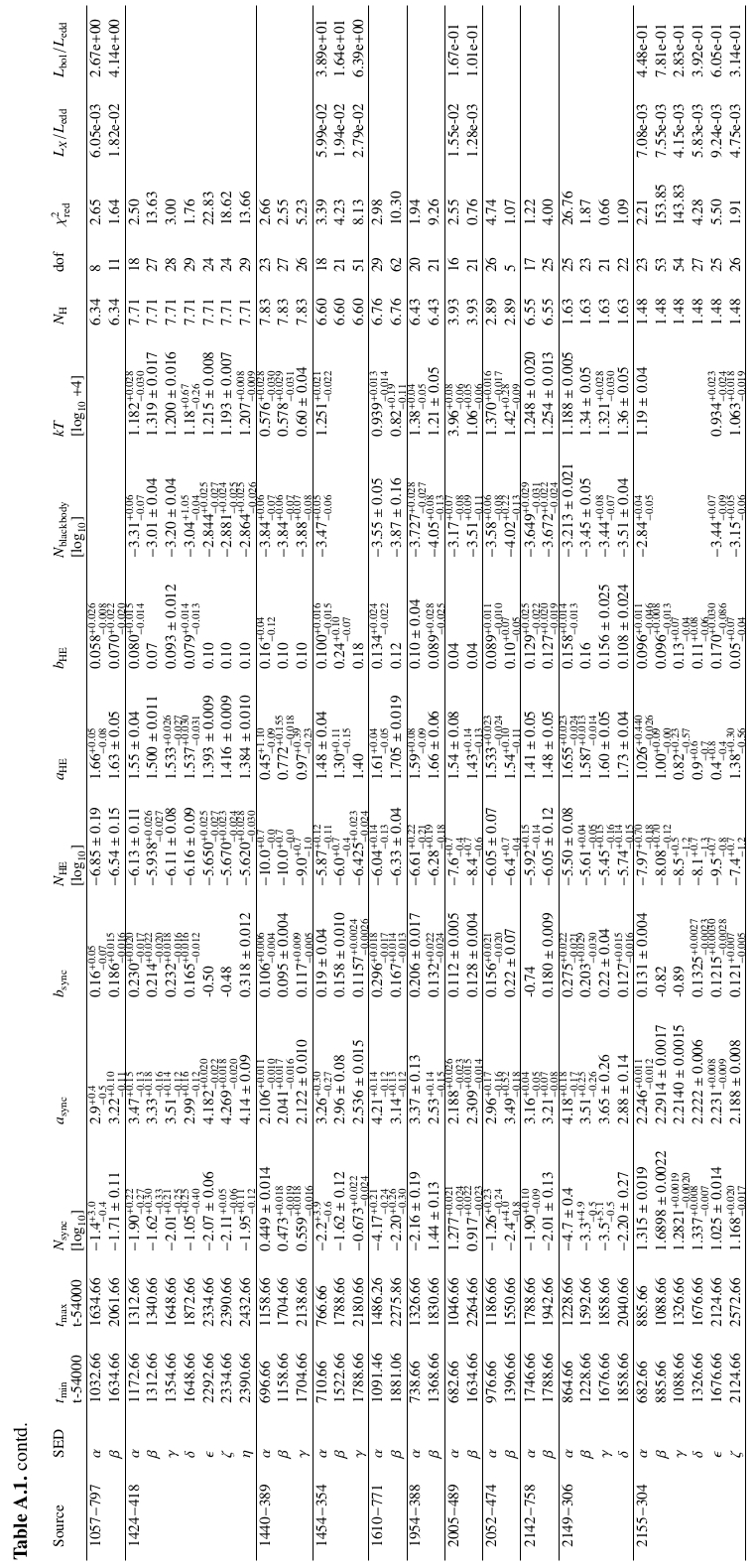} 
\end{figure}

\end{appendix}
\end{document}